\pdfoutput=1
\documentclass[a4paper,fleqn]{sc}
\usepackage[authoryear]{natbib} 
\usepackage{array}
\usepackage{longtable}
\usepackage{bbding}

\usepackage{amssymb}
\usepackage{graphicx}
\usepackage{enumerate}
\usepackage{algorithmic}
\usepackage[ruled, vlined]{algorithm2e}
\usepackage{tikz}
\usepackage{pgfplots}
\usepackage{float}
\usepackage{lipsum}
\usepackage{verbatim}
\pgfplotsset{compat=1.17}
\usepackage{amsmath}
\usepackage{caption}
\usepackage{subcaption}
\usepackage{booktabs}
\usepackage{threeparttable}
\usepackage{multirow}
\usepackage{tikz}
\usepackage{lineno}
\usepackage{xcolor}
\usepackage{bbding}
\usepackage{pifont}

\def\tsc#1{\csdef{#1}{\textsc{\lowercase{#1}}\xspace}}

\tsc{WGM}
\tsc{QE}
\tsc{EP}
\tsc{PMS}
\tsc{BEC}
\tsc{DE}

\begin{document}
\let\WriteBookmarks\relax
\def\floatpagepagefraction{1}
\def\textpagefraction{.001}

\shorttitle{A Multimodal Data Retrieval Platform with Query-aware Feature Representation and Learned Index Based on Data Lake}

\shortauthors{M. Sheng et~al.}

\title [mode = title]{MQRLD: A Multimodal Data Retrieval Platform with Query-aware Feature Representation and Learned Index Based on Data Lake}                      
\tnotemark[1]
\cortext[cor1]{Corresponding author}







\affiliation[1]{organization={School of Computer Science and Technology, Beijing Institute of Technology},
    addressline={Beijing 100081}, 
    country={China}}

\affiliation[2]{organization={BNRist, DCST, RIIT, Tsinghua University},
    addressline={Beijing 100084}, 
    country={China}}

\affiliation[3]{organization={School of Artificial Intelligence, Henan University},
    addressline={Zhengzhou 450046}, 
    country={China}}

\affiliation[4]{organization={School of Artificial Intelligence, University of Chinese Academy of Sciences},
    addressline={Beijing 101408}, 
    country={China}}
    
%
\author[1]{Ming Sheng}[orcid=0009-0002-8813-4558]

\ead{shengming@bit.edu.cn}
\credit{Conceptualization, Methodology, Writing - Original Draft, Writing - Review \& Editing, Formal Analysis, Investigation, Project Administration}

\author[1]{Shuliang Wang}[orcid=0000-0001-5326-7209]
\cormark[1]
\ead{slwang2011@bit.edu.cn}
\credit{Methodology, Writing - Review \& Editing, Funding Acquisition}

\author[2]{Yong Zhang}[orcid=0000-0001-8803-2055]
\cormark[1]
\ead{zhangyong05@tsinghua.edu.cn}
\credit{Conceptualization, Methodology, Writing - Original Draft, Writing - Review \& Editing, Project Administration}

\author%
[3]{Kaige Wang}[orcid=0009-0009-8101-2030]
\ead{wkg@henu.edu.cn}
\credit{Software, Validation, Visualization, Writing - Review \& Editing}

\author%
[1]{Jingyi Wang}
\ead{jwang2024@bit.edu.cn}
\credit{Methodology, Writing - Original Draft, Writing - Review \& Editing, Investigation}

\author%
[1]{Yi Luo}[orcid=0000-0003-0518-1608]
\ead{luoyi@bit.edu.cn}
\credit{Writing - Original Draft, Writing - Review \& Editing, Investigation}

\author%
[4]{Rui Hao}
\ead{haorui24@mails.ucas.ac.cn}
\credit{Writing - Review \& Editing}





\begin{abstract}
Multimodal data has become a crucial element in the realm of big data analytics, driving advancements in data exploration, data mining, and empowering artificial intelligence applications. 
To support high-quality retrieval for these cutting-edge applications, a robust multimodal data retrieval platform should meet the challenges of transparent data storage, rich hybrid queries, effective feature representation, and high query efficiency. However, among the existing platforms, traditional schema-on-write systems, multi-model databases, vector databases, and data lakes, which are the primary options for multimodal data retrieval, make it difficult to fulfill these challenges simultaneously.
Therefore, there is an urgent need to develop a more versatile multimodal data retrieval platform to address these issues.


In this paper, we introduce a \underline{\textbf{M}}ultimodal Data Retrieval Platform with \underline{\textbf{Q}}uery-aware Feature \underline{\textbf{R}}epresentation and \underline{\textbf{L}}earned Index based on \underline{\textbf{D}}ata Lake (\textbf{MQRLD}). 
It leverages the transparent storage capabilities of data lakes, integrates the multimodal open API to provide a unified interface that supports rich hybrid queries, introduces a query-aware multimodal data feature representation strategy to obtain effective features, and offers high-dimensional learned indexes to optimize data query.
We conduct a comparative analysis of the query performance of MQRLD against other methods for rich hybrid queries. 
Our results underscore the superior efficiency of MQRLD in handling multimodal data retrieval tasks, demonstrating its potential to significantly improve retrieval performance in complex environments. 
We also clarify some potential concerns in the discussion.
\end{abstract}


\begin{highlights}
\item The study offers a multimodal data retrieval platform, which supports transparent data storage by using data lake technology, enables rich hybrid queries through the usage of a multimodal open API, and provides a query-aware mechanism to optimize retrieval.
\item The study proposes a multimodal feature representation technique that converts raw multimodal data into representative features and optimal data layouts, improving the effective retrieval of multimodal data.
\item The study introduces a high-dimensional learned index that can adaptively optimize its inner structure, enhancing efficient retrieval of multimodal data.
\end{highlights}

\maketitle

\section{Introduction}\label{section:intro}
Multimodal data, including both structured and unstructured data, continuously influxes in vast volumes from various sources. These data often need to be integrated to extract comprehensive and meaningful insights \citep{intro1}.
This makes "Multimodal Data Retrieval" a core discipline of the more general domain of "Big data" research. For instance, in power systems, various devices continuously collect multimodal data (supervisory control and data acquisition logs, meteorological data, satellite remote sensing images, surveillance videos, etc) from diverse sources, and the growing interactive demand response requires power companies to efficiently extract target data from massive datasets \citep{electronic}. Similarly, large e-commerce platforms like Taobao and Amazon can provide millions of users with billions of items (the products' names, prices, images, promotional videos, etc), and the users usually search for their desired products on these platforms \citep{E-commercerecommendation, heterogeneous, rasappan2024transforming}.
The efficient handling of large-scale multimodal data is crucial for digging deeper insights, such as big data analysis, data exploration and data mining, and driving innovation across diverse domains, including recommendation \citep{wu2024promise}, disease prediction in healthcare \citep{predicting_disease,Thukralknowledge}, fake news detection in law enforcement \citep{xue2021detecting}, etc. 
Moreover, with the evolution and advancement of large-scale models, especially multimodal large-scale language models, AI is endowed with the ability to accept multi-sensory inputs like humans and provide more human-like interactions.

All these applications heavily rely on multimodal data retrieval,  as it can significantly impact their real-world performance.
In multimodal data retrieval, the data can be handled as multimodal objects (MMOs), which consist of complex information that combines structured data (usually represented as numerical attributes) and unstructured data (usually represented as vector features). 
As the real-world scenario shown in Fig \ref{fig1}, multimodal data retrieval platforms are required to support complex joint queries (we call them rich hybrid queries), with the query results returned in the form of MMOs. However, existing multimodal data retrieval platforms, such as traditional schema-on-write systems, multi-model databases, vector databases, and data lakes, exhibit limited capabilities in achieving transparent storage and supporting rich hybrid queries simultaneously. Moreover, their query efficiency and accuracy tend to degrade when handling large-scale datasets. Therefore, the main objective of research on multimodal data retrieval platforms is to develop a platform that can efficiently and effectively retrieve MMOs within large-scale multimodal data through rich hybrid queries along with the concern of transparent data storage \citep{Big-data-analytics}. To achieve this, retrieval platforms must overcome the following challenges:


\begin{figure}[h!]
\begin{center}
\includegraphics[width=0.8\linewidth]{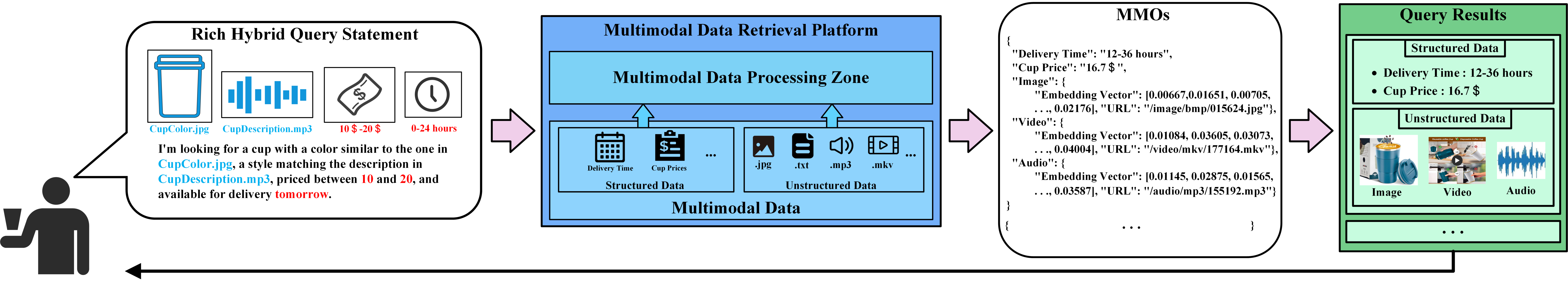}
\caption{Querying in multimodal data, including structured query attributes ("10\$-20\$" and "0-24 hours") and vector query attributes ("CupColor.jpg" and "CupDescription.mp3").} \label{fig1}
\end{center}
\end{figure}

(1)\textit{Storage layer: Transparent data storage}. Users usually aim to retrieve target MMOs through query statements. As shown in Fig \ref{fig1}, the user is more concerned with obtaining all the information of his desired products (MMOs), which includes the products' names, prices, images, promotional videos, etc. However, multimodal data often involves heterogeneous sources, varied modalities such as text, images, and videos, and complex interrelationships among different types of information, lacking a fixed organization or predefined format. This makes it challenging to store and access these data in a seamless manner without requiring users to understand the underlying complexities \citep{multimodal_data}. To address this limitation, a multimodal data retrieval platform should enable the conversion and tracking of information within MMOs, thereby supporting transparent data storage.

(2)\textit{Query layer: Rich hybrid queries and query-aware mechanism}. The richness of multimodal data exploration and mining depends largely on the query functionality, while the retrieval performance relies on the ability of the platform to perceive query behaviors. Compared to the hybrid queries proposed in existing research \citep{wei2020analyticdb}, which support queries containing a single vector and a set of structured attributes, many real-world scenarios require more complex joint queries. These queries are referred to as rich hybrid queries, which support more flexible combinations of a broader spectrum of "attributes", whether they can be structured or vector attributes (as detailed in Section \ref{section:hybridQuery}).
For instance, in Fig \ref{fig1} rich hybrid query statement, "priced between 10 and 20" and "delivered tomorrow" serve as structured query attributes, while the "CupColor.jpg" and "CupDescription.mp3" as vector query attributes. Additionally, real-world query behavior is often skewed rather than average. To enhance the performance of multimodal data retrieval, platforms must also be aware of these query behaviors. Therefore, a multimodal data retrieval platform should support rich hybrid queries and be capable of perceiving query behaviors to provide effective and efficient responses.

(3) \textit{Feature representation layer: Effective feature representation}. 
A common approach to multimodal data retrieval is to  "perceive" and "understand" multimodal data by representing it as features and obtaining data with corresponding features \citep{multimodal_retrieval}. Given the vast amount of multimodal data, where different types emphasize distinct information, flexible feature embedding methods and unified measurement techniques are essential to accommodate varying datasets. Additionally, query behaviors often prioritize specific features, impacting retrieval precision and recall. This requires a data-aware and query-aware feature enhancement approach to dynamically adjust data layouts. Therefore, a multimodal data retrieval platform should have a unified framework for feature embedding, measurement, and enhancement to achieve effective feature representation.

(4)\textit{Index layer: High query efficiency}. 
Efficient multi-dimensional and high-dimensional indexes are a key solution for supporting efficient rich hybrid queries on MMOs. Therefore, a multimodal data retrieval platform needs an indexing structure capable of handling rich hybrid queries that include both vector features and structured attributes. Moreover, different users usually have diverse query patterns. 
For instance, on an e-commerce platform, new parents typically search for baby products, whereas for students, their queries are typically focused on study supplies. Consequently, designing a high-dimensional indexing structure tailored for rich hybrid queries, along with mechanisms to adaptively update the index structure based on query workloads to reduce computational costs, has become a significant challenge.


For fulfilling the storage and retrieval functions for multimodal data, data platforms have been evolving to meet the growing need for inferring meaningful insights from these data. 
Existing retrieval platforms can be categorized into traditional schema-on-write systems, multi-model databases, vector databases, and data lakes. 
(1) \textit{Traditional schema-on-write} systems such as PostgreSQL \citep{obe2017postgresql} and MySQL \citep{mysql}, store data in predefined schemas and are commonly used for structured data management and transactional operations. These systems are often inflexible when handling multimodal data, as they are constrained by their fixed schema, making it difficult to realize transparent storage and support complex hybrid queries.
(2) \textit{Multi-model databases} have been proposed to integrate several schema models within a single, integrated backend, allowing them to handle several data formats within one unified system \citep{Multi-model_Databases, X-stor}. For example, MongoDB \citep{mongodb} can offer transparent storage for certain data modalities, but its storage capabilities are limited when dealing with wilder range of multimodal data, such as video and audio. 
Additionally, their functionality and efficiency are constrained when handling rich hybrid queries, especially for high-dimensional data.
(3) \textit{Vector databases} are designed specifically for efficient retrieval of high-dimensional vector data through vector similarity searches\citep{vdbms}. These platforms, such as Faiss \citep{Faiss}, are specialized in managing vector data, which is often used for machine learning or AI-powered applications. However, they typically focus only on vector queries and neglect the original multimodal data storage and feature representation process. While highly effective for tasks involving a single type of data (e.g., image retrieval or document search), vector databases may struggle with rich hybrid queries that involve multiple data attributes, such as a combination of vectors and structured attributes \citep{vdbms2}.
(4) \textit{Data lakes}, such as Apache Hudi \citep{Hudi}, are large repositories that transparently store data in its original, raw format sourced from diverse origins\citep{datalake2018,hai2datalake}. Data lakes are commonly seen as a cost-effective solution for storing vast amounts of raw data, and support adding additional computational layers, but most of them lack support for rich hybrid queries\citep{HMDFF,MHDP,MHDML}. 
Furthermore, the efficiency of queries in data lakes can be compromised due to insufficient attention to feature representation and indexing of raw data. 
Therefore, in this study, we explore how to leverage the potential of the data lake to propose a more full-fledged multimodal data retrieval platform. 
In addition to providing infrastructural transparent storage and rich hybrid queries, we focus on improving the effectiveness and efficiency of the data retrieval platform.

For improving the effectiveness of multimodal data retrieval, feature representation is critical. Effective feature representation involves converting raw data into representative features and optimal data layout. Therefore, a retrieval platform must convert data into optimal representative features through feature embedding and measurement, followed by feature enhancement to optimize the data layout. However, most existing multimodal data retrieval platforms rely on fixed methods for feature embedding and layout adjustments, failing to integrate feature embedding, measurement, and enhancement into a unified framework for achieving effective feature representation. Specifically, some progresses have been made in feature embedding, measurement, and enhancement separately.
Current research has proposed various embedding methods,  ranging from single-modality embedding methods like doc2vec \citep{doc2vec} and graph2vec \citep{graph2vec} to advanced multi-modality embedding techniques such as CLIP \citep{clip}. Existing feature measurement and enhancement methods primarily focus on optimizing features for building machine learning models, whereas feature measurement techniques aim to identify features best suited for model construction, and feature enhancement methods, such as imputation \citep{imputation}, discretization \citep{Discretization}, and feature creation \citep{creation,falcon2022feature}, target at improving model training performance. 
Although these methods perform well on specific scenarios, they are not well-suited for enabling more effective retrieval on multimodal data retrieval platforms.

To improve the efficiency of multimodal data retrieval, the index must be query-aware and support rich hybrid queries. Current indexing methods mainly include traditional multi-dimensional indexes, vector similarity indexes, and multi-dimensional learned indexes.
(1) \textit{Traditional multi-dimensional indexes} such as grid file \citep{grid-file} and R-tree \citep{R_tree} are designed to efficiently support numeric queries, particularly on low-dimensional data, while having difficulties when faced with rich hybrid queries and lacked query-aware mechanism. 
(2) \textit{Vector similarity indexes} such as ANNOY \citep{annoy} and HNSW \citep{HNSW}, excel in the realm of high-dimensional vector data indexing through distance computation to find similar vectors. 
Nevertheless, they cannot handle rich hybrid queries performed on MMOs, and increasing data records can significantly decrease their query efficiency. (3) \textit{Multi-dimensional learned indexes} extends traditional multi-dimensional structures by employing machine learning models to quickly find expected results \citep{high_learened_index}. 
This method offers a query-aware mechanism for indexes to optimize their internal structures by learning data layout and query behaviors. 
But still, current works in learned indexes primarily focus on efficiently proving low-dimensional queries, failing to achieve rich hybrid queries. 
Current data retrieval platforms typically achieve rich hybrid queries by constructing two separate indexes: a high-dimensional vector index for unstructured data and a multi-dimensional index for structured data \citep{wei2020analyticdb}. 
So far, exploring an efficient query method that supports rich hybrid queries with a query-aware mechanism, remains a challenge yet to be fully resolved.

Therefore, we summarize the key problems of multimodal data retrieval platforms to be solved in this paper as:
\begin{itemize}
\item[1. ]
How to achieve transparent data storage and rich hybrid queries, where user queries involve flexible combinations of structured and unstructured attributes, and the final results are returned as MMOs?
\item[2. ]
How to implement a unified feature representation technique for feature embedding, measurement, and enhancement to convert multimodal data into optimal features and data layouts, thereby improving the effectiveness of query results?
\item[3. ]
How to implement a query-aware high-dimensional index that manages massive vectors and structured data, supporting efficient multimodal data retrieval?
\end{itemize}

To address these problems, we present a multimodal data retrieval platform with query-aware feature representation
and learned index based on data lake (MQRLD), which supports transparent data storage and rich hybrid queries empowered by our high-dimensional learned index that benefits from effective multimodal data feature representation and a query-aware mechanism to deliver optimal performance. 
The main contributions are as follows:
\begin{enumerate}
\itemsep=0pt
\item We provide a seamless and flexible access method for multimodal data, including the utilization of data lake technology to support transparent multimodal data storage, and the usage of multimodal open API to support rich hybrid queries which return query results in the form of MMOs, based on which a query-aware mechanism is built to enhance query performance. (Section \ref{section:backbone})
\item We propose a unified multimodal data feature representation technique, which is designed to convert multimodal raw data into representative features and optimal data layout by perceiving and understanding both data characteristics and query behaviors, in order to facilitate index building, thereby enhancing the query effectiveness. (Section \ref{section:dataPreparation})
\item We present an efficient high-dimensional learned index built on feature representation that can adaptively optimize its inner structure under different feature data layouts and various task scenarios to support flexible and efficient data retrieval of multimodal data. (Section \ref{section:highDimensional})
\item We conduct extensive performance evaluations using real and synthetic datasets. The results demonstrate that MQRLD outperforms other methods in multimodal data retrieval tasks, showing its ability to significantly enhance retrieval performance in complex environments. (Section \ref{section:experiments})
\end{enumerate}

\section{Related Work} \label{RelatedWork}
In view of the challenges mentioned above, we summarize the current works on data retrieval platforms, feature representation techniques, and typical multi-dimensional and high-dimensional indexes used in data retrieval platforms. 
We first enumerate data retrieval platforms in Section \ref{section:RWplatform}, focusing on architecture and functionality which include storage, query options, feature representation functionality, and index type. 
Then in Section \ref{section:RWfr}, we introduce the existing techniques in feature representation domain. 
Finally, we conduct a more detailed comparison of multi-dimensional and high-dimensional indexes in Section \ref{section:RWindex}, focusing on Structure and Functionality, which include index structure, query types, and whether data-aware or query-aware techniques are applied to improve index efficiency.

\subsection{Related Work on Data Retrieval Platform}\label{section:RWplatform}
Table \ref{table:DB_platforms} lists existing data retrieval platforms that support multimodal data retrieval. 
We can classify them into traditional schema-on-write systems, multi-model databases, vector databases, and data lakes. 

\begin{table}
\caption{Comparison of data retrieval platforms on Type, Storage Layer, Query Layer, Feature Representation Layer, and Index Layer.}
\label{table:DB_platforms}
\scriptsize 
\begin{tabular*}{\tblwidth}{m{45pt}<{\centering}m{45pt}<{\centering}m{35pt}<{\centering}m{25pt}<{\centering}m{35pt}<{\centering}m{30pt}<{\centering}m{35pt}<{\centering}m{40pt}<{\centering}m{60pt}<{\centering}}
\toprule
\multicolumn{2}{c}{~} & Storage Layer & \multicolumn{2}{c}{Query Layer} & \multicolumn{3}{c}{Feature Representation Layer} & Index Layer\\
\cmidrule(lr){3-3}
\cmidrule(lr){4-5}
\cmidrule(lr){6-8}
\cmidrule(l){9-9}

Name & Type & Transparent Data Storage & Rich Hybrid Queries & Query\-aware Mechanism & Embedding & Measurement &  Enhancement & Index \\ 
\midrule
Ours & data lake& \large\ding{51} & \large\ding{51} & \large\ding{51} & \large\ding{51} & \large\ding{51} & \large\ding{51} & high-dimensional learned index 
\\ 
\midrule
PostgreSQL \citep{obe2017postgresql} & schema-on-write system & \XSolidBrush &
\textcolor{black}{\large\ding{51}}\textsuperscript{\textcolor{black}{\kern-1.1em\small\ding{55}}}
& \XSolidBrush & \XSolidBrush & \XSolidBrush & \XSolidBrush & multi-dimensional index/ vector similarity index
\\ 
\midrule
ArangoDB \\ \citep{ArangoDB} & multi-model database & \textcolor{black}{\large\ding{51}}\textsuperscript{\textcolor{black}{\kern-1.1em\small\ding{55}}} & \textcolor{black}{\large\ding{51}}\textsuperscript{\textcolor{black}{\kern-1.1em\small\ding{55}}} & \XSolidBrush & \XSolidBrush & \XSolidBrush & \XSolidBrush & multi-dimensional index
\\ 
\midrule
Azure Cosmos DB \citep{CosmosDB} & multi-model database & \textcolor{black}{\large\ding{51}}\textsuperscript{\textcolor{black}{\kern-1.1em\small\ding{55}}} & \textcolor{black}{\large\ding{51}}\textsuperscript{\textcolor{black}{\kern-1.1em\small\ding{55}}} & \XSolidBrush & \XSolidBrush & \XSolidBrush & \XSolidBrush & multi-dimensional index
\\ 
\midrule
OrientDB \citep{orientdb} & multi-model database & \textcolor{black}{\large\ding{51}}\textsuperscript{\textcolor{black}{\kern-1.1em\small\ding{55}}} & \textcolor{black}{\large\ding{51}}\textsuperscript{\textcolor{black}{\kern-1.1em\small\ding{55}}} & \XSolidBrush & \XSolidBrush & \XSolidBrush & \XSolidBrush & multi-dimensional index
\\ 
\midrule
MongoDB \citep{mongodb} & multi-model database & \textcolor{black}{\large\ding{51}}\textsuperscript{\textcolor{black}{\kern-1.1em\small\ding{55}}} & \textcolor{black}{\large\ding{51}}\textsuperscript{\textcolor{black}{\kern-1.1em\small\ding{55}}} & \XSolidBrush & \XSolidBrush & \XSolidBrush & \XSolidBrush & multi-dimensional index
\\ 
\midrule
Faiss \citep{Faiss} & vector database & \XSolidBrush  & \XSolidBrush & \XSolidBrush & \XSolidBrush & \XSolidBrush & \XSolidBrush & vector similarity index \\ 
\midrule
Pinecone \citep{Pinecone} & vector database & \XSolidBrush & \XSolidBrush & \XSolidBrush & \XSolidBrush & \XSolidBrush & \XSolidBrush & vector similarity index \\ 
\midrule
ADAMpro \\ \citep{Giangreco2016} & vector database &\XSolidBrush & \XSolidBrush & \XSolidBrush & \large\ding{51} & \XSolidBrush & \XSolidBrush & vector similarity index  \\ 
\midrule
Cottontail DB \citep{CottontailDB} & vector database & \XSolidBrush & \textcolor{black}{\large\ding{51}}\textsuperscript{\textcolor{black}{\kern-1.1em\small\ding{55}}} & \XSolidBrush & \large\ding{51} & \XSolidBrush & \XSolidBrush & vector similarity index 
\\ 
\midrule
Milvus \citep{wang2021milvus} & vector database & \XSolidBrush & \textcolor{black}{\large\ding{51}}\textsuperscript{\textcolor{black}{\kern-1.1em\small\ding{55}}} & \XSolidBrush & \large\ding{51} & \XSolidBrush & \XSolidBrush & vector similarity index
\\ 
\midrule
HMDFF \citep{HMDFF} & data lake & \large\ding{51} & 
\textcolor{black}{\large\ding{51}}\textsuperscript{\textcolor{black}{\kern-1.1em\small\ding{55}}} 
& \XSolidBrush & \large\ding{51} & \XSolidBrush & \XSolidBrush & multi-dimensional index  \\ 
\midrule
MHDP \citep{MHDP} &data lake & \large\ding{51} & \XSolidBrush & \XSolidBrush & \large\ding{51} & \XSolidBrush & \XSolidBrush & N/A \\ 
\midrule
MHDML \citep{MHDML} & data lake & \large\ding{51} & \XSolidBrush & \XSolidBrush & \XSolidBrush & \XSolidBrush & \XSolidBrush & N/A \\ 
\bottomrule
\end{tabular*}
\end{table}

Traditional schema-on-write systems require a fixed schema for data storage, organize data in columns, and utilize traditional one-dimensional or multi-dimensional indexes for data retrieval. 
This structure is well-suited for managing structured data, for example, PostgreSQL performs effectively in supporting bank transactions \citep{cloud_post}. 
However, this fixed schema approach is less adaptable to high-dimensional data. 
Although plugins like pgvector \citep{pgvector} have been introduced to handle high-dimensional vector data processing and similarity search, they are unable to deliver the high performance and scalability required for large-scale applications.

Multi-model databases aim to accommodate different data types by designing separate schema models (e.g., graph model for network data, document model for JSON and XML data, key-value model for ID-identified data, etc.).
While this targeted design provides centralized management for various data types, it lacks generality and struggles to accommodate the increasing variety of data modalities. 
Furthermore, although most multi-model databases can support simple hybrid queries such as text content search combined with numeric search, they are typically confined to specific data modalities. 
Examples include ArangoDB \citep{ArangoDB}, OrientDB \citep{orientdb}, and CosmosDB \citep{CosmosDB}. 
Due to the absence of feature representation functionality, these databases are incapable of supporting semantic searches across different data modalities, thus limiting their effectiveness in multimodal data retrieval.

Standard vector databases, such as Faiss \citep{Faiss} and Pinecone \citep{Pinecone}, focus solely on the storage and retrieval of vector-format data and use only vector similarity indexes. 
This limitation prevents them from executing rich hybrid queries. 
Extensions like ADAMpro \citep{Giangreco2016}, Cottontail DB \citep{CottontailDB}, and Milvus \citep{wang2021milvus} offer additional functionalities, such as simple hybrid queries(e.g. a structured attribute and a vector feature) and feature embedding. 
However, they exhibit significant flaws: 1. Firstly, they neglect the storage of raw data and need to artificially distinguish between the processing of different data types.
2. Secondly, they lack MMO management processing, making it explicit to track back original data.
3. Thirdly, despite providing basic feature embedding functionality, they cannot guarantee the quality of features, thus impacting search effectiveness.
4. Finally, although they can support simple hybrid queries, this comes at the expense of significantly reduced query performance.
This heavy query process leads to exponential degradation in performance as the data records increase, making it difficult to handle large-scale data. 

Data lake \citep{datalake2010}, introduced around 2010, is more flexible for different types of analyses, as no decisions regarding the modeling and processing of data have to be made in advance. 
Due to their low operational costs, high scalability, and flexibility, distributed file systems or object storages, such as the Hadoop Distributed File System (HDFS) \citep{hdfs}, are commonly employed within data lakes for storing raw data \citep{lakehouseSurvey}. 
However, this flexibility comes at the cost of lower robustness, as the raw data can barely be validated on ingestion. 
In addition, most data lakes lack structural management of raw data and effective index structure, limiting search performance for large data volumes. 
Furthermore, many works built on data lake are often tailored for domain-specific data, such as HMDFF \citep{HMDFF}, MHDP \citep{MHDP}, and MHDML \citep{MHDML} focus on medical or healthcare data, lacking generality. 

Existing data retrieval platforms perform well within their specific application areas; however, none of them are likely to solve all the challenges related to multimodal data retrieval. 
Moreover, none of these platforms improve query performance using query-aware or learned index techniques, resulting in suboptimal query efficiency.

\subsection{Related Work on Feature Representation}\label{section:RWfr}
To achieve effective multimodal data retrieval, a multimodal data retrieval platform needs to incorporate advanced multimodal feature representation techniques.
Multimodal feature representation can be categorized into three stages: feature embedding, feature measurement, and feature enhancement. Each stage involves different techniques and methods, addressing distinct aspects of feature representation. Feature embedding and measurement convert multimodal data into representative features, which facilitate subsequent tasks on the platform. Feature enhancement optimizes the layout of these representative features, thereby improving query effectiveness across different workloads.

Although some platforms support a variety of embedding methods, these methods are typically chosen manually and may not be the most suitable for some datasets or query workloads. Furthermore, these platforms often fail to apply feature measurement and enhancement techniques to improve feature fidelity and generalization, limiting the robustness and adaptability of their embedding methods across different retrieval scenarios. While existing multimodal data retrieval platforms lack a unified technique for multimodal feature representation, researchers have made some progress in feature embedding, measurement, and enhancement separately.

The existing feature embedding technique embeds the semantic information from raw data into a unified representation space, allowing data from different modalities to be compared and analyzed within the same space.
Feature embedding has been widely studied and has evolved from simple methods like BoW and TF-IDF to advanced techniques such as Word2Vec \citep{word2vec}, GloVe \citep{Glove}, and CLIP \citep{clip}, providing diverse options across different data modalities. 
These developments have enabled practical applications like YouTube's content recommendation \citep{Youtub}. 

Feature measurement involves the assessment and evaluation of multimodal feature embedding. Currently, feature measurement is mainly used to assess the quality, importance, and relevance of embedding features in predictive models. Commonly used metrics like Matthews Correlation Coefficient (MCC), F-value, and variance threshold evaluate a feature’s ability to distinguish between classes, while metrics such as accuracy and recall reflect the overall performance of predictive models.

Feature enhancement refers to the further optimization and enhancement of existing embeddings, aiming to improve their expressive power or adaptability to downstream tasks. Key techniques include dimensionality reduction methods (e.g., principal component analysis (PCA), t-SNE \citep{t_NSE}, Linear Discriminant Analysis (LDA)), scaling, imputation, discretization, and feature creation.

However, existing feature embedding, measurement, and enhancement techniques still face challenges in meeting the demands for accuracy and efficiency in multimodal data retrieval tasks. 
For embedding methods, some are general-purpose but lack generalization, while others, although more accurate, are specific to certain data types, making it challenging to adapt them to a variety of multimodal data retrieval tasks. Similarly, many existing measurement metrics oversimplify the evaluation of multimodal data embedding models and overlook the specific requirements of query tasks. Furthermore, current measurement methods often neglect the high fidelity needed in feature representations, which is crucial for capturing the intricate relationships between multimodal data. Regarding feature enhancement techniques, many are not tailored specifically to the multimodal data retrieval scenario. Methods primarily focused on improving clustering algorithms may enhance retrieval efficiency but can be limited by clustering performance bottlenecks. These approaches frequently disregard optimization at the data level and fail to address how data layout and query-specific information nature can improve clustering performance. Therefore, a multimodal data retrieval platform requires a multimodal feature representation technique which can convert multimodal data into representative features and optimal data layouts.


\subsection{Related Work on Indexes for Multimodal Data Retrieval}\label{section:RWindex}
Table \ref{table:index} lists typical indexes used in multimodal data retrieval. 
We can categorize them into vector similarity indexes, traditional multi-dimensional indexes, and multi-dimensional learned indexes.

Vector similarity indexes treat data as points in hyperspace and return the top-k nearest neighbors by comparing "distance" between points. 
They are primarily classified into four categories: table-based index (e.g., E2 LSH \citep{E2LSH}, IVFADC \citep{IVFADC}), tree-based index (e.g., FLANN \citep{FLANN}, RPTree \citep{RPTree}), graph-based index (e.g., KNNGraph \citep{KGraph}, HNSW) and hybrid index (e.g., DB\_LSH). 
Although these indexes are effective for vector queries, they struggle to search combinations of structured attributes and vector features and lack data-aware and query-aware mechanisms, thus limiting their effectiveness in querying MMOs and under different query scenarios. Moreover, their query performance declines sharply with an increase in data records.

Traditional multi-dimensional indexes, such as tree structures (R-tree), grid structures (Grid File), and SFC (space-filling curve)-based structures (z-order \citep{z-order}), are designed for structured data queries on multi-dimensions, for example, looking up a location on a map by its X and Y coordinates. 
Their effectiveness is confined to queries of structured attributes on small datasets, as they are functionally unable to support complex queries and cannot adjust their index structure according to query workload and data layout.

Multi-dimensional learned indexes have emerged as a prominent topic in recent years. 
\citet{learned-index} proposed that an index can be treated as a model that learns to locate the expected result, making it particularly suitable for accelerating retrieval in large datasets. 
Specifically, multi-dimensional learned indexes build upon traditional multi-dimensional index structures and construct a learning model to enhance or replace the inner structure of the index by learning data layout (e.g., LISA \citep{LISA}, RSMI \citep{RSMI}) and query workload patterns (e.g., Qd-tree \citep{Qd-tree}, Flood \citep{Flood}, Tsunami \citep{Tsunami}). 
Although some multi-dimensional learned indexes support numeric and vector queries separately (e.g., ML \citep{Ml-index}, LISA), they do not support rich hybrid queries that simultaneously search for several numeric and vector data.
Moreover, while these indexes demonstrate good performance on multi-dimensional data, they still suffer from the "curse of dimensionality" when applied to higher-dimensional datasets. Additionally, these indexes struggle to balance the trade-off between prediction model accuracy and complexity. High-accuracy models tend to become more complex, resulting in increased time and space costs for index construction.

Consequently, the widely used multi-dimensional and high-dimensional indexes in existing data retrieval platforms still have limitations in supporting multimodal data retrieval, including:
(1)\textit{Lack of versatile functionality of rich hybrid query.}
Existing indexes fail to support rich hybrid query processing. 
Although some multi-dimensional learned indexes accommodate multiple query types, they execute these queries sequentially, rather than simultaneously. These indexes typically create separate index structures for high-dimensional vector data and multi-dimensional structured data, preventing them from effectively handling the complex demands of multimodal queries.
(2) \textit{Inefficient performance.}
As shown in Table \ref{table:index}, most indexing methods lack data-aware and query-aware capabilities, which prevents them from maintaining stable performance across different datasets and query workloads, leading to decreased query efficiency.  Although some index structures can be manually monitored and adjusted to fit varying query demands, this dependence on human intervention is impractical and error-prone, especially in dynamic or large-scale environments where query workloads frequently change.
In contrast, learned indexes can adjust their structure based on data-aware and query-aware mechanisms. However, they still suffer from the curse of dimensionality when dealing with high-dimensional data, rendering them unsuitable for multimodal data retrieval scenarios. Therefore, a multimodal data retrieval platform requires a unified query-aware high-dimensional index that supports rich hybrid queries.

\begin{table}
\caption{Comparison of multi-dimensional and high-dimensional indexes on structure and functionality.}
\label{table:index}
\scriptsize 
\begin{tabular*}{\tblwidth}{m{2.25cm}<{\centering}m{0.8cm}<{\centering}m{1cm}<{\centering}m{1cm}<{\centering}m{1cm}<{\centering}m{1cm}<{\centering}m{1cm}<{\centering}m{1cm}<{\centering}m{1cm}<{\centering}m{1cm}<{\centering}m{1cm}<{\centering}}
\toprule
 & Structure & Learned & Multi-dimensional Index & High-dimensional Index & Data-aware & Query-aware & Numeric Query & Vector Query & Hybrid Query & Rich Hybrid Query \\
\midrule
Ours& Tree & \large\ding{51} & \XSolidBrush & \large\ding{51} & \large\ding{51} & \large\ding{51} &\large\ding{51}&\large\ding{51} &\large\ding{51} &\large\ding{51} \\
\midrule
E2 LSH \citep{E2LSH} & Table & \XSolidBrush & \XSolidBrush & \large\ding{51} & \XSolidBrush & \XSolidBrush & \XSolidBrush & \large\ding{51} & \XSolidBrush & \XSolidBrush  \\
\midrule
IVFADC \citep{IVFADC} & Table & \XSolidBrush & \XSolidBrush & \large\ding{51} & \XSolidBrush & \XSolidBrush & \XSolidBrush & \large\ding{51} & \XSolidBrush & \XSolidBrush  \\
\midrule
FLANN \citep{FLANN} & Tree & \XSolidBrush & \XSolidBrush & \large\ding{51} & \XSolidBrush & \XSolidBrush & \XSolidBrush & \large\ding{51} & \XSolidBrush & \XSolidBrush \\
\midrule
RPTree \citep{RPTree} & Tree & \XSolidBrush & \XSolidBrush & \large\ding{51} & \XSolidBrush & \XSolidBrush & \XSolidBrush & \large\ding{51} & \XSolidBrush & \XSolidBrush \\
\midrule
KGraph \citep{KGraph} & Graph & \XSolidBrush & \XSolidBrush & \large\ding{51} & \XSolidBrush & \XSolidBrush & \XSolidBrush & \large\ding{51} & \XSolidBrush & \XSolidBrush \\
\midrule
HNSW \citep{HNSW} & Graph & \XSolidBrush & \XSolidBrush & \large\ding{51} & \XSolidBrush & \XSolidBrush & \XSolidBrush & \large\ding{51} & \XSolidBrush & \XSolidBrush \\
\midrule
DB-LSH \citep{DB_LSH} & Table, Tree & \large\ding{51} & \XSolidBrush & \large\ding{51} & \XSolidBrush & \large\ding{51} & \XSolidBrush & \large\ding{51} & \XSolidBrush & \XSolidBrush \\
\midrule
ZM \citep{ZM} & SFC & \large\ding{51} & \large\ding{51} & \XSolidBrush & \XSolidBrush & \XSolidBrush & \large\ding{51} & \XSolidBrush & \XSolidBrush & \XSolidBrush \\
\midrule
ML \citep{Ml-index} & Grid & \large\ding{51} & \large\ding{51} & \XSolidBrush & \XSolidBrush & \XSolidBrush & \large\ding{51} & \large\ding{51} & \XSolidBrush & \XSolidBrush \\
\midrule
LISA \citep{LISA} & Grid & \large\ding{51} & \large\ding{51} & \XSolidBrush & \large\ding{51} & \XSolidBrush & \large\ding{51} & \large\ding{51} & \XSolidBrush & \XSolidBrush \\
\midrule
RSMI \citep{RSMI} & Grid, SFC & \large\ding{51} & \large\ding{51} & \XSolidBrush & \large\ding{51} & \XSolidBrush & \large\ding{51} & \XSolidBrush & \XSolidBrush & \XSolidBrush \\
\midrule
Qd-tree \citep{Qd-tree} & Tree & \large\ding{51} & \large\ding{51} & \XSolidBrush & \large\ding{51} & \large\ding{51} & \large\ding{51} & \XSolidBrush & \XSolidBrush & \XSolidBrush \\
\midrule
Flood \citep{Flood} & Grid & \large\ding{51} & \large\ding{51} & \XSolidBrush & \large\ding{51} & \large\ding{51} & \large\ding{51} & \XSolidBrush & \XSolidBrush & \XSolidBrush \\
\midrule
Tsunami \citep{Tsunami} & Grid, Tree & \large\ding{51} & \large\ding{51} & \XSolidBrush & \large\ding{51} & \large\ding{51} & \large\ding{51} & \XSolidBrush & \XSolidBrush & \XSolidBrush \\
\midrule
COAX \citep{COAX} & Grid, Tree & \large\ding{51} & \large\ding{51} & \XSolidBrush & \large\ding{51} & \large\ding{51} & \large\ding{51} & \XSolidBrush & \XSolidBrush & \XSolidBrush \\
\midrule
SPRIG \citep{SPRIG} & Grid & \large\ding{51} & \large\ding{51} & \XSolidBrush & \large\ding{51} & \large\ding{51} & \large\ding{51} & \XSolidBrush & \XSolidBrush & \XSolidBrush \\
\midrule
PAW \citep{PAW} & Grid & \large\ding{51} & \large\ding{51} & \XSolidBrush & \large\ding{51} & \large\ding{51} & \large\ding{51} & \XSolidBrush & \XSolidBrush & \XSolidBrush\\
\midrule
LIMS \citep{LIMS} & Cluster & \large\ding{51} & \large\ding{51} & \XSolidBrush & \large\ding{51} & \XSolidBrush & \XSolidBrush & \large\ding{51} & \XSolidBrush & \XSolidBrush \\
\midrule
LMSFC \citep{LMSFC} & SFC & \large\ding{51} & \large\ding{51} & \XSolidBrush & \large\ding{51} & \large\ding{51} & \large\ding{51} & \XSolidBrush & \XSolidBrush & \XSolidBrush \\
\midrule
ELSI \citep{ELSI} & Tree & \large\ding{51} & \large\ding{51} & \XSolidBrush & \large\ding{51} & \large\ding{51} & \large\ding{51} & \XSolidBrush & \XSolidBrush & \XSolidBrush \\
\bottomrule
\end{tabular*}
\end{table}

\section{Framework}
\begin{figure}[pos=h!]
    \centering
    \includegraphics[width=0.6\textwidth]{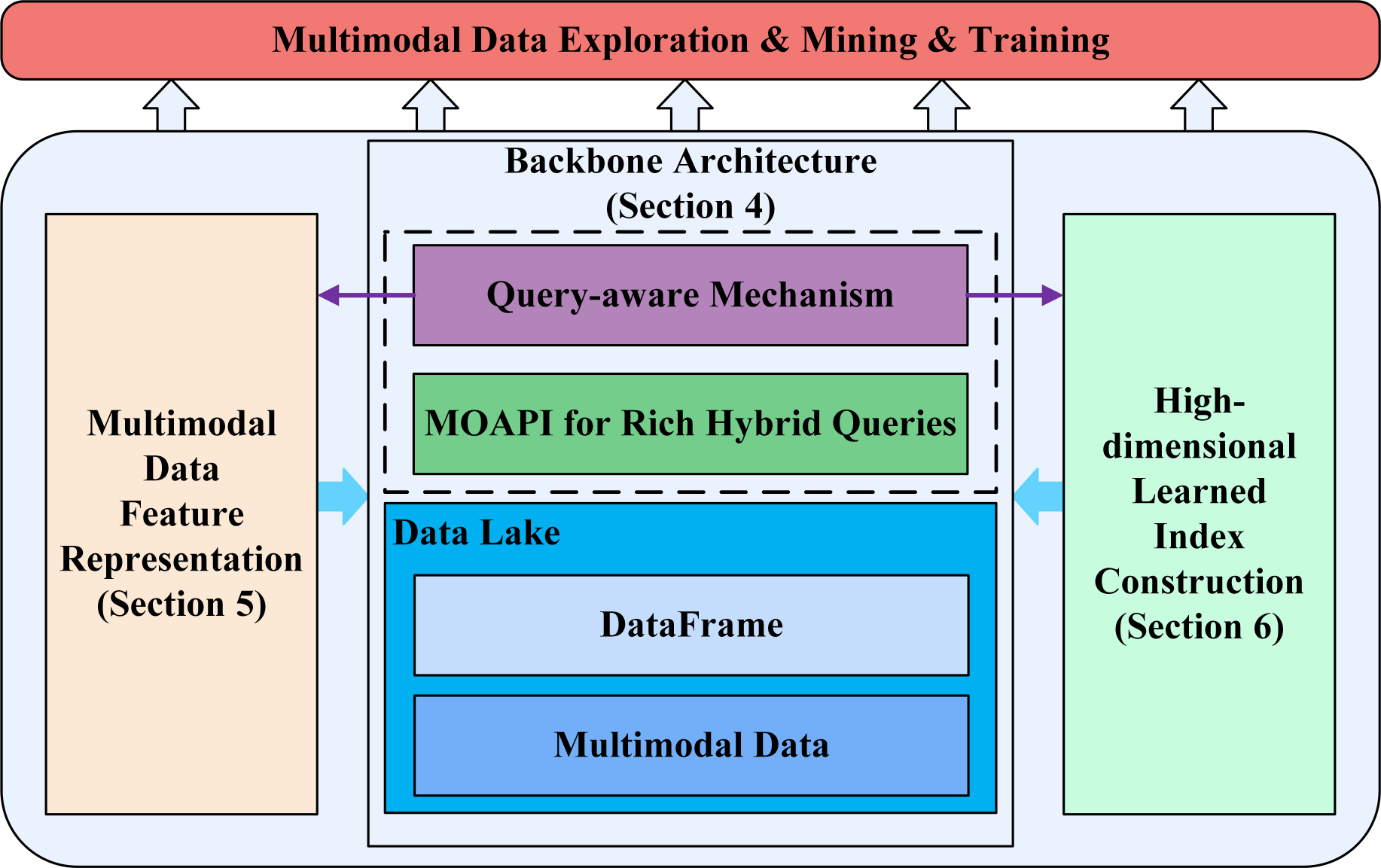}
    \caption{MQRLD framework.}
    \label{fig2}
\end{figure}

The MQRLD platform fulfills efficiently and effectively multimodal data retrieval by supporting transparent data storage and rich hybrid queries. 
It leverages our innovative multimodal data feature representation technique to enhance the learned index structure, thereby improving query performance. 
As illustrated in Fig \ref{fig2}, the overall framework comprises three core modules: the backbone architecture (Section \ref{section:backbone}), a unified multimodal data feature representation technique (Section \ref{section:dataPreparation}), and query-aware high-dimensional learned index supporting rich hybrid queries(Section \ref{section:highDimensional}). 
The backbone architecture not only provides a standardized storage format and query interface for multimodal data but also preserves quantified records for multimodal query behavior.
To further improve retrieval operations, we implement two key methods to flesh out the backbone architecture. 
First, we employ a multimodal data feature representation technique to convert multimodal raw data into an optimal format suitable for indexing. 
Second, we design a query-aware high-dimensional learned index structure to efficiently retrieve MMOs.

\begin{figure}[pos=h!]
    \centering
    \includegraphics[width=0.8\textwidth]{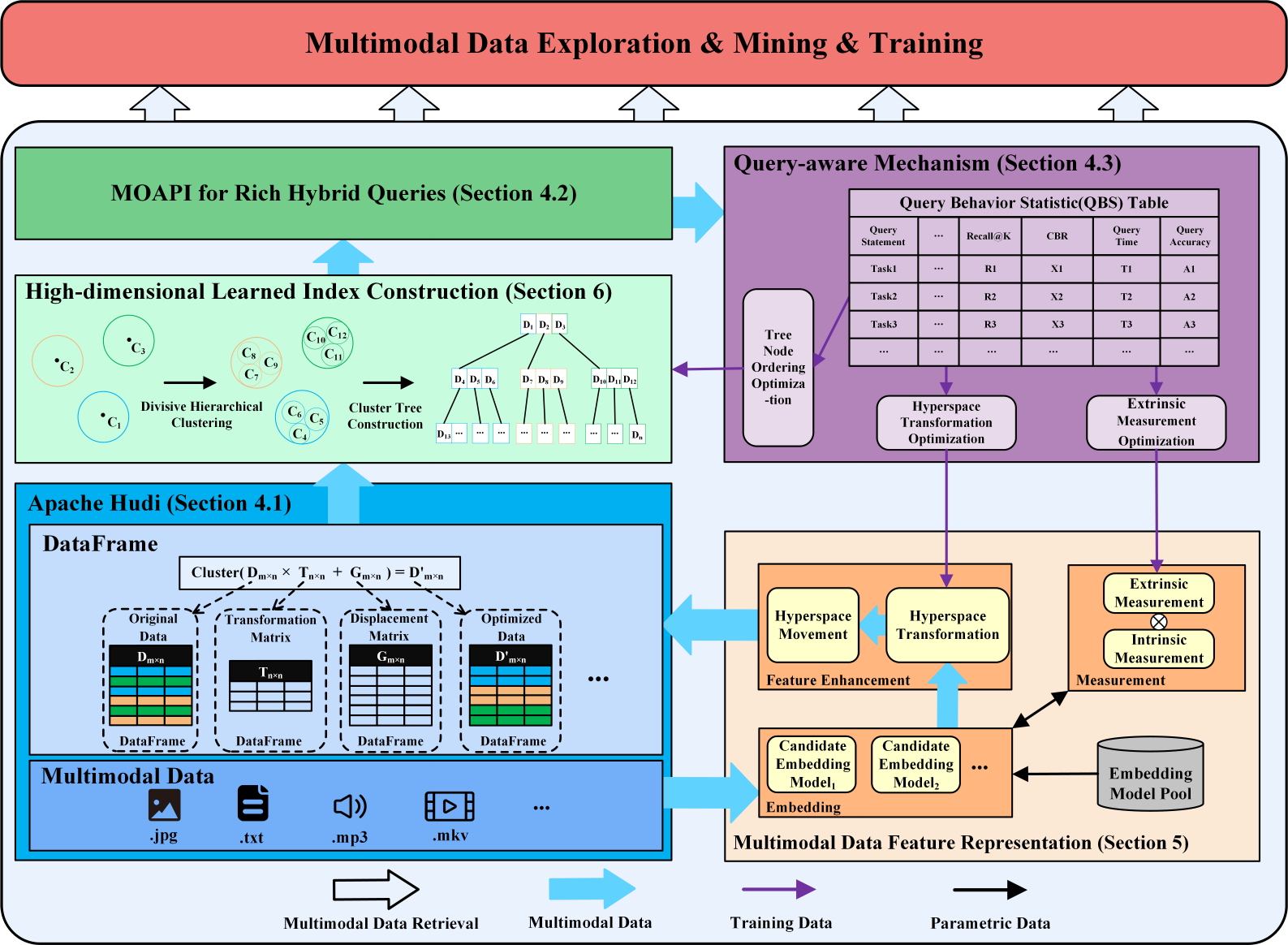}
    \caption{MQRLD workflow.}
    \label{fig3}
\end{figure}

Fig \ref{fig3} shows the main highlight and overall workflow of MQRLD. 
We use a data lake to seamlessly store and manage multimodal data, serving as the backend storage of our platform (Section \ref{section:storage}). 
Building upon it, we provide multimodal open API (MOAPI) to offer a unified query interface for rich hybrid queries (Section \ref{section:hybridQuery}). 
While executing each query, we record a series of meaningful statistic data from query behaviors to build the query-aware mechanism (Section \ref{section:queryAware}), which is used to refine subsequent feature representation (Section \ref{section:dataPreparation}) and high-dimensional learned index construction (Section \ref{section:highDimensional}). 
The multimodal data feature representation (Section \ref{section:dataPreparation}) acts as a bridge between multimodal raw data and data retrieval, facilitating the conversion of multimodal raw data into representative features through steps of feature embedding, measurement, and enhancement. 
The embedding and measurement (Section \ref{section:encoding&eval}) aim at discerning representative features for multimodal data on different scenarios. 
The embedded features then undergo the enhancement process (Section \ref{section:enhance}), which can be seen as an optimization process including transformation and movement in hyperspace based on data patterns and query behaviors, resulting in optimal features and data layout. 
Finally, the high-dimensional learned index construction (Section \ref{section:highDimensional}) fully utilizes the feature representation results to build a cluster tree structure to support rich hybrid query (Section \ref{section:indexConstruction}). To further enhance performance, we optimize index's inner structure based on query behaviors (Section \ref{section:IndexOptimize}), making the index efficient across different query workloads for rich hybrid queries. 

\section{MQRLD Backbone Architecture}\label{section:backbone}
The MQRLD backbone architecture, based on a data lake, functionally supports multimodal data transparent storage, rich hybrid queries, and query behavior recording. 
We introduce transparent storage for multimodal data in Section \ref{section:storage}, MOAPI for rich hybrid queries in Section \ref{section:hybridQuery}, and statistic table for query-aware mechanism in Section \ref{section:queryAware}.
\subsection{Transparent Storage for Multimodal Data}\label{section:storage}
Multimodal data retrieval requires storing large amounts of data, each with different modalities that need to be identified and retrieved.
Therefore, the primary challenge of storage lies in how to manage this large-scale multimodal data through a transparent process, ensuring that the final query results are returned in the form of MMOs.

A data lake is a prominent and well-established kind of data platform that stores multimodal data in MMOs, thereby preserving all flexible options for future multimodal data retrieval. 
MQRLD integrates popular open-source data lake Apache Hudi \citep{Hudi}, which can be integrated with Apache Spark \citep{shaikh2019apache}, enabling parallel processing of large-scale data. 
It utilizes DataFrame, which is conceptually equivalent to a table in a relational database, corresponding to a MMOs. 
As shown in Fig \ref{fig4}, DataFrame columns represent the attributes (such as image texture, text semantics...) of the MMO, embedded by different feature vectors. 
Each MMO can be represented as various vectors and numeric data according to different query tasks, allowing MQRLD to facilitate versatile query tasks across different modalities. 
For each attribute, the DataFrame also records the embedding models for unstructured data and the HDFS path of the original raw data, enabling quick tracing back to MMO.

\subsection{Multimodal Open API (MOAPI) for Rich Hybrid Queries}\label{section:hybridQuery}
To achieve our goal of conducting rich hybrid queries on MQRLD, a versatile and easy-to-use query interface is essential. 
The Jina API \citep{Jina} is a well-behaved open API that is specifically designed for interpreting and interacting with multimodal data, providing a uniform interface specification for querying multiple data types. 
Therefore, our MOAPI utilizes Jina API as the query interface and defines four basic query types, including:
\begin{itemize}
\item Numeric Equal (N.E) Query: Returns results that are equal to the given value of an attribute.
\item Numeric Range (N.R) Query: Returns results where the value of an attribute falls within the given range.
\item Vector KNN (V.K) Query: Given a set $P$, a query object $q\in P$, and a positive integer $k$, returns $k$ objects in $P$, denoted as V.K($q$,$k$), such that $\forall p\in \text{V.K}(q,k), p'\in P\backslash \text{V.K}(q,k), dist(q,p)\leq dist(q,p')$.
\item Vector Range (V.R) Query: Given a set $P$, a query object $q\in P$, and a query radius $r\geq 0$, returns all objects in $P$ within the distance $r$ of $q$, i.e., V.R($q$,$r$)=$\{p\in P|dist(p,q)\leq r\}$.
\end{itemize}

\begin{figure}[pos=!h]
    \centering
    \includegraphics[width=0.9\textwidth]{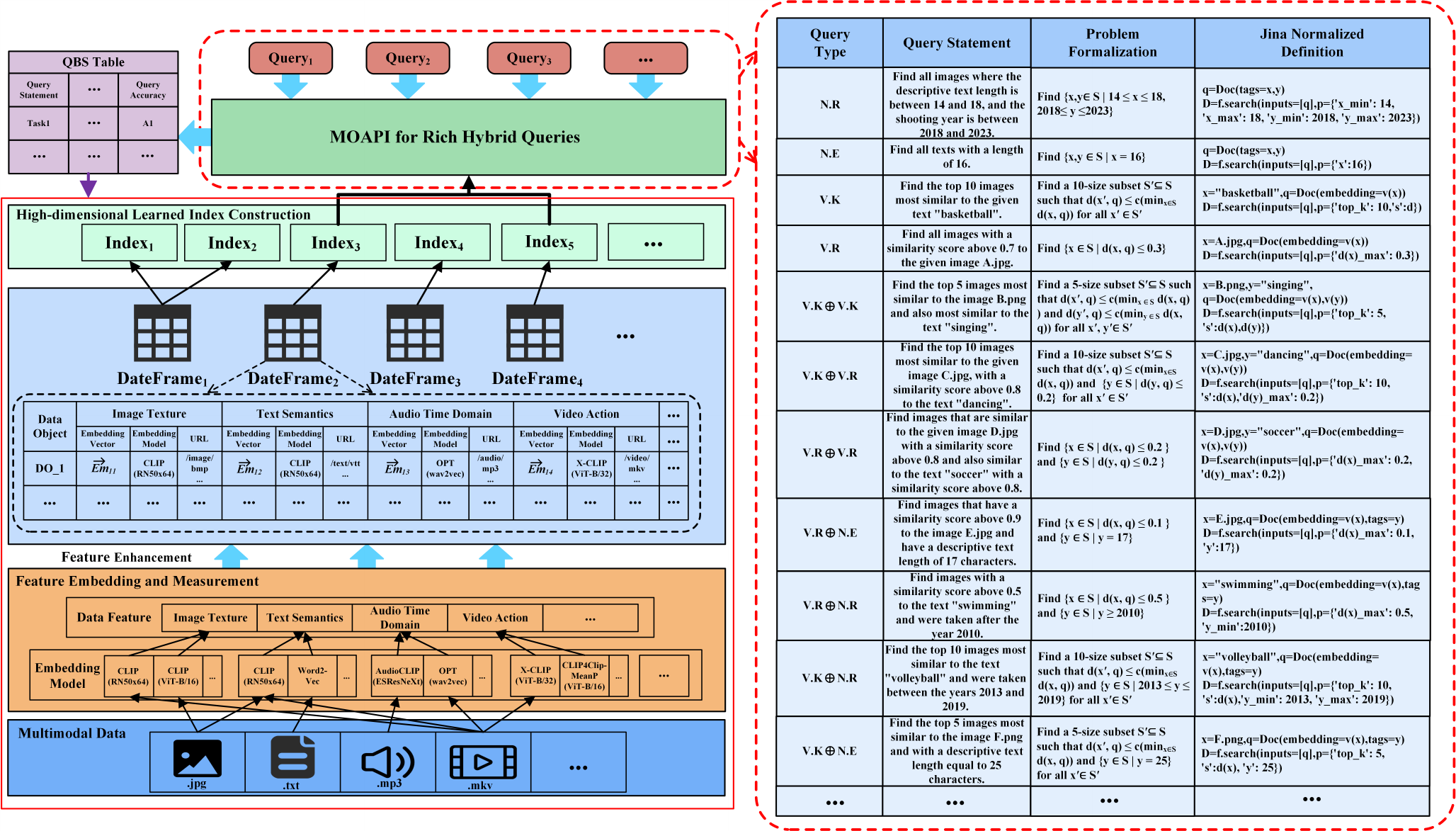}
    \caption{Transparent data storage and rich hybrid queries in MQRLD.}
    \label{fig4}
\end{figure}

Our rich hybrid queries can be defined as combinations of these basic queries, i.e., $q_1\bigoplus q_2\bigoplus...\bigoplus q_n$, where $q_i \in \{\text{N.E, N.R, V.R, V.K}\}$, and $\bigoplus$ is the "combine" operation that can either be $\bigcap$ or $\bigcup$. 
$\bigcap$ denotes an intersection of all the query requirements, e.g., finding images contains a player in red jersey and a player in white jersey, and $\bigcup$ denotes a union of all the query requirements, e.g. finding images contains a player in red jersey or a player in white jersey. 
Fig \ref{fig4} shows some of the typical rich hybrid queries, which can efficiently handle complex data retrieval requirements.


\subsection{Statistic Table for Query-aware Mechanism}\label{section:queryAware}
We establish the query-aware mechanism by building a Query Behavior Statistic (QBS) table and recording a series of insightful variables during the query execution. This query-aware mechanism impacts the optimization of both feature representation and index structure via the QBS table, which plays a vital role in constructing MQRLD through (1) reflecting the performance of query tasks to facilitate feature measurement (Section \ref{section:eval}), (2) identifying the feature of multimodal data representation in different query contexts to improve data layout (Section \ref{section:trans}), and (3) dynamically adjusting and improving index structure to provide faster and more accurate retrieval (Section \ref{section:IndexOptimize}). 
\begin{table}[pos=h!]
\caption{Query behavior statistic table.}
\label{table:qs}
\footnotesize
\renewcommand\arraystretch{1.5}
\belowrulesep=0pt
\aboverulesep=0pt
\begin{tabular*}{\tblwidth}{|m{2cm}<{\centering}|m{2cm}<{\centering}|m{2cm}<{\centering}|m{1.5cm}<{\centering}|m{1.5cm}<{\centering}|m{1cm}<{\centering}|m{1.5cm}<{\centering}|m{1.45cm}<{\centering}|}
\toprule
Query Statement & Query Multimodal Object Set & Query Attributes  & Query Type & Recall@K & CBR & Query Time & Query Accuracy \\
\midrule
Task1 & DF1 & Em1, Em2  & NE,VK & R1 & X1 & T1 & A1 \\
\midrule
Task2 & DF2 & Em3  & VR & R2 & X2 & T2 & A2 \\
\midrule
… & … & … &  …  & … & … & … & … \\
\bottomrule
\end{tabular*}
\end{table}

As shown in Table \ref{table:qs}, the statistic variables recorded in the QBS table are as follows:


\begin{itemize}
\item Query Statement: The input query statement.
\item Query Multimodal Object Set: The DataFrame involved multimodal objects in the query statement.
\item Query Attributes: The columns involved in the query statement. 
In hybrid queries, achieving accurate results often requires utilizing two data attributes, while in rich hybrid queries, we may introduce more data attributes into the query process.
\item Query Type: The four basic query types performed on each attribute.
\item Recall@K: The proportion of correct matches among the top K retrieved results.
\item CBR (Cross Bucket Rate): The evaluation metric for the index effectiveness, akin to the "Inter-bucket Traversal Rate" or "Cross-partition Traversal Rate". 
We utilize the DataFrames partitions to correspond to the "buckets" or "partitions".
\item Query Time: The average query execution time for queries of the specific task.
\item Query Accuracy: The ratio between the query results and the ground truth.
\end{itemize}

The QBS table is populated in two ways: logs from MOAPI, and statistical data from the Spark engine. MOAPI primarily handles query interpretation, supporting a wide range of rich hybrid queries that can be covered by the four basic query types provided. As a result, its logs are comprehensive enough to fully capture these rich hybrid queries' characteristics. From its logs, we can extract detailed information about the Query Statement, Query Multimodal Object Set, Query Attributes, and Query Type. The Spark engine is responsible for executing queries, from which we can obtain statistical information such as Query Time, the number of buckets traversed by the query, query results, and calculated metrics including Recall@K, CBR, and Query Accuracy. The data of new queries can be continuously appended to the QBS table, and different combinations of columns in the QBS table can be used as distinct training datasets, supporting the query-aware feature measurement (Section \ref{section:eval}), feature enhancement (Section \ref{section:trans}), and index optimization (Section \ref{section:IndexOptimize}).

\section{Multimodal Data Feature Representation}\label{section:dataPreparation}
\begin{figure}[pos=!h]
    \centering
    \includegraphics[width=0.8\textwidth]{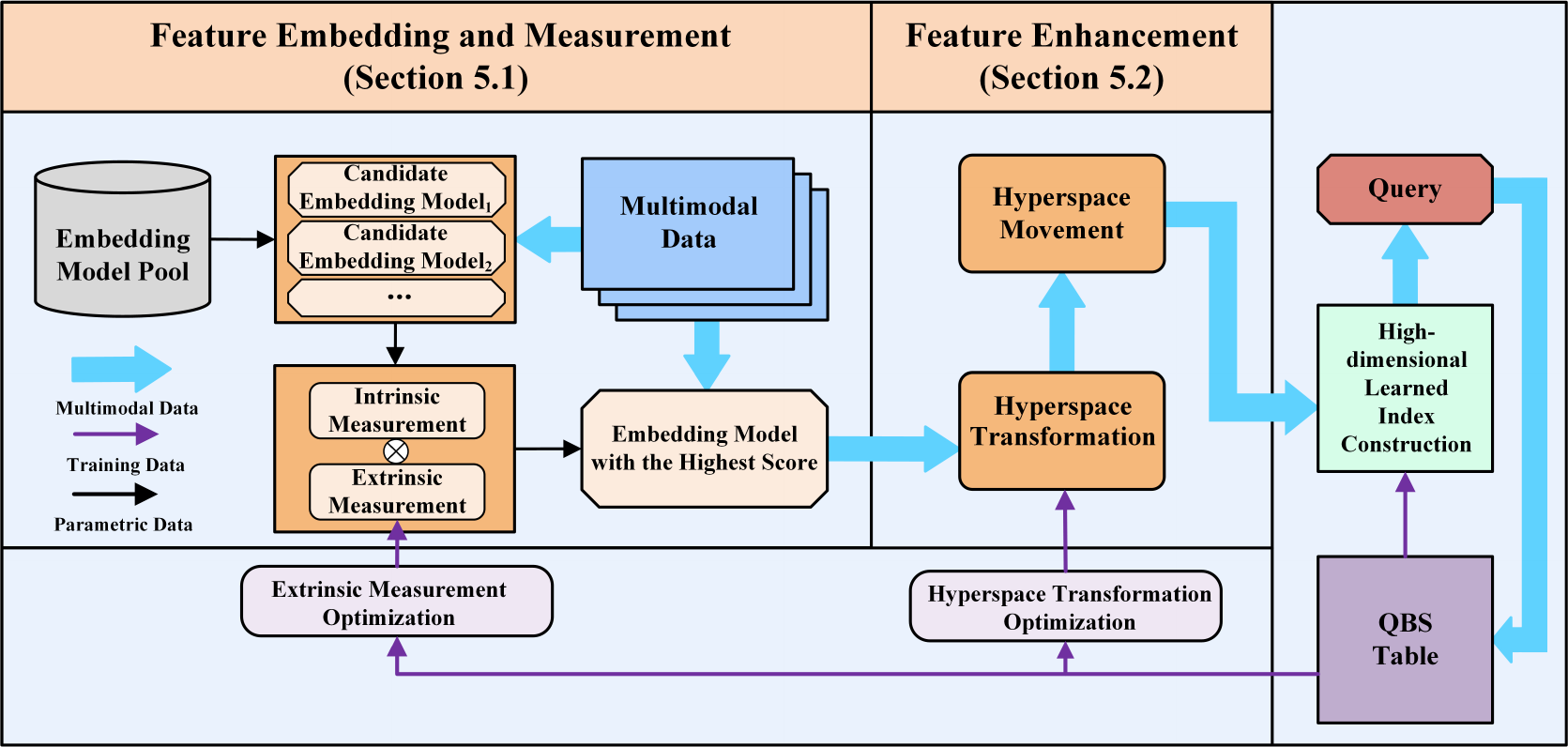}
    \caption{Overview of multimodal data representation, including feature embedding and measurement, feature enhancement, and optimization based on the query-aware mechanism.}
    \label{figure5}
\end{figure}

Current multimodal data retrieval platforms often fail to fully exploit the synergistic capabilities of the three feature representation stages—embedding, measurement, and enhancement—to optimize data itself for the purpose of enhancing query performance. To fill this gap, our objective in feature representation is to derive the most representative features and optimized data layout for multimodal data, based on insights from both the data itself and the queries executed upon it, to refine the quality of the high-dimensional learned index and thereby ultimately enhance query performance.
As shown in Fig \ref{figure5}, this process can be divided into two parts:
(1)\textit{Feature Embedding and Measurement}: We obtain optimal features of multimodal data using our unique measurement process by concerning intrinsic and extrinsic metrics to select the embedding model with the highest score. This serves as the foundation for subsequent optimization of data layout.
(2)\textit{Feature Enhancement}: We optimize overall data layout through hyperspace transformation and movement by analyzing data patterns and learning from query behaviors, making index and rich hybrid queries more effective and efficient.

\subsection{Feature Embedding and Measurement}\label{section:encoding&eval}


\begin{figure}[pos=!h]
    \centering
    \includegraphics[width=.8\textwidth]{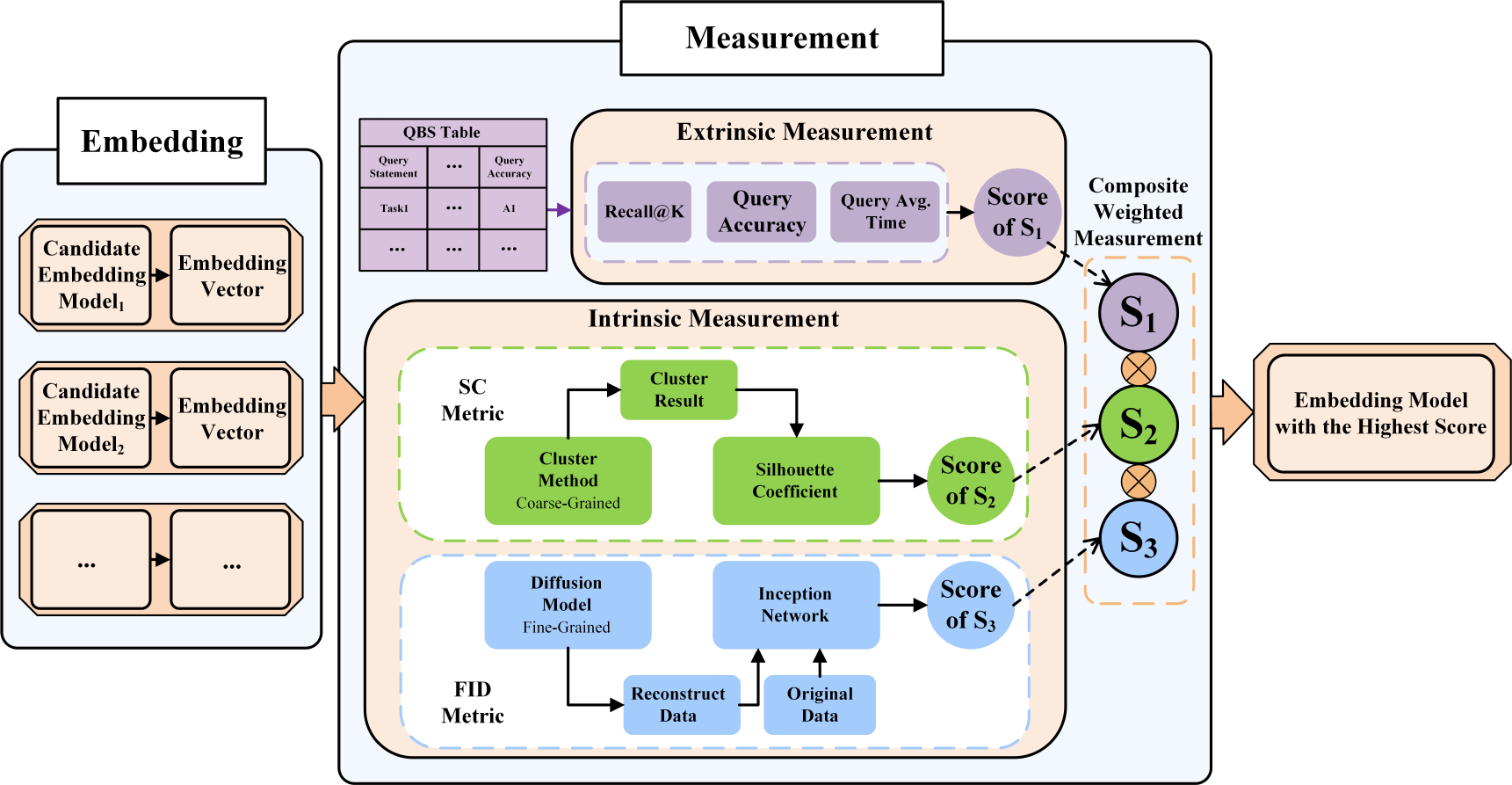}
    \caption{Workflow of feature embedding and measurement. Embedding vector from each model is input into the measurement process consisting of extrinsic and intrinsic metrics to calculate a score, and select the model with the highest score.}
    \label{fig6}
\end{figure}

Feature embedding is the process of converting multimodal raw data into a set of features, while the measurement procedure serves as an assessment to determine the effectiveness of the embedding model, aiding in the selection of the more representative features for subsequent retrieval tasks. 
The overall workflow of feature embedding and measurement is shown in Fig \ref{fig6}. 
In the feature embedding process, MQRLD offers an embedding model pool for embedding the given multimodal raw data. 
The embedded features are then subjected to the measurement process, where we score each model using a data-aware intrinsic metric and a query-aware extrinsic metric to select the embedding model with the highest score.

\subsubsection{Embedding}\label{section:encoding}
In the field of feature embedding domain, numerous existing works have proposed diverse techniques for efficiently representing raw data. 
Table \ref{table:model_pool} lists some of the notable works that have proven successful in multimodal data feature embedding, which we have collected and consolidated into the embedding model pool to support the embedding of wide-ranging attributes of multimodal data. 
Nevertheless, there is still a critical need to select a suitable model for different scenarios, as the performance of these embedding models can vary significantly depending on factors such as data characteristics and query requirements. 
For instance, compared to CLIP4Clip, X-CLIP uses temporal embedding models to capture dynamic actions and interactions over time, making it more suitable for the task of embedding a soccer match video. 
Conversely, CLIP4Clip can effectively capture static, frame-level information, which is more appropriate for life recordings that consist of fragmented information with weak temporal correlations. 
Consequently, measurement metrics are needed to measure the embedding features, demonstrating the effectiveness of each embedding model.

\begin{table}[width=.9\linewidth,cols=6]
\caption{Pre-trained cross-modal embedding models in embedding model pool.}\label{table:model_pool}
\footnotesize
\begin{tabular*}{\tblwidth}{m{140pt}<{\centering}m{50pt}<{\centering}m{50pt}<{\centering}m{50pt}<{\centering}m{50pt}<{\centering}m{10pt}<{\centering}}
\toprule
\textbf{Embedding Model} & Image Embedding & Text Embedding & Video Embedding & Audio Embedding & ...\\
\midrule
CLIP \citep{clip} & \large\ding{51} & \large\ding{51} & \XSolidBrush & \XSolidBrush  & ... \\
CLIP4Clip \citep{CLIP4Clip} & \large\ding{51} & \large\ding{51} & \large\ding{51} & \XSolidBrush & ...\\
X-CLIP \citep{Xclip} & \large\ding{51}  & \large\ding{51} &  \large\ding{51} & \XSolidBrush & ... \\
AudioCLIP \citep{audioclip} & \large\ding{51} & \large\ding{51} &  \XSolidBrush & \large\ding{51} & ...\\
OPT \citep{opt} & \large\ding{51} & \large\ding{51} &  \XSolidBrush & \large\ding{51} & ...\\
...&...&...&...&...&...\\
\bottomrule
\end{tabular*}
\end{table}

\subsubsection{Measurement}\label{section:eval}
Evaluating feature embedding models for multimodal data is a challenging task because it requires consideration of generalization, fidelity, and support for high-quality downstream query tasks. 
In this paper, we propose a combined measurement metric: a widely-used extrinsic metric, complemented by our innovative intrinsic metric, to provide a coarse-to-fine-grained evaluation paradigm. 
As shown in Fig \ref{fig6}, we define a scoring system to evaluate the effectiveness of each modal, which is formulated as:
\begin{equation}
\label{featureEvaluation}
Score=w_1S_1+w_2S_2+w_3S_3
\end{equation}
where the weights ${w_1, w_2, w_3}$ are used to control the influence of different metrics. 
$S_1$ is an extrinsic measurement metric, focusing on the performance in downstream tasks, while $S_2$ and $S_3$ are intrinsic measurement metrics, emphasizing generalization and fidelity ability, respectively. 

To be specific, extrinsic metric $S_1$ is obtained after executing the specific downstream task by measuring three metrics: Recall@K, Query Accuracy, and Query Time, which are recorded in the QBS table during the query process. 
In contrast, intrinsic metrics $S_2$ and $S_3$ are obtained by analyzing the dataset itself, by calculating Silhouette Coefficient (SC) \citep{rousseeuw1987silhouettes} and Fréchet distance (FID) \citep{heusel2017gans}, respectively. 

The remainder of this section will introduce these two metrics in detail, and finally present experimental validations demonstrating that our measurement pipeline improves the selection of feature embedding models.
\\

\noindent \textbf{Silhouette Coefficient (SC)} 

SC is measured from the perspective of the whole dataset. 
Effective feature embeddings should exhibit well-clustered characteristics, meaning the intra-class distance should be small and the inter-class distance should be large. 
Therefore, we introduce the SC metric to assess the quality of clusters formed by embedding models.
Specifically, we embed the features using $n$ candidate embedding models:
\begin{equation}
\label{eq-extract}
X_n=\{\text{Model}_1(I),\text{Model}_2(I),.....\text{Model}_n(I)\},
\end{equation}
where $I$ is the input data, and $X_n$ is the $n$ sets of features embedded from $n$ candidate models. Then we cluster each set of features into $k$ classes:\begin{equation}
\label{eq-cluster}
   \text{Cluster}(X_n,k)=\{\{x_1^i\}_{i=0}^k,\{x_2^i\}_{i=0}^k,...,\{x_n^i\}_{i=0}^k\},
\end{equation}
where $\text{Cluster}$ is a common clustering method, which can be K-means \citep{kmeans} or GMM \citep{GMM}, and $ \{x_j^i\}_{i=0}^k $ is the clustered features encoded by $Model_j$. 
Next, we obtain the score of $S_2$ for each model by calculating the SC value of the clustered features:
\begin{equation}
\label{eq-sc}
   \text{Score of }S_2 \text{ for } Model_j=SC(\{x_j^i\}_{i=0}^k),
\end{equation}
where $SC()$ is the function to calculate the SC value of a given cluster results.
\\

\noindent \textbf{Fréchet distance (FID)}

Unlike the global consideration of SC, FID focuses on feature quality for each individual MMO. 
Recent advancements in text-to-image diffusion models have yielded impressive results in generating realistic and diverse images, building a bridge between text and image. These studies use FID to reflect the difference between the original image and the generated image. 
Therefore, We use FID to judge the fidelity of feature embedding for each MMO. 
Specifically, for embedding model $Model_j$, we input the embedded features into the same pre-trained high-performance generative model like Stable Diffusion \citep{rombach2022high} to get the reconstructed images $\hat{I}_{j}^{'}$, $\hat{I}_{j}^{'}=\text{Diffusion}(X_n)$. Then we input reconstructed images in turn with the original images $\hat{I}_{j}$ into the inception network to get FID value $\text{FID}_j= \text{Inception}(\hat{I}_j,\hat{I}_{j}^{'})$, and the score of $S_3$ can be represented as:
 \begin{equation}
 \label{eq-s3}
 \text{Score of } S_3 \text{ for } Model_j= 1-\text{Normalization}(\text{FID}_j)
 \end{equation}
 
A small FID value between the generated image and the original image indicates that the embedding model can effectively learn the key information from the original data.

Integrating both coarse-grained and fine-grained measurements, we derive the final inner score (combination of $S_2$ and $S_3$). 
This inner score is then fused with the external score, allowing for a comprehensive selection of the embedding model for different scenarios.\\


\noindent \textbf{Experimental Validations} 

To demonstrate the effectiveness of our measurement method, we select a set of image feature embedding models for evaluation. 
The experiments utilize a real dataset AI Challenger, with evaluation methods including the SC (Silhouette Coefficient), IN (Intrinsic Measurement), and IN + EX (a combination of Intrinsic Measurement and Extrinsic Measurement). Note that IN itself can function independently in feature measurement, making it applicable for cold start.
The scoring results are calculated based on different evaluation methods:
\begin{equation}
{ Score }=\left\{\begin{array}{ll}
S_{2} & \text { if method }=\mathrm{SC} \\
w_{2} S_{2}+w_{3} S_{3} & \text { if method }=\mathrm{IN} \\
w_{1} S_{1}+w_{2} S_{2}+w_{3} S_{3} & \text { if method }=\mathrm{IN}+\mathrm{EX}
\end{array}\right.
 \end{equation}
where the weights for the IN are \( w_2 = 0.3 \) and \( w_3 = 0.7 \). 
For the IN + EX, the weights are \( w_1 = 0.2 \), \( w_2 = 0.3 \) and \( w_3 = 0.5 \).

\begin{figure}[pos=!h]
    \centering
    \includegraphics[width=0.7\textwidth]{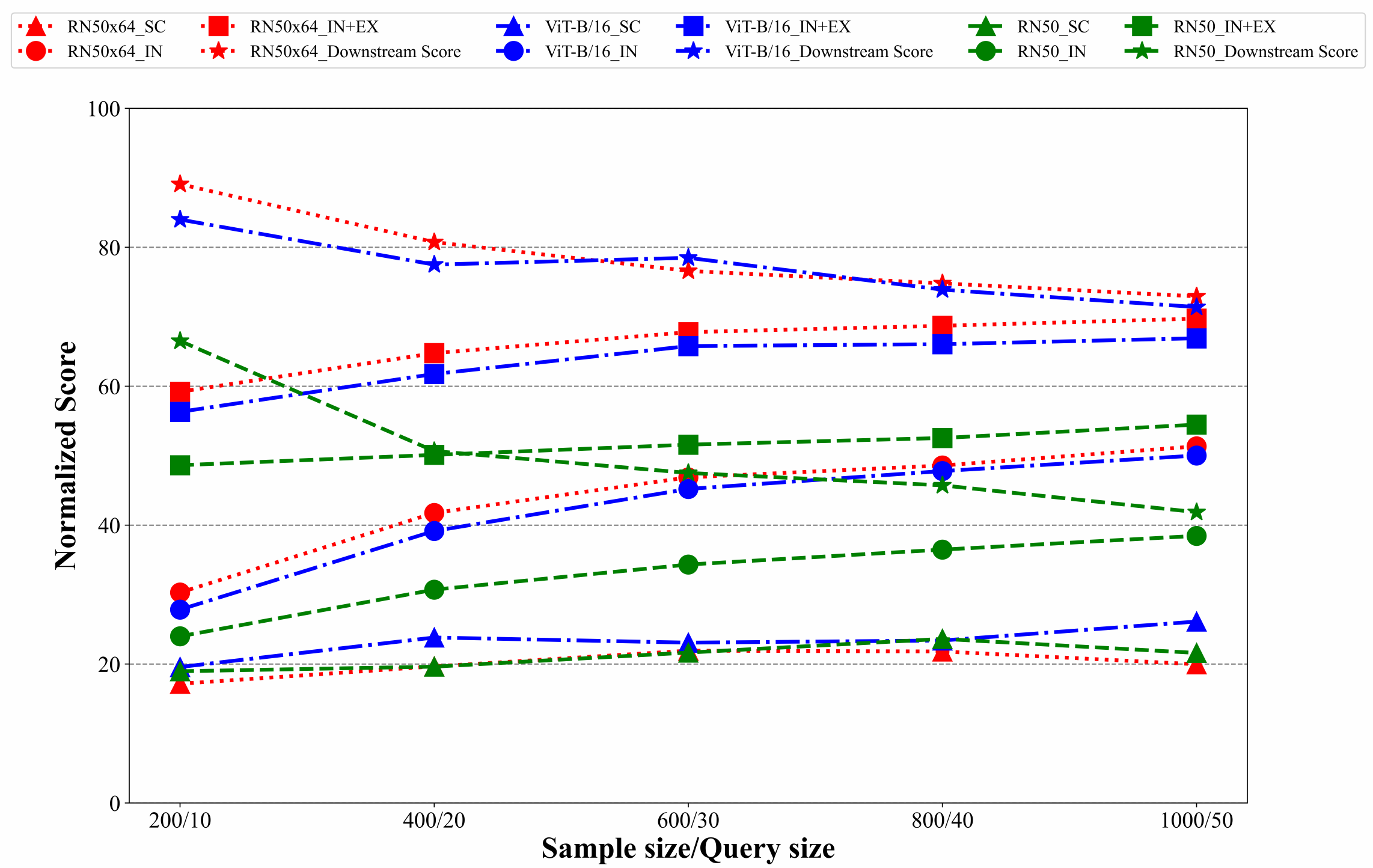}
    \caption{
    Comparative evaluation results of different embedding models. The combination of IN and EX can best simulate the score value of downstream tasks.}
    \label{fig:score}
\end{figure}

In the experiment, we select the image feature embedding models RN50, ViT-B/16, and RN50x64 as experimental benchmarks. 
Fig \ref{fig:score} illustrates the scoring result of different feature embedding models across various evaluation methods compared with the result of downstream query tasks. 
The x-axis represents different sizes of image samples and downstream query tasks in the experiment, and the y-axis represents the scoring results after normalization. 
The scoring results for the downstream query task show that RN50x64 achieves the highest score, followed by ViT-B/16 and RN50, reflecting the actual effectiveness of their respective features in the real query task after embedding. 
This scoring method effectively distinguishes the performance of different embedding models for downstream query tasks and aligns with the actual scoring results. 
Comparing the scoring results from different evaluation methods (SC, IN, IN + EX) with those of the downstream query tasks, we find that the SC scoring results show more generalization, as they always fluctuate in a small interval. On the other hand, scoring results IN and IN + EX show more variation and align closely with the scoring results of downstream query tasks.
This differentiation arises because SC emphasizes the generalization ability of the embedding model, while IN integrates FID which measures the fidelity of image features, improving evaluation accuracy compared to SC. Compared to the SC and IN evaluation, we find that IN + EX provides a more accurate assessment of each embedding model, closely reflecting their actual effectiveness. This indicates that the proposed query-ware mechanism, which facilitates the calculation of EX, significantly impacts the optimization of feature representation. Furthermore, although the IN evaluation alone is less precise, it remains capable of selecting relatively suitable embedding models. Therefore, in the absence of a query-aware mechanism (e.g., during platform cold starts), the platform may experience a decline in its capacity but can still maintain normal functionality and deliver reliable performance.

\subsection{Feature Enhancement}\label{section:enhance}
\begin{figure}[pos=!h]
    \centering
    \includegraphics[width=0.8\textwidth]{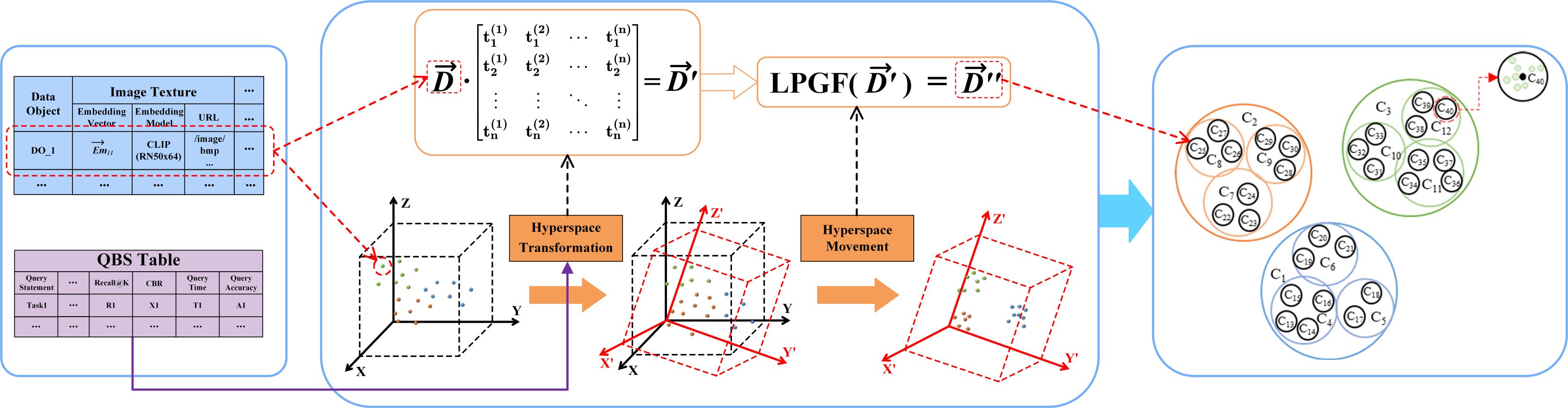}
    \caption{An overview of feature enhancement. Features from Dataframe are represented as a matrix and undergo hyperspace transformation and movement with query-awareness, ultimately being organized into an optimal data layout that exhibits efficient clustering performance.}
    \label{featureAware}
\end{figure}

A well-designed data layout of multimodal data is crucial for reducing index scans and thus accelerating queries. The primary objective of indexing is to enable effective and efficient data retrieval, and an optimized data layout is fundamental to this process. By physically arranging similar or related data points, such layouts reduce the search scope and significantly enhance indexing performance, particularly in high-dimensional or large-scale scenarios.
Currently, most existing approaches, focus on optimizing the data layout through clustering methods. However, these methods typically treat original data as input, neglecting intrinsic data characteristics and their potential impact on layout optimization.
In contrast, we propose a novel approach that transforms original data in hyperspace to intrinsically improve the optimization process of obtaining the optimal data layout. 
To be specific, as shown in Fig \ref{featureAware}, we treat a high-dimensional vector as a data point in hyperspace, and then relocate it to an optimal new location through a unique "projection", considering both the dataset characteristics and query behaviors (Section \ref{section:awareness}). 
The specific method of this projection includes Hyperspace Transformation (Section \ref{section:trans}) and Hyperspace Movement (Section \ref{section:movement}). 
This projection allows the most discriminative dimensions of the data to be emphasized, and similar feature vectors are clustered closely together. 
Then, we provide evidence (Section \ref{section:representionEval}) demonstrating that our feature representation process significantly improves clustering performance. This, in turn, validates the effectiveness of our enhancements in optimizing the data layout. Additionally, we show that both learned indexes and vector similarity indexes achieve improved performance when operating on the optimized data layout.

\subsubsection{Awareness of Projection}\label{section:awareness}
There are two aspects we take into account in the projection process, which are the dataset characteristics and query behaviors. From the perspective of dataset, we have the following considerations: (1)\textit{Identify and retain information-rich dimensions of certain attributes}; (2)\textit{Discern underlying distribution of entire dataset}; (3)\textit{Utilize relationships between MMOs}.

The first two considerations are applied during the hyperspace transformation phase in Section \ref{section:trans}, while the third consideration is applied during the hyperspace movement phase in Section \ref{section:movement}. 
From the perspective of query behaviors, different queries may focus on different features when searching for multimodal data. 
Therefore, we add an optimization step to the process of hyperspace transformation to facilitate locating relevant data.

\subsubsection{Hyperspace Transformation}\label{section:trans}
The hyperspace transformation converts all data points (corresponding to vectors) into new, optimal positions by considering the data characteristics. This is achieved through a transformation matrix $T$, which can be derived and optimized in four steps, as detailed below:\\

\noindent\textbf{Step1:\hspace{0.5em}Represent Data with Matrix $D$}

\begin{figure}[pos=!h]
    \centering
    \includegraphics[width=0.8\textwidth]{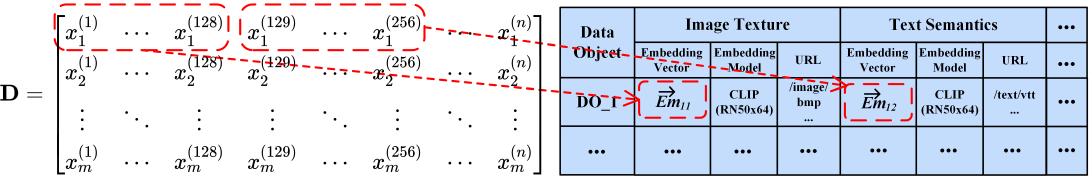}
    \caption{Elements in matrix D correspond to features in DataFrame.}
    \label{figd}
\end{figure}

All the MMOs are managed in DataFrame, from where we select columns of DataFrame to be indexed and represent them with a matrix $D$, as shown in Fig \ref{figd}, where $D\in\mathbb{R}^{m\times n}$, with each row in matrix $D$ corresponding to an MMO. 
$m$ is the total number of records in the DataFrame, and $n$ is the dimension number of all selected columns after embedding. 
To enhance the features of multimodal data without further compressing the information and maintain traceability to the original dataset, we define an $n\times n$ transformation matrix $T$ that establishes a one-to-one mapping. 
Then, we multiply the original matrix $D$ by the $T$ to get enhanced feature vectors denoted as $D_T$, where $D_T\in\mathbb{R}^{m\times n}$, and $D_T = DT$.\\

\noindent\textbf{Step2:\hspace{0.5em}Compute the Covariance Matrix $C$}

We then compute the covariance matrix $C$ ($C\in\mathbb{R}^{n\times n}$) of $D$. 
The covariance matrix $C$ not only reflects the joint distribution characteristics of the data but also reveals the importance of each dimensions in feature vectors. 
The rows of matrix $C$ capture the distribution pattern of data records. Minimal fluctuations in the rows lead to small deviations from the mean, resulting in low covariance. The columns of matrix $C$ represent the importance of each dimension, preserving the complete dimensional information for accurate mapping of the selected DataFrame columns.
\\

\noindent\textbf{Step3:\hspace{0.5em}Define the Transformation Matrix $T$}

Through the decomposition of $C$, we can construct the transformation matrix $T$ applied to the feature vectors. 
$T$ comprises a rotation matrix $R$ and a scaling matrix $S$, i.e., $T = RS$. 
The rotation matrix $R$ indicates the direction of data transformation in hyperspace, while $S$ is a diagonal matrix reflecting the importance of each vector dimension through the scaling of its diagonal elements, thus stretching each dimension of feature vectors. 
$R$ and $S$ are calculated though eigen decomposition on the covariance matrix $C$, $C=V\Lambda V^T$.

According to the linear transformation relationship of matrices, we know that the eigenvectors in $V$ serve as the basis for the linear transformation of hyperspace coordinates, and the square root of each eigenvalue in $\Lambda$ represents the scaling factor for each dimension in hyperspace. 
Therefore, we can obtain the rotation matrix $R = V$ and scaling matrix $S=\sqrt{\Lambda}$ from the eigen decomposition results of $C$, and the transformation matrix $T=RS$. 

The invertibility of the hyperspace transformation is an innegligible issue, because after performing indexing, the indexed data needs to be re-transformed into the original vector to support query operations on MMOs. 
Therefore, to ensure the reversibility of matrix $T$ and maintain the linearity of data transformation properties, we impose the following constraints on $R$ and $S$:
\begin{equation}
\label{eq:cons}
\begin{aligned}
&(1)R,S\in\mathbb{R}^{n\times n}. \\
&(2)\forall v_i,v_j\in R,||v_i||_2=1,v_i^Tv_j=0,\forall i,j\in\{1,2,\ldots,n\},i\neq j. \\
&(3)S=diag(a_{11},a_{22},\cdots,a_{nn}),a_{ii}>0,\forall i\in\{1,2,\ldots,n\}.
\end{aligned}   
\end{equation}
Constraint (1) indicates that after the transformation, the dataset retains $n$ dimensions in the hyperspace with no loss of dimensional information, ensuring that the original dataset is easily regained through one-to-one mapping.
Constraint (2) guarantees that $R$ is an orthonormal matrix, ensuring that the columns of the matrix transformed by $R$ are independent of other columns, allowing us to focus on specific dimensions corresponding to the feature vectors during query-aware optimization without affecting other dimensions.
Constraint (3) specifies that matrix $S$ must be a positive definite matrix because we only consider scaling the feature vectors in the positive direction.
Therefore, after calculating $T$, we can get enhanced feature vectors $D_T$, where $D_T = DT$.
These three constraints ensure that matrix $T$ is invertible through the inverse operation by multiplying $D_T$ by $T^{-1}$.\\

\noindent\textbf{Step4:\hspace{0.5em}Optimize $T$ by Query Awareness}

To improve query efficiency and accuracy under different query scenarios, we adjust the hyperspace transformation process by optimizing $T$ through learning from query workloads. 
To simplify this optimization process, we decompose the objective of optimizing $T$ into optimizing $R$ and $S$.

For a multimodal dataset $D$ and a given query workload $Q$, we aim to achieve high query accuracy, minimal query time, and low CBR when executing queries. Specifically, we seek to identify the optimal $R^*$, $S^*$ to optimize these objectives as much as possible. Therefore, we formulate this as a multi-objective optimization problem with the objective function:
\begin{equation}\label{eq:opt}
\begin{aligned}
\min _{R, S} & {\left[f_{\text {time }}(D, q ; R, S), f_{C B R}(D, q ; R, S),-f_{\text {acc }}(D, q ; R, S)\right] } \\
\text { s.t. } & \text { Formula }(7)
\end{aligned}
\end{equation}
The constraints of this problem are identical to those in equation (\ref{eq:cons}). In this multi-objective optimization problem, there are three objective functions:\\
(1) $f_{time}(D,q;R,S)$: Query time, which we aim to minimize.\\
(2) $f_{CBR}(D,q;R,S)$: CBR, which we aim to minimize.\\
(3) $f_{acc}(D,q;R,S)$: Query accuracy, which we aim to maximize.\\
Selecting a multi-objective optimization method involves addressing two key challenges. First, the objective functions are multiple competing black-box objectives, each with high evaluation costs. Second, both $R$ and 
$S$ are high-dimensional hyperparameters, and their search space is also high-dimensional. To address these challenges, we leverage the Bayesian multi-objective optimization algorithm, MORBO \citep{MORBO}. MORBO employs surrogate models to approximate the relationship between parameters and the actual values of the objective functions, thereby reducing evaluation costs. It also designs a coordinated strategy that enables parallel exploration of multiple local regions within the high-dimensional hyperparameter search space to identify global optima. Building on the advantages of the MORBO algorithm, we propose a matrix optimization algorithm, as described in Algorithm \ref{al:1}.
First, we initialize multiple trust regions based on the records of Query Accuracy, Query Time, and CBR in the QBS table. Then, within each trust region, we train a local Gaussian process model, select new candidate points, and evaluate new observations. Subsequently, we dynamically update each trust region based on the latest candidate points and observations. Finally, we approximate the Pareto Front (PF) using the new observations and extract the approximate Pareto-optimal solution set P.
To obtain unique $R^*$ and $S^*$ for our feature enhancement process, we assign appropriate weights to the various multi-objective observations based on task-specific requirements. The optimal solution is determined by computing the weighted cumulative sum.\\
\begin{algorithm}
    \caption{Matrix Optimization Based On MORBO}
    \label{al:1}
    \renewcommand{\algorithmicrequire}{\textbf{Input:}} 
    \renewcommand{\algorithmicensure}{\textbf{Output:}} 
    \begin{algorithmic}[1]
        \REQUIRE dataset matrix $D$, workload $Q$, objective functions $f=\{f_{acc},f_{time},f_{CBR}\}$, max iteration $k$, trust region parameters:\{$n,L_{init},L_{min}$\}
        \ENSURE approximate Pareto optimal solution set $P$
        
        \STATE Initialize the trust regions $T = \{T_1,...,T_n\}$, $X_{0}=\emptyset,Y_{0}=\emptyset$.
        \FOR{i in [0,k]}
        \STATE Fit local gaussian process models $LGP=\{lgp_{1},...lgp_{n}\}$
        \STATE $(R_{next}^{t},S_{next}^{t})_{t\in[1,n]}=SelectNext(LGP,T)$ /*Select new candidates in each trust region*/
        \STATE $((Y_{time}^t,Y_{CBR}^t,Y_{acc}^t)_{t\in[1,n]})=\text{BatchEval}( f, D, Q; ( R_t^{next}, S_t^{next} )_{t \in [1,n]}$
        \STATE $X_0.append((R_{next}^{t},S_{next}^{t})_{t\in[1,n]})$,$Y_0.append ((Y_{time}^t,Y_{CBR}^t,Y_{acc}^t)_{t\in[1,n]})$
        \FOR{j in [0, n]}
        \STATE $T_j,L_j=Updata(T_j,(R_{next}^j,S_{next^{\prime}}^jY_{time^{\prime}}^jY_{CBR^{\prime}}^jY_{acc}^j))$
        \IF{$L_j<L_{\min}$}
        \STATE Terminate $T_j$
        \STATE $T_j,X_0,Y_0=Reinitialize(L_{init},X_0,Y_0)$ /*Add new region centroid and objective function values into $X_0$, $Y_0$*/
        \ELSE
        \STATE update $T_j$'s region center
        \ENDIF
        \ENDFOR 
        \ENDFOR 
        \STATE $P=SelectPF(X_k,Y_k)$
        \RETURN $P$
    \end{algorithmic}
\end{algorithm}

\noindent\textbf{Hyperspace Transformation Evaluation}

\begin{figure}[pos=!h]
	\begin{minipage}{0.49\linewidth}
		\vspace{3pt}
		\centerline{\includegraphics[width=\textwidth]{5.2.2_1.pdf}}
		\centerline{(a)}
	\end{minipage}
	\begin{minipage}{0.49\linewidth}
		\vspace{3pt}
		\centerline{\includegraphics[width=\textwidth]{5.2.2_2.pdf}}
		\centerline{(b)}
	\end{minipage}
	\caption{The cost time of T construction and data transformation with different number of tasks and dataset sizes. (a)Dataset size = 100M, (b) Task Number = 40.}
	\label{fig:computation overhead}
\end{figure}

The hyperspace transformation process in MQRLD is executed offline, utilizing large-scale of historical data. To provide a comprehensive analysis of this process, we evaluate both the complexity and the effectiveness of hyperspace transformation. The complexity is assessed through theoretical and experimental analysis, and the effectiveness is assessed by comparing hyperspace transformation with other similar approaches.

First, we evaluate the complexity of hyperspace transformation. In Step 1, the matrix \( D \) is obtained, $D\in\mathbb{R}^{N\times d}$, where $N$ is the number of data points, ranging from 0.30M to 210M in our experiments, and $d$ is the dimensionality of each data points, ranging from 3 to 1026. Storing the matrix \( D \) requires space complexity of \( O(N \times d) \). Step 2 involves the computation and storage of the covariance matrix \( C \), which has a time complexity of \( O(N \times d^2) \) and a space complexity of \( O(d^2) \). In Step 3, the matrices \( R \) and \( S \) are derived through eigen decomposition on the covariance matrix \( C \), which requires \( O(d^3) \) time and \( O(d^2) \) space using the Jacobi method. In Step 4, the time complexity of the MORBO algorithm in each iteration is \( O(n_{QBS}^3) \), where $n_{QBS}$ is the number of records in the QBS table. Since optimization of matrix \( T \) requires \( k \) iterations, the overall time complexity for Step 4 is bounded by \( O(k \times n_{QBS}^3) \). The space complexity for Step 4 is bounded by \( O(d \times n_{QBS}) \).  After obtaining the optimal matrix \( T \), the matrix multiplication \( D \times T \) is performed, requires \( O(N \times d^2) \) time for computation and \( O(N \times d) \) space for storing the results. Consequently, the overall time complexity of hyperspace transformation is bounded by \( O(N) \) + \( O(n_{QBS}^3) \), and overall space complexity is bounded by \( O(N) \) + \( O(n_{QBS}) \). As shown in Fig~\ref{fig:computation overhead}, we conduct experiments to evaluate the computational overhead of hyperspace transformation. The \( T \) construction process includes steps 1 to 4, while the data transformation refers to the process of \( D \times T \). The experimental results in Fig~\ref{fig:computation overhead}(a) demonstrate that we can further accelerate the hyperspace transformation by increasing parallelism. Fig~\ref{fig:computation overhead}(b) indicates that, as the dataset size increases, the hyperspace transformation process does not incur high computational overhead in Hudi's parallel computing environment.

\begin{table*}[width=\linewidth,cols=6]
\caption{
Feature Selection and Feature Scaling Methods}
\label{table:hyperspace trans}
\footnotesize
\begin{threeparttable} 
\begin{tabular*}{\tblwidth}{m{60pt}<{\centering}m{40pt}<{\centering}m{30pt}<{\centering}m{50pt}<{\centering}m{40pt}<{\centering}m{40pt}<{\centering}m{60pt}<{\centering}m{60pt}<{\centering}}
\toprule
Methods & Type & Invertibility & Optimization based on Query Aware & Clustering Improvement& "Last-mile" Model Improvement & Time Complexity & Space Complexity\\
\midrule
XGBoost\citep{xgboost} & Feature Selection & \XSolidBrush & \XSolidBrush & \XSolidBrush & \XSolidBrush  & $O(KhNlogN)$ & $O(KN)$ \\
LDA\citep{LDA} & Feature Selection &\XSolidBrush & \XSolidBrush & \XSolidBrush & \XSolidBrush & $O(N)$ & $O(N)$ \\
PCA\citep{PCA} & Feature Selection &\XSolidBrush  & \XSolidBrush &  \XSolidBrush & \XSolidBrush & $O(N)$ & $O(N)$ \\
CCNF\citep{CCNF} & Feature Scaling &\large\ding{51} & \XSolidBrush &  \XSolidBrush & \XSolidBrush & $O(KhNlogN)$ & $O(KN)$ \\
DTization\citep{dtization} & Feature Scaling &\large\ding{51} & \XSolidBrush &  \XSolidBrush & \XSolidBrush & $O(NlogN)$ & $O(N)$ \\
Ours&Hyperspace Transformation&\large\ding{51}&\large\ding{51}&\large\ding{51}&\large\ding{51}& Init\_T:$O(N)$ Opt\_T:$O(N)+O({n^3_{QBS}})$ & Init\_T:$O(N)$ Opt\_T:$O(N)+O({n_{QBS}})$ \\
\hline
\end{tabular*}
\begin{tablenotes}
\footnotesize
\item[a] K is the number of tree.
\item[b] h is the depth of tree.
\item[c] Init\_T(Initialized\_T) represents the initialization process of matrix T in hyperspace transformation, which corresponds to Steps 1 - 3.
\item[d] Opt\_T(Optimized\_T) represents the initialization and optimization process of matrix T in hyperspace transformation which corresponds to Steps 1 - 4.
\end{tablenotes}
\end{threeparttable}
\end{table*}

\begin{figure}[pos=!h]
	\begin{minipage}{0.49\linewidth}
		\vspace{3pt}
		\centerline{\includegraphics[width=\textwidth]{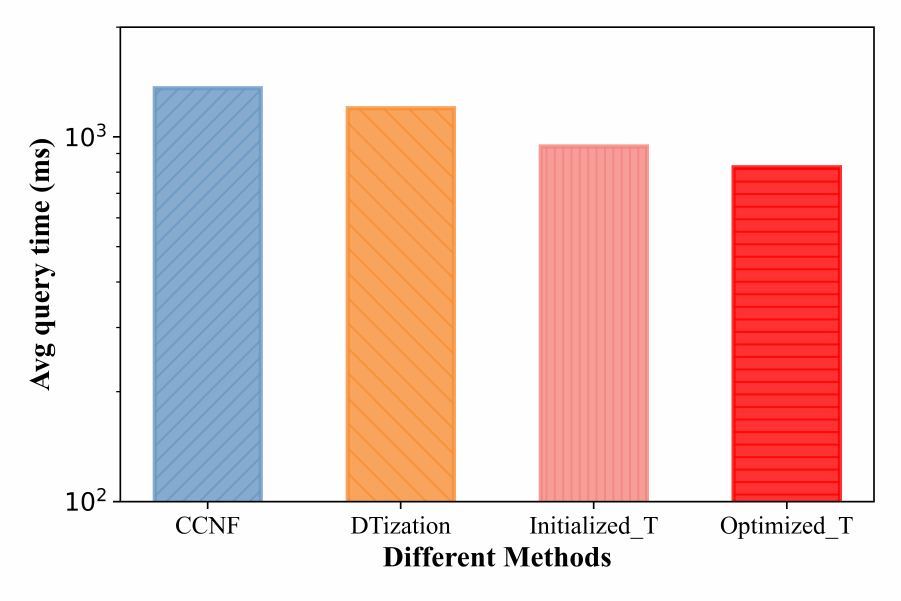}}
		\centerline{(a)}
	\end{minipage}
	\begin{minipage}{0.49\linewidth}
		\vspace{3pt}
		\centerline{\includegraphics[width=\textwidth]{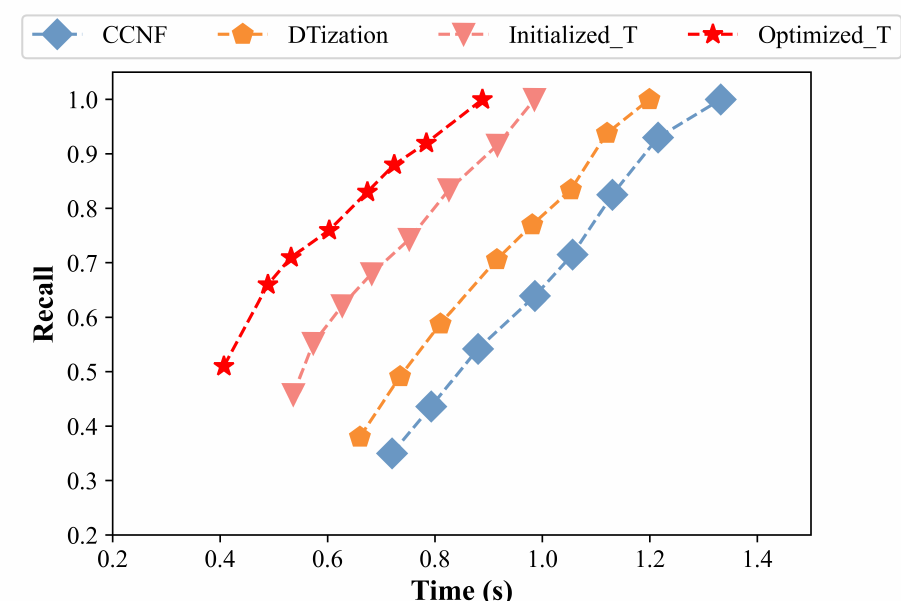}}
		\centerline{(b)}
	\end{minipage}
	\caption{The average query times(a) and recall-time curves(b) of different methods in MQRLD.}
	\label{fig:scaling}
\end{figure}

Next, we evaluate the effectiveness of hyperspace transformation. 
Our analytical framework includes two parts: theoretical analysis and experimental analysis.
(1) \textit{Theoretical analysis}: As shown in Table \ref{table:hyperspace trans}, we compare hyperspace transformation with several feature selection and feature scaling approaches, which share similarities with our hyperspace transformation process but also have notable differences. Feature selection is a dimensionality reduction technique aimed at selecting the most relevant features to enhance the performance and accuracy of machine learning algorithms \citep{featureselection}. Typical feature selection methods, such as XGBoost, PCA, and LDA, reduce dimensions and may eliminate crucial numerical features necessary for responding to specific queries, thus preventing the retrieval of certain information. On the other hand, feature scaling involves rescaling all features to a new scale, improving outcomes and speeding up computations in data processing and machine learning tasks \citep{featurescaling}. Typical feature scaling methods, such as CCNF and DTizatoin, can support invertibility by using an \( O(N) \)  space to store feature transformation information and could potentially replace our hyperspace transformation process.
However, these feature scaling methods are not well-integrated with query-aware processes, cannot support clustering improvement, or facilitate high-quality "last-mile" model training. Additionally, they still exhibit higher time complexity compared to our approach.
(2) \textit{Experimental analysis}: We conduct experiments on these feature scaling methods compared with hyperspace transformation. As shown in Fig \ref{fig:scaling}. The Optimized\_T achieves the best average query time and recall rate, followed by Initialized\_T, with the feature scaling methods CCNF and DTization performing worse than the hyperspace transformation, showing that our hyperspace transformation can better improve the efficiency and accuracy of multimodal data retrieval than other methods.

\subsubsection{Hyperspace Movement}\label{section:movement}
Hyperspace movement further optimizes the feature vectors on the result of previous hyperspace transformation by moving similar data points in hyperspace together to get $\hat{D}$, where similar data can be placed in adjacent places or the same cluster. 
Thus it could benefit index construction and multimodal data retrieval. 

\begin{figure}[pos=!h]
    \centering
    \includegraphics[width=0.8\textwidth]{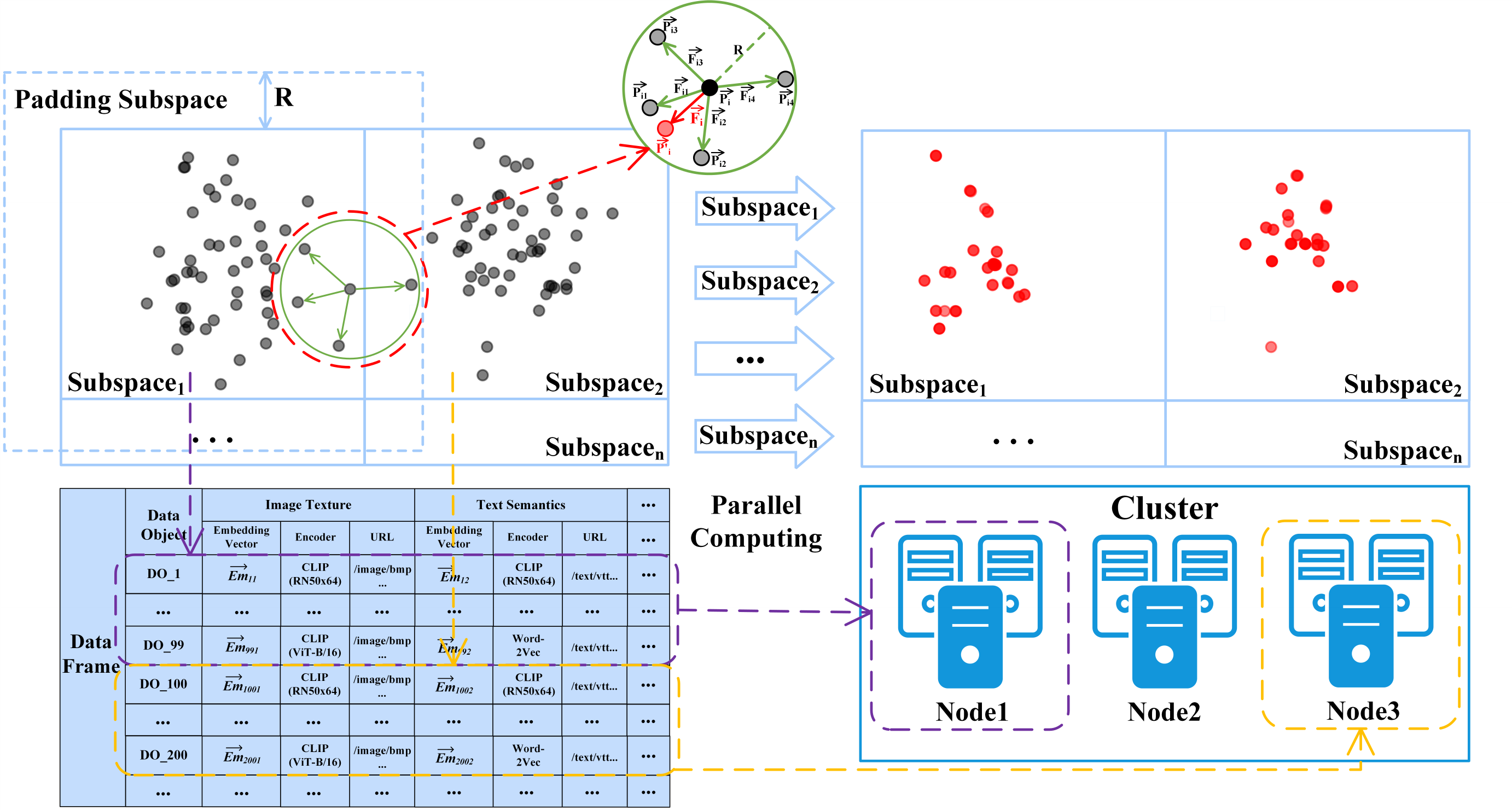}
    \caption{Overview of hyperspace movement. Force computation of each point is calculated within radius R. 
All data points are diveded into n subspace, corresponding to n segments of Dataframe, and computed parallel in different nodes of cluster in Apach Hudi.}
    \label{fig:hyperspaemovement}
\end{figure}

To achieve this objective, we propose a new data movement method called Local Parallelized Gravitational Field (LPGF) which improves on an existing data movement method HIBOG \citep{HIBOG}. 
HIBOG posits that in hyperspace, a data point is attracted by its nearest $K$ similar data points, causing it to move towards the resultant force direction of these attractions. 
Though HIBOG shows effectiveness in gathering similar data, it suffers the following drawbacks:
(1) \textit{Computationally expensive}: Finding the nearest $K$ data points for force calculation requires sorting all data points, which requires huge computation, especially for large datasets. 
Additionally, the parameter $K$ needs to be set and adjusted, further increasing the computation time. 
(2) \textit{Anomalies of movement}: The calculation of forces between data points may not be suitable for all scenarios. 
Sometimes, this force can cause anomalies in the movement of tightly clustered data. 
(3) \textit{Weak parallel capability}: The spatial range of the nearest $K$ neighbors is not predetermined, leading to the inability to perform grid partitioning and parallel processing of data. 
To overcome these shortcomings, we introduce the pipeline of LPGF, which consists of three steps to improve the above shortcomings, as shown in Fig \ref{fig:hyperspaemovement}.\\

\noindent \textbf{Step1:\hspace{0.5em}Boundary of Force Area}

To lower computationally expensive, we define a radius $R$ to calculate the resultant force applied to a single point instead of using the nearest $K$ points. 
For each data point $\overrightarrow{P_i}$, we only consider the forces exerted on it by other points within $R$, eliminating the influence of points outside $R$. 
The size of $R$ is related to the cluster computing power and typically set between 5$G$ and 10$G$, where $G$ is the average distance from each point in the dataset to its nearest neighbor, and  $G=\frac1m\sum_{i=1}^m\parallel\overrightarrow{P_{i1}}-\overrightarrow{P_i}\parallel_2$ ($\overrightarrow{P_{i1}}$ is the nearest point of $\overrightarrow{P_i}$). 
This improvement avoids the need for extensive calculations and removes dependence on the value of $K$, thereby reducing calculation time.\\

\noindent \textbf{Step2:\hspace{0.5em}Distinct Regions of Force}

\begin{figure}[pos=!h]
    \centering
    \includegraphics[width=0.8\textwidth]{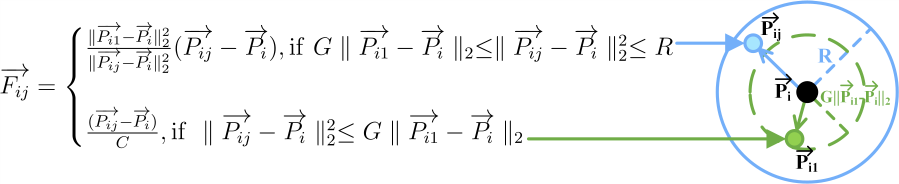}
    \caption{Force calculation for each point is dependent on all points within its radius R, and the applying force varies for points at different distances.}
    \label{fig:Fcalculate}
\end{figure}

To avoid anomalies of movement, we make the strength and direction of applied force on point $\overrightarrow{P_i}$ depend on the distance between $\overrightarrow{P_i}$ and all the other points within radius $R$. 
To be specific, for each $\overrightarrow{P_i}$, the resultant force $\overrightarrow{F_i}$ is calculated by summing the applied force from all relevant points $\overrightarrow{P_i}$ in $R$.
$\overrightarrow{F_{ij}}$ applied by $\overrightarrow{P_{ij}}$ depends on the distance between $\overrightarrow{P_i}$ and $\overrightarrow{P_{ij}}$, which is calculated as Fig \ref{fig:Fcalculate}, where $C$ is a constant slightly larger than 1.
For example, taking a value of $1+10^{-1}$. 
During the calculation of gravitational forces, data points are represented in vector form. 
Within the radius $R$ of data point $\overrightarrow{P_i}$, the resultant force $\overrightarrow{F_{i}}$ exerted on $\overrightarrow{P_i}$ is determined by 
$\overrightarrow{F_i}=\sum_{j=1}^N\overrightarrow{F_{ij}}$, where N is the number of data points contained within the radius 
$R$ of $\overrightarrow{P_i}$.\\

\noindent \textbf{Step3:\hspace{0.5em}Parallelization of Force Computation}

To improve parallel capability, we split the dataset into multiple regions and utilize the partitioning mechanism of Apache Hudi to perform the computation in parallel within each region. 
This is achievable because we define a padding area with radius $R$ to limit the range for processing each point, ensuring that the calculation area for each data point is confined to a fixed region. 
After applying LPGF, we can obtain a displacement matrix $M$, $M\in\mathbb{R}^{m\times n}$, which represents the displacement of each data point.

\subsubsection{Feature Representation Evaluation}\label{section:representionEval}
The effectiveness of the feature representation is demonstrated through clustering performance, and clustering improves data layout's separability and compactness. Meanwhile, superior data layout optimization can further enhance indexing performance, thereby benefiting efficient and effective queries.  To validate this, we conduct three experiments. The first experiment compares the clustering performance before and after our feature representation process (Evaluation 1). The second and third experiments assess the query performance of learned indexes (Evaluation 2) and vector similarity indexes (Evaluation 3) on the optimized data layout, respectively.\\

\noindent \textbf{Evaluation 1:\hspace{0.5em}Clustering Performance Evaluation}

In this evaluation, we show that the data layout after feature representation significantly enhances clustering performance.
We compare five feature representation methods for enhancing clustering performance, which are T, HIBOG, LPGF, T + HIBOG, and T + LPGF.
T represents Hyperspace Transformation.
HIBOG and LPGF denote Hyperspace Movement.
T + HIBOG as well as T + LPGF represent the combination of Hyperspace Transformation and Hyperspace Movement. 
The experiments are conducted using the real dataset Flame \citep{fu2007flame} and employ three well-known clustering methods, which are K-means, Agglomerative \citep{agglomerative}, and Density Peaks Clustering (DPC) \citep{peak}, to represent diverse clustering paradigms. 
The selection of these methods is not meant to imply exclusivity but rather to demonstrate the generalization ability across various clustering algorithms. 
To comprehensively evaluate the impact of different representation methods on clustering performance, we use the Silhouette Coefficient, Calinski-Harabasz Index, and Normalized Mutual Information as experimental metrics. 
These metrics assess clustering effectiveness from different perspectives.

\begin{table*}[width=1.0\linewidth]
\caption{Clustering enhancement results of feature representation}
\label{table:Clustering Enhancement Evaluation}
\footnotesize
\begin{threeparttable}
\begin{tabular*}{\tblwidth}{m{80pt}<{\centering}m{80pt}<{\centering}m{80pt}<{\centering}m{80pt}<{\centering}m{80pt}<{\centering}}
\toprule
Clustering Method & Optimization Method & Silhouette Coefficient & Calinski-Harabasz Index & Normalized Mutual Information\\
\midrule
\multirow{6}*{K-means} & Unoptimized & 0.412 & 202.695 & 0.476 \\
~ & T & 0.425 & 214.568 & 0.506\\
~ & HIBOG & 0.448 & 222.804 & 0.523\\
~ & LPGF & 0.455 & 228.109 & 0.541\\
~ & \textit{T + HIBOG} & \textit{0.472} & \textit{245.784} & \textit{0.576}\\
~ & \textbf{T + LPGF} & \textbf{0.503} & \textbf{260.556} & \textbf{0.603}\\
\bottomrule
\multirow{6}*{Agglomerative} & Unoptimized & 0.329 & 122.782 & 0.330 \\
~ & T & 0.342 & 129.652 & 0.377\\
~ & HIBOG & 0.363 & 144.331 & 0.425\\
~ & LPGF & 0.364 & 144.868 & 0.425\\
~ & \textit{T + HIBOG} & \textit{0.395} & \textit{151.546} & \textit{0.489}\\
~ & \textbf{T + LPGF} & \textbf{0.423} & \textbf{159.651} & \textbf{0.541}\\
\bottomrule
\multirow{6}*{DPC} & Unoptimized & 0.338 & 133.615 & 0.413 \\
~ & T & 0.345 & 135.781 & 0.614\\
~ & HIBOG & 0.351 & 138.855 & 0.890\\
~ & LPGF & 0.410 & 169.101 & 0.936\\
~ & \textit{T + HIBOG} & \textit{0.455} & \textit{176.252} & \textit{0.958}\\
~ & \textbf{T + LPGF} & \textbf{0.491} & \textbf{184.451} & \textbf{0.992}\\
\bottomrule
\end{tabular*}
\begin{tablenotes}
\footnotesize
\item[a] Bold font indicates the best result, and italic font indicates the second-best result.
\item[b] T: Apply hyperspace Transformation for feature enhancement.
\item[c] HIBOG: Apply HIBOG in hyperspace movement for feature enhancement.
\item[d] LPGF: Apply LPGF in hyperspace movement for feature enhancement.
\end{tablenotes}
\end{threeparttable}
\end{table*}

Table \ref{table:Clustering Enhancement Evaluation} presents the experimental results of different feature representation methods. 
Comparing the results of LPGF and HIBOG, we observe that LPGF significantly enhances clustering performance across various algorithms, suggesting that optimizations in gravitational field design in LPGF improve intra-class compactness and inter-class separation more effectively than HIBOG. 
Additionally, T + HIBOG and T + LPGF demonstrate that the hyperspace transformation matrix T not only enhances the discriminative features of multimodal data but also substantially improves clustering effectiveness.\\

\noindent \textbf{Evaluation 2:\hspace{0.5em}Multi-dimensional Learned Index Query Performance Enhancement Evaluation}

The core concept of the learned index is to use a learned model to narrow down queries to a "leaf node", followed by "last-mile" search within each leaf node for the final retrieval step. However, existing learned indexes face limitations in terms of both model accuracy and complexity. By optimizing the data layout, we improve clustering performance, which not only accelerates the "last mile" search but also improves both precision and recall. Additionally, the optimized data layout results in a better distribution of the data points, reducing the cost of model training, and thereby accelerating index construction. We design two experiments to validate the effectiveness of our proposed feature representation method in enhancing the performance of high-dimensional learned indexes. The first experiment shows that our approach enables the training of a "last mile" model which facilitates accelerated query processing and index construction, while the second evaluates its impact on shortening the query time and index construction time of current state-of-the-art multi-dimensional learned indexes. The experiments utilize a generated dataset comprising 2M records, with data distributed according to a Gaussian mixture distribution.

\begin{figure}[h]
    \centering
    \includegraphics[width=0.8\textwidth]{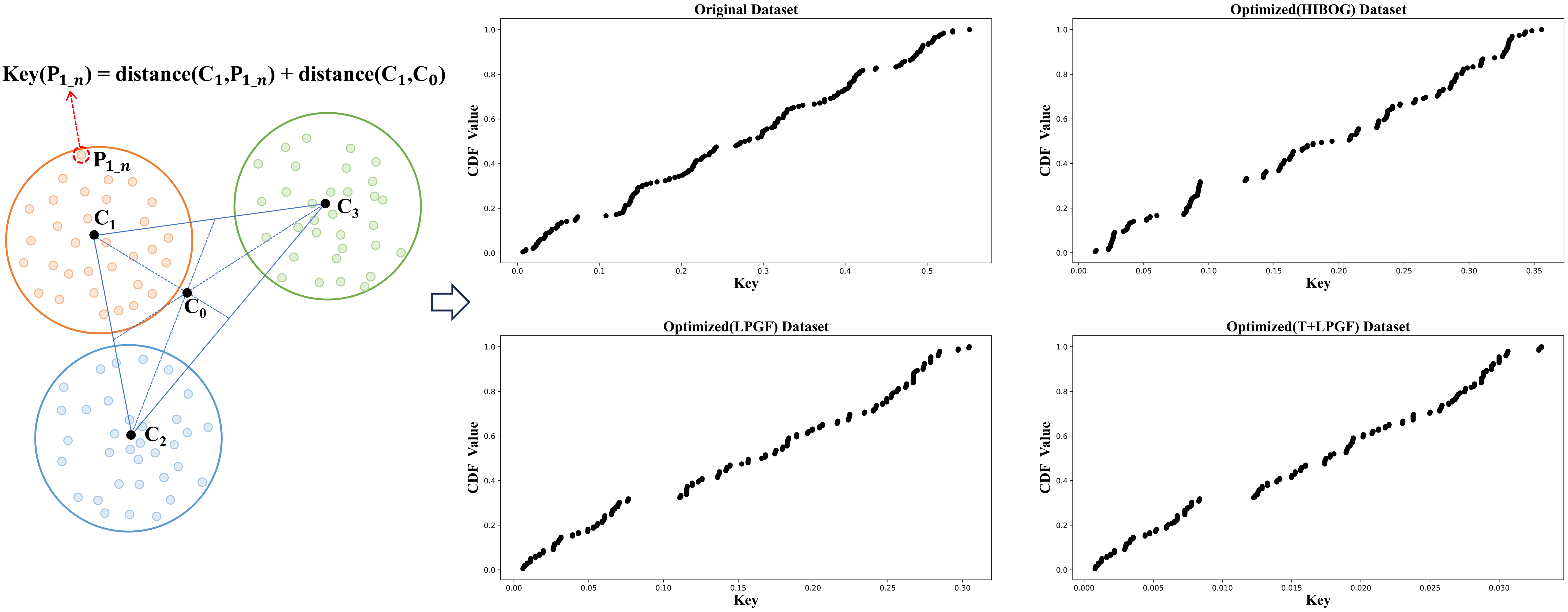}
    \caption{The CDF curves of different feature representation methods. Keys denote training labels of the "last-mile" search model. The CDF curve after T+LPGF is the smoothest, indicating that the "last-mile" model can be trained in a simpler, faster, and more accurate manner.}
    \label{fig:CDF}
\end{figure}

In the first experiment, we predict the performance of the "last mile" model by computing the cumulative distribution function (CDF) of training labels for the "last mile" model, which we denote as keys. As shown in Fig \ref{fig:CDF}, the key of a data point $P_{1_n}$ is computed as the sum of two distances. The first distance is from $P_{1_n}$ to its cluster's centroid $C_1$, and the second distance is from $C_1$ to $C_0$ which is the barycenter of all clusters' centroids. A smooth CDF curve suggests that the "last mile" model can effectively learn with high accuracy without the need for complex strategies, thereby reducing the query time as well as index construction time. We compare the CDF curves of four datasets, which are Original Dataset, HIBOG Optimized Dataset, LPGF Optimized Dataset, and T + LPGF Optimized Dataset. 
As shown in Fig \ref{fig:CDF}, by comparing the smoothness of the CDF curves for each dataset, we find that T + LPGF achieves the highest improvement in the smoothness among all optimized methods. 
Compared to HIBOG, LPGF significantly smoothens the global CDF curve, because the force calculation method in LPGF reduces the probability of anomalous data movement, resulting in tighter data clustering. 
T + LPGF not only retains the advantages of LPGF but also highlights critical clustering features through matrix T, further enhancing the clustering capabilities. 
Overall, the T + LPGF method results in tighter multimodal data clustering, improving the smoothness of the CDF curve, and effectively reducing the complexity of the training process involved in index construction.

\begin{figure}[pos=!h]
	\begin{minipage}{0.49\linewidth}
		\vspace{3pt}
		\centerline{\includegraphics[width=\textwidth]{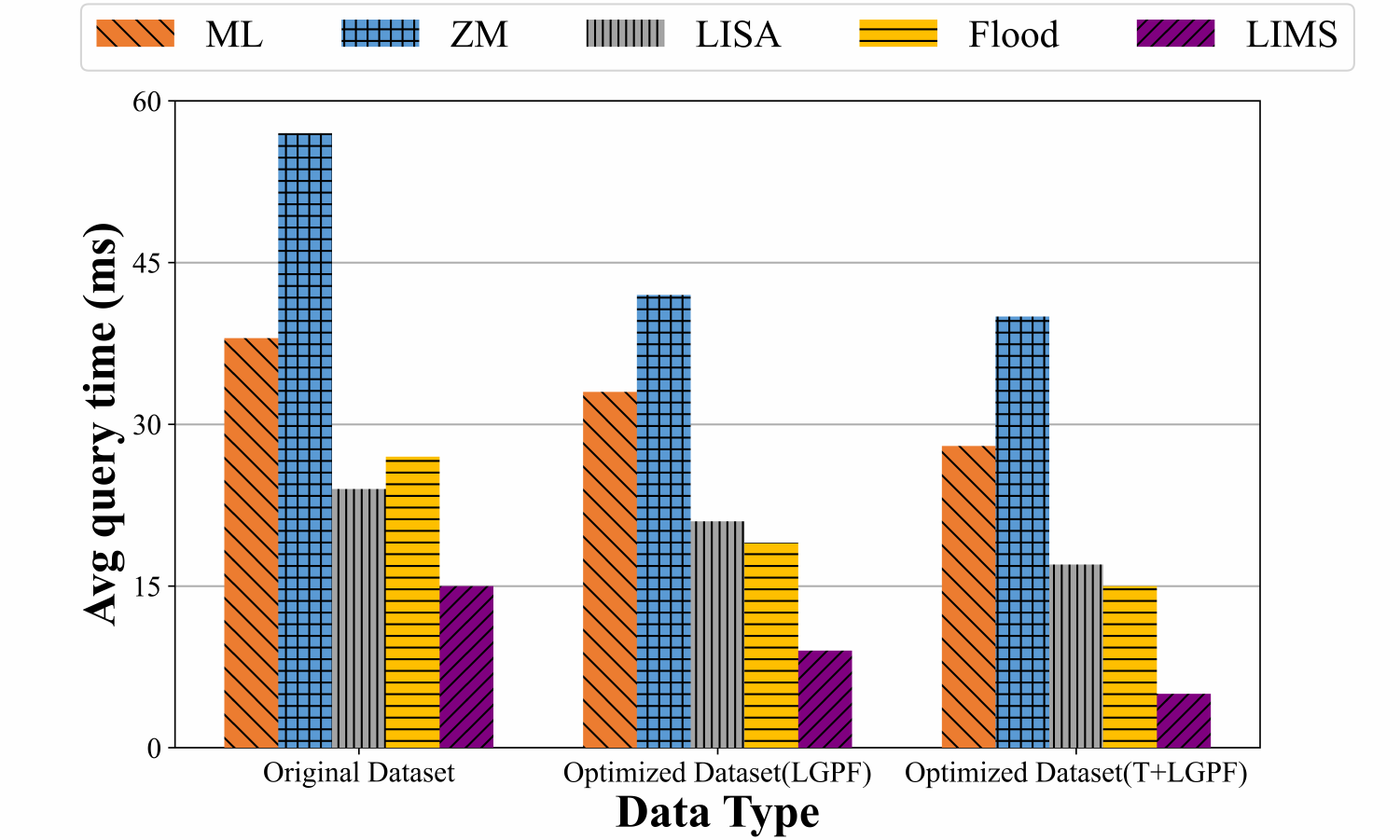}}
		\centerline{(a)}
	\end{minipage}
	\begin{minipage}{0.49\linewidth}
		\vspace{3pt}
		\centerline{\includegraphics[width=\textwidth]{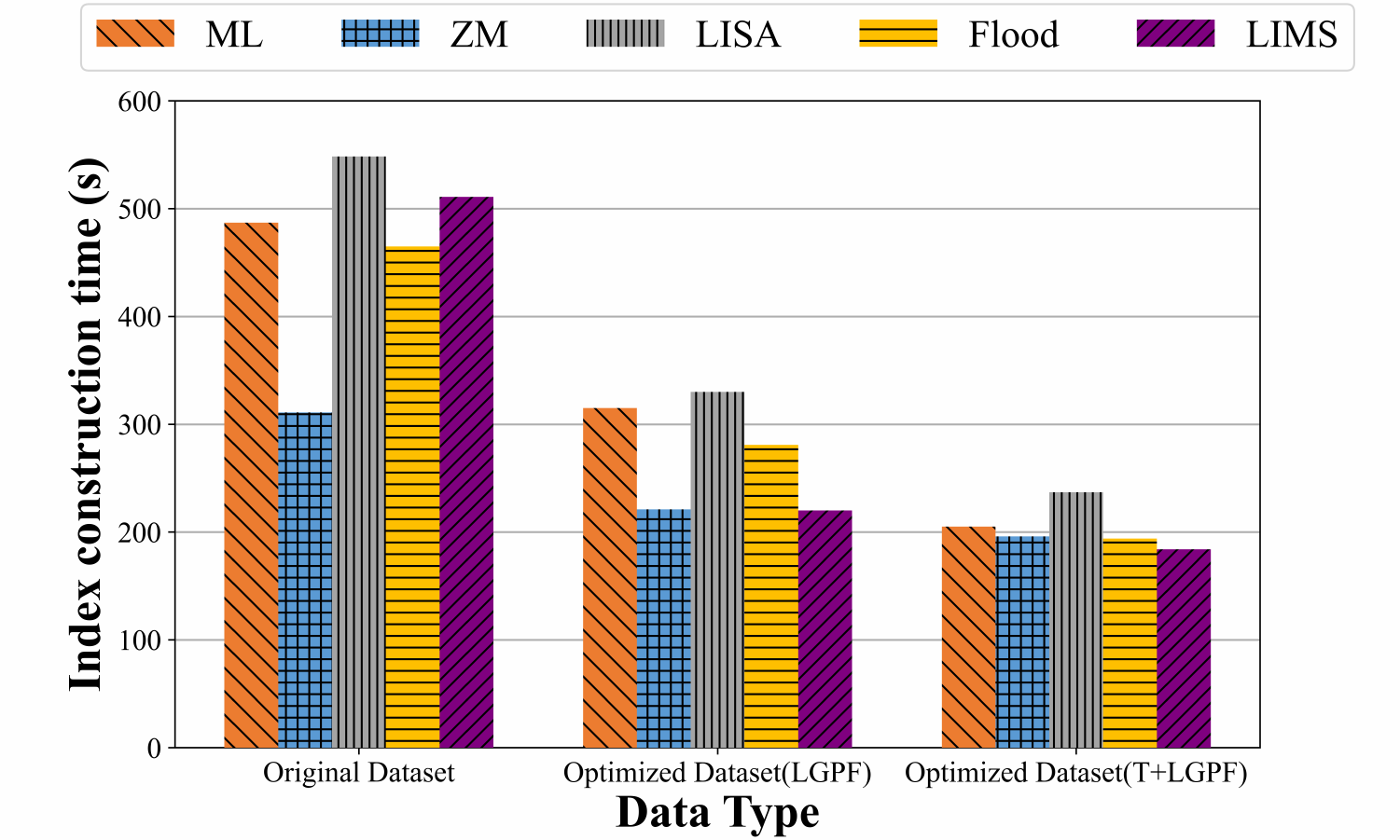}}
		\centerline{(b)}
	\end{minipage}
	\caption{The average query times(a) and index construction times(b) of different multi-dimensional indexes on original dataset and optimized dataset.}
	\label{fig:querytime}
\end{figure}

In the second experiment, we demonstrate the effectiveness of our method in improving the performance of typical multi-dimensional learned indexes by comparing average query times and index construction times.  
We select five typical multi-dimensional learned indexes, which are ML, ZM, LISA, Flood, and LIMS. 
These learned indexes all share the characteristic that their query performance is significantly influenced by the smoothness of the CDF curve which directly impacts "last-mile" searching.
As shown in Fig \ref{fig:querytime}(a) and Fig \ref{fig:querytime}(b), we compare the average query times and index construction times of different indexes on the original, the LPGF optimized, and the T + LPGF optimized datasets. 
The experimental results indicate that both LPGF and T + LPGF methods reduce query time and index construction time for these indexing methods, with T + LPGF exhibiting the most substantial improvements. On average, the query time of T + LPGF is 39.6\% faster than that of the original data set, and the model training time is reduced by 56.7\%.
The average query times of ML, ZM, and LIMS have significantly improved because these index structures heavily depend on data distribution and clustering results, which confirms the effectiveness of our method in enhancing data clustering performance. LISA and Flood, being variants of grid indexes, benefit from our method’s ability to make data more tightly in each grid, resulting in faster indexing. For all the indexes, our method improves the data layout, leading to a smoother CDF curve that accelerates training time and ultimately reduces index construction time.\\

\begin{figure}[pos=!h]
	\begin{minipage}{0.41\linewidth}
		\vspace{3pt}
		\centerline{\includegraphics[width=\textwidth]{14_1.pdf}}
		\centerline{(a)}
	\end{minipage}
	\begin{minipage}{0.58\linewidth}
		\vspace{3pt}
		\centerline{\includegraphics[width=\textwidth]{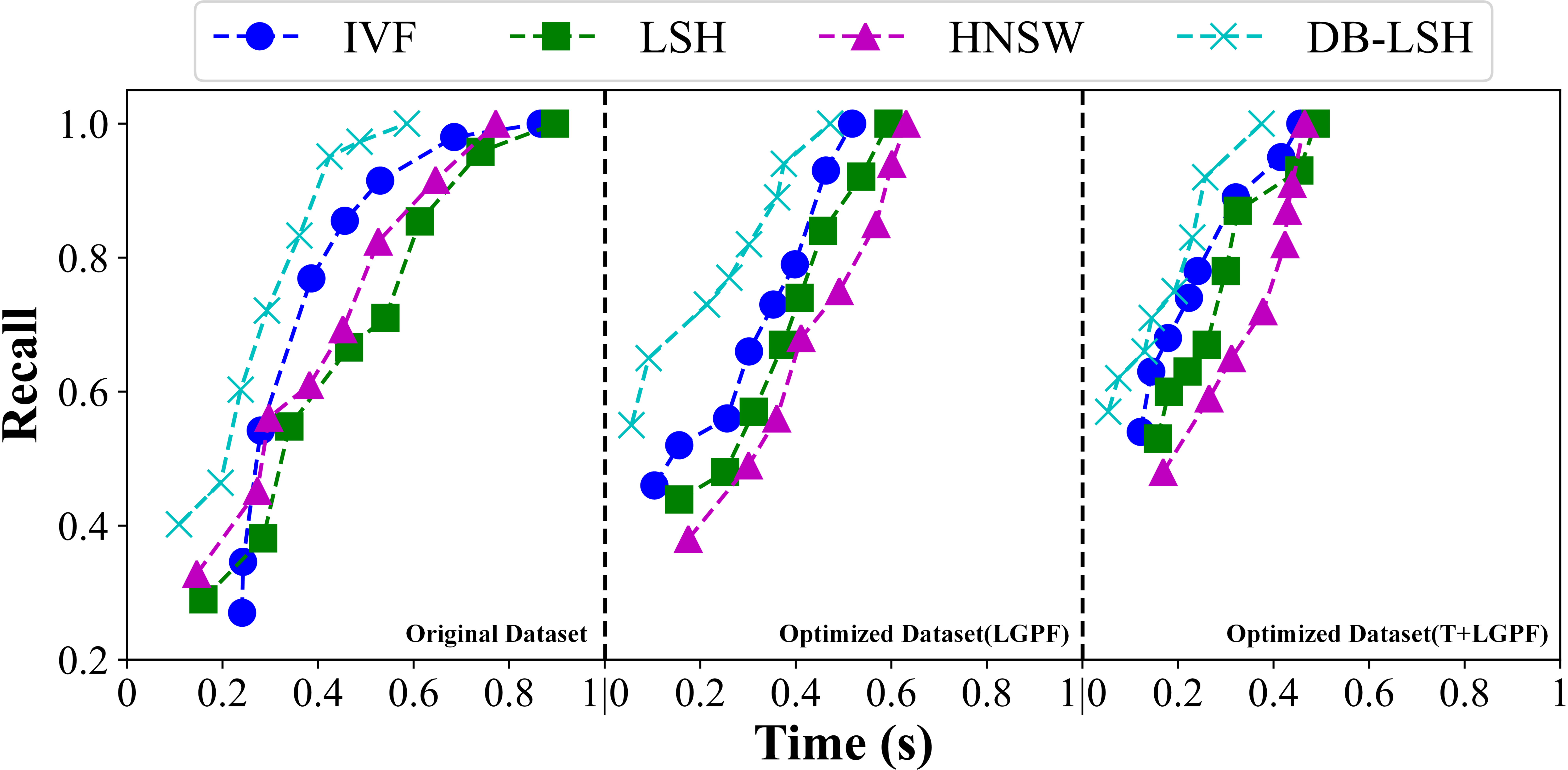}}
		\centerline{(b)}
	\end{minipage}
	\caption{The average query times(a) and recall-time curves(b) of different vector similarity indexes on the original dataset and optimized dataset. Fig (a) shows that the query time of each index exhibits the most significant decrease after applying T+LGPF. Fig (b) shows that T+LGPF achieves the fastest attainment of 100\% recall for each index.}
	\label{fig:querytime2}
\end{figure}
\noindent \textbf{Evaluation 3:\hspace{0.5em}Vector Similarity Index Query Performance Enhancement Evaluation}

Vector similarity index identifies similar data points by comparing their distances in space. Our feature representation process brings similar data points closer together in the space, leading to better clustering results and improved performance of vector similarity indexes.
To demonstrate the effectiveness of our method in enhancing vector index performance, we compare the average query time and the Time-Recall curve. We select four vector indexes: DB-LSH, IVF, HNSW, and LSH. The performance of these indexes is significantly influenced by the data distribution, with datasets exhibiting clear clustering features showing substantial improvements in index performance. As shown in Fig \ref{fig:querytime2}(a) and Fig \ref{fig:querytime2}(b), we compare the average query time and recall rate for queries on the original, LPGF-optimized, and T + LPGF-optimized datasets. From the comparison results, we observe that all vector similarity indexes show the greatest improvements on the T + LPGF-optimized dataset, both in terms of average query time and the time required to achieve 100\% recall. T + LPGF reduces the query time by an average of 42.8\% compared to the original dataset, and the time to reach 100\% recall is shortened by 55.8\%. IVF shows the most significant improvement, as the T + LPGF method groups similar data points within the same cluster as well as clarifies the boundaries of different clusters. This enables the recall rate to reach 100\% in a shorter time. DB-LSH, a hybrid index, relies on the hash function to partition data based on the hash values. The T + LPGF method improves this by tightly clustering similar data points, effectively reducing the error bucket rate caused by hash collisions. Additionally, the advantage of T + LPGF makes the centroids of different clusters more distinct. Using these centroids to build the HNSW graph structure significantly reduces the complexity of each layer, enhancing the data navigation efficiency.

\section{High-dimensional Learned Index}\label{section:highDimensional}
\begin{figure}[pos=!h]
    \centering
    \includegraphics[width=0.8\textwidth]{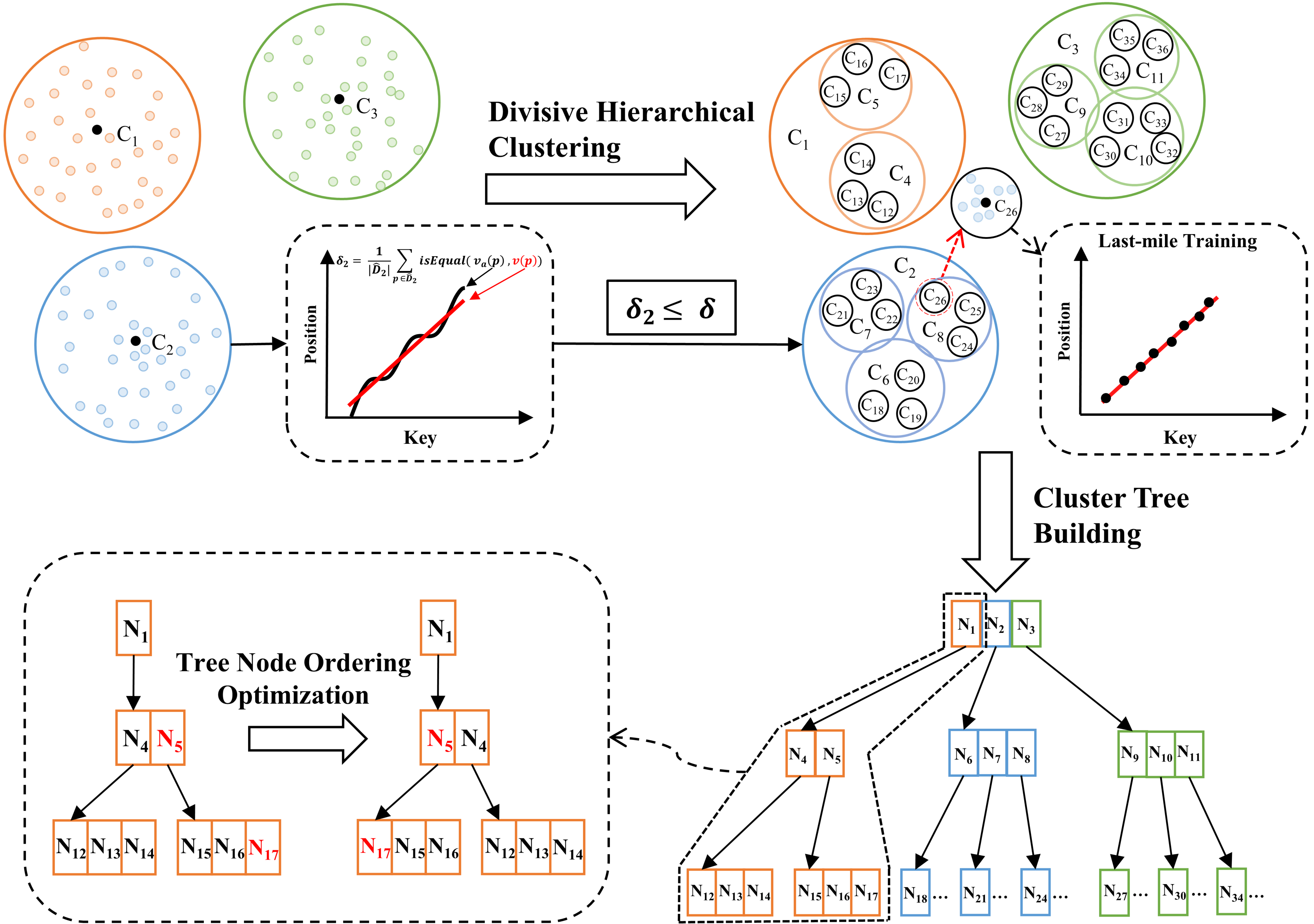}
    \caption{The index construction process consists of divisive hierarchical clustering, cluster tree building, and query-aware optimization.}
    \label{fig:index}
\end{figure}

To fully leverage the advantages of the "Multimodal Data Feature Representation" introduced in Section \ref{section:dataPreparation} and further enhance multimodal data retrieval efficiency, we propose a high-dimensional learned index. Unlike multi-dimensional learned indexes like Flood and Tsunami, etc. which are restricted to multiple numeric queries, or ML and LIMS, etc. which only support single numeric queries and vector queries, our index is designed for rich hybrid queries and performs better on both numeric and vector queries than these indexes. Compared to vector similarity indexes such as FLANN and HNSW, etc. which focus on vector queries on high-dimensional data, our index provides more versatile query options and offers better efficiency and effectiveness on large datasets.
Our index hierarchically organizes clusters and their sub-clusters to form a cluster tree, where the index of the leaf nodes is established through a "last-mile" training approach. 
Furthermore, the index's inner structure can be optimized through our query-aware mechanism. 
An overview of our index is illustrated in Fig \ref{fig:index}, detailing both the construction and optimization processes. 
During the construction phase, divisive hierarchical clustering is employed to form clusters at various levels, corresponding to different levels within the tree structure, with the leaf nodes representing the most granular, indivisible sub-clusters, thereby substantially enhancing the precision of the "last-mile" training for the index. 
Concurrently, the optimization phase employs a query-aware mechanism to refine the internal structure of the index, improving its adaptability to diverse query demands. 
The detailed processes of index construction and optimization are elaborated in Section \ref{section:indexConstruction} and Section \ref{section:IndexOptimize}, respectively.
\subsection{Index Construction}\label{section:indexConstruction}
The process of constructing the index comprises two primary stages, which are divisive hierarchical clustering with "last-mile" training and cluster tree construction.

\subsubsection{Divisive Hierarchical Clustering with "Last-mile" Training}\label{section:hierarchicalClustering}
This stage splits data points through recursive partitioning, creating nested clusters that progressively divide the dataset into smaller and more refined subsets, delineating the hierarchical level of each subset. 
This nested clustering approach enables a hierarchical narrowing of the search scope, followed by "last-mile" training in the most granular subsets for faster lookup. 
The entire Divisive Hierarchical Clustering algorithm can be described as an iterative step of "Division" and "Training based Evaluation", as outlined in Algorithm \ref{clusterAlg}. \\

\noindent \textbf{Step 1:\hspace{0.5em}Division}

In the division step, the intermediate result set from the previous iteration is further divided into smaller subsets, while the training-based evaluation step assesses whether the subsets obtained from the previous iteration have reached an optimal distribution. 
If optimal distribution is reached, the division stops, and the training results for each optimal subset are saved for the "last-mile" search. 
Otherwise, the division continues. 
The iteration stops when all subsets from the previous iteration have achieved an optimal distribution, resulting in the final clustering hierarchy and completing the "last-mile" training for all minimal subsets.

We begin by treating the entire dataset $\hat{D}$, processed through the previous multimodal data feature representation process, as a single cluster. Using the DPC algorithm, we then partition this dataset into initial sub-clusters $\{\hat{D_i}\}$. The selection of the DPC algorithm is based on evaluations of multiple metrics, including data division time, depth of index tree, and average query time. As presented in Table \ref{table:Data Division Results}, different clustering methods are employed in Algorithm 2. Our findings indicate that the DPC method aligns most suitably with our data division strategy. While the K-means algorithm achieves a lower data division time at K=4 compared to the DPC algorithm, it relies on a manually defined parameter K. Conversely, although the Agglomerative algorithm does not necessitate manual parameter setting, its computational complexity is higher as evidenced by significantly longer data partitioning time when compared to DPC.The DPC algorithm is capable of automatically ascertaining the optimal number of sub-clusters, and determining clusters' centroids jointly by density and distance. Moreover, DPC utilizes the results of LPGF where data points are more densely grouped, enhancing the clustering outcomes and ensuring more uniform data distribution within subsets. \\

\begin{table*}[width=1.0\linewidth]
\caption{Data division results of different cluster methods}
\label{table:Data Division Results}
\footnotesize
\begin{threeparttable}
\begin{tabular*}{\tblwidth}{m{80pt}<{\centering}m{80pt}<{\centering}m{80pt}<{\centering}m{80pt}<{\centering}m{80pt}<{\centering}}
\toprule
Cluster Method & Number of Clusters & Data Division Time(s) & Depth of Index Tree & Avg Query Time(ms)\\
\midrule
\multirow{3}*{K-means} 
  & K = 2 & 1276.92 & 5 & 72.06 \\
~ & K = 3 & 877.06 & 3 & 26.55\\
~ & K = 4 & 611.45 & 2 & 9.11\\

\multirow{1}*{Agglomerative} & - & 1135.66 & 2 & 7.14 \\

\multirow{1}*{DPC} & - & 674.54 & 2 & 6.42 \\
\bottomrule
\end{tabular*}
\begin{tablenotes}
\footnotesize
\item[a] "-" indicates that the number of clusters is determined by the algorithm itself.
\end{tablenotes}
\end{threeparttable}
\end{table*}

\noindent \textbf{Step 2:\hspace{0.5em}Training based Evaluation} 

\begin{figure}[pos=!h]
	
	\begin{minipage}{0.32\linewidth}
		\vspace{3pt}
		\centerline{\includegraphics[width=\textwidth]{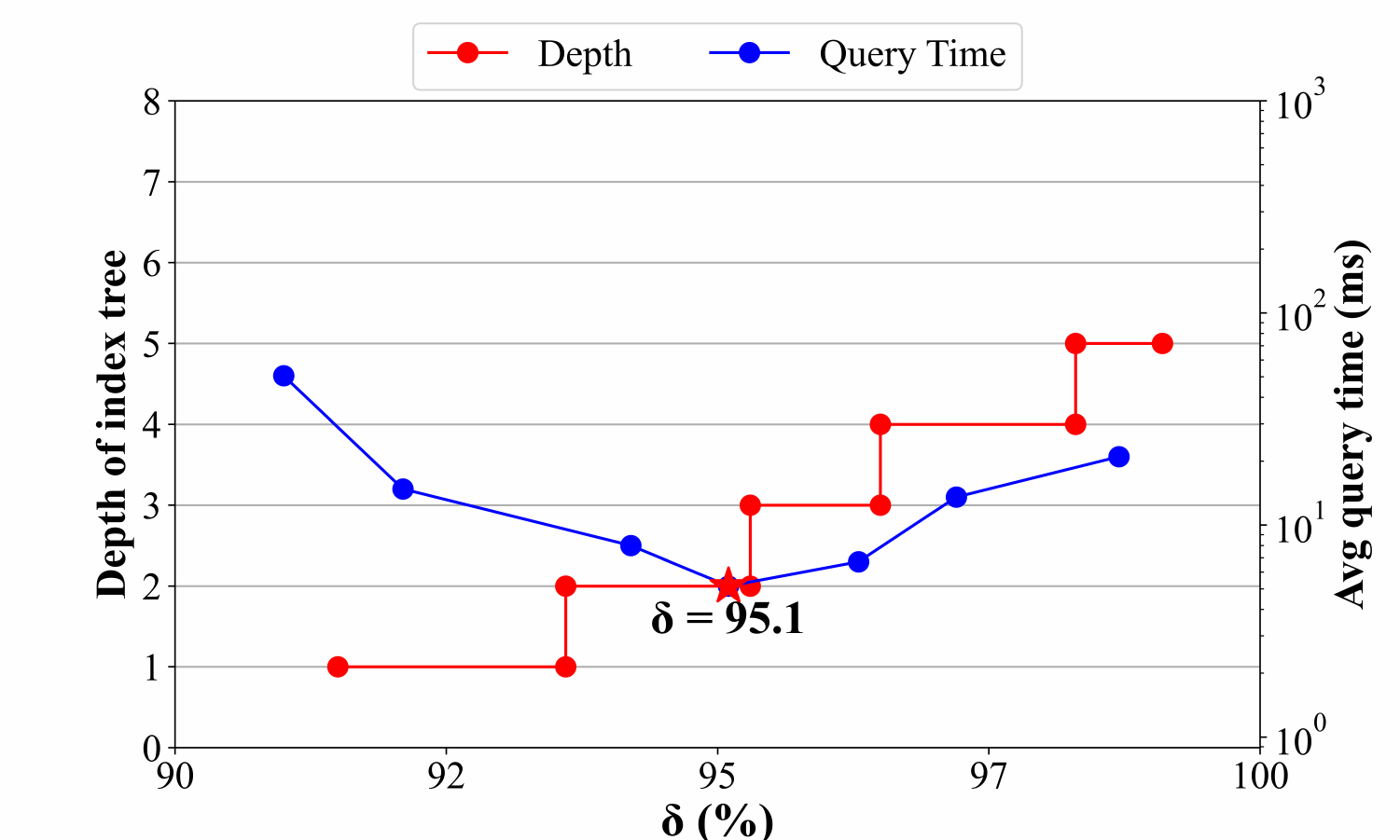}}
		\centerline{(a)}
	\end{minipage}
	\begin{minipage}{0.32\linewidth}
		\vspace{3pt}
		\centerline{\includegraphics[width=\textwidth]{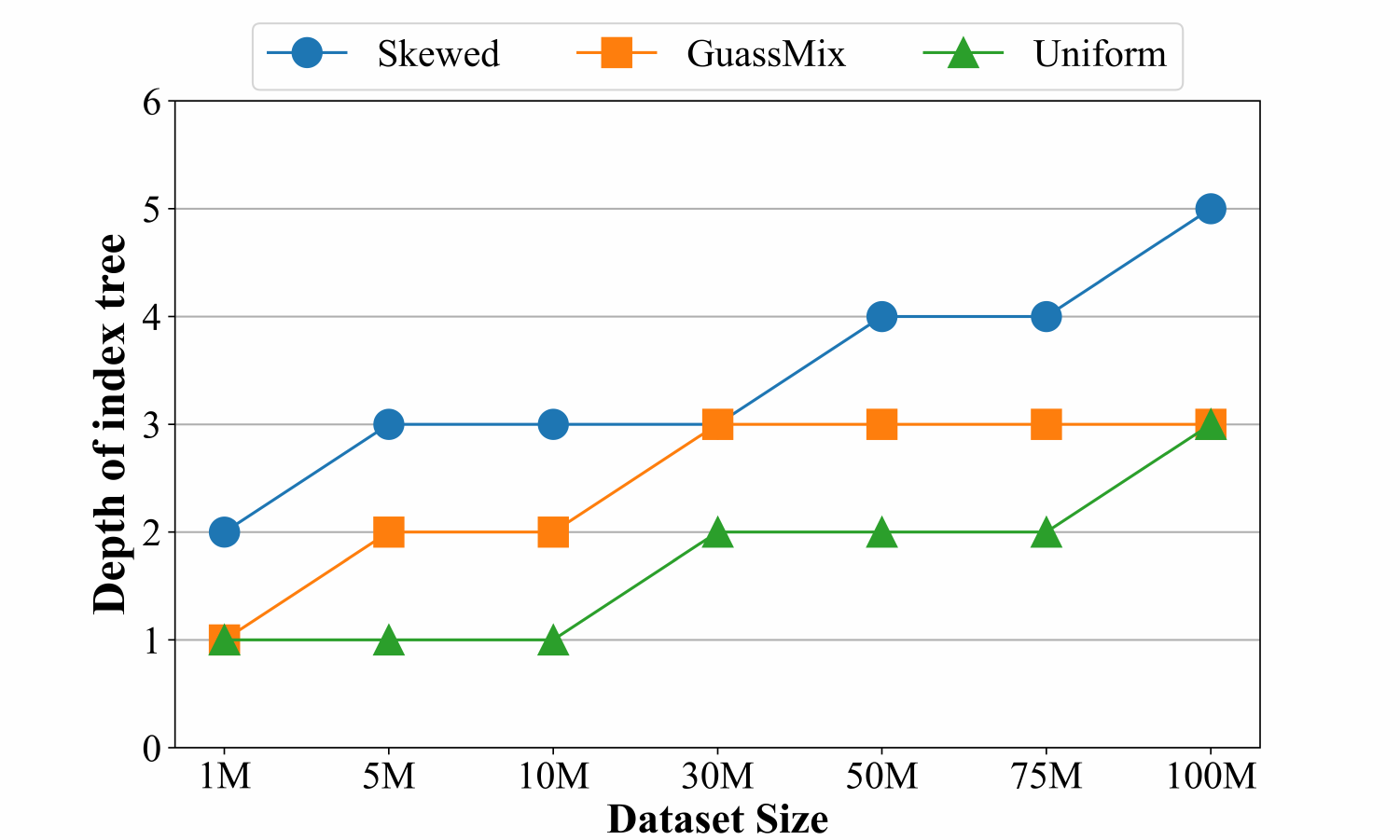}}
		\centerline{(b)}
	\end{minipage}
	\begin{minipage}{0.32\linewidth}
		\vspace{3pt}
		\centerline{\includegraphics[width=\textwidth]{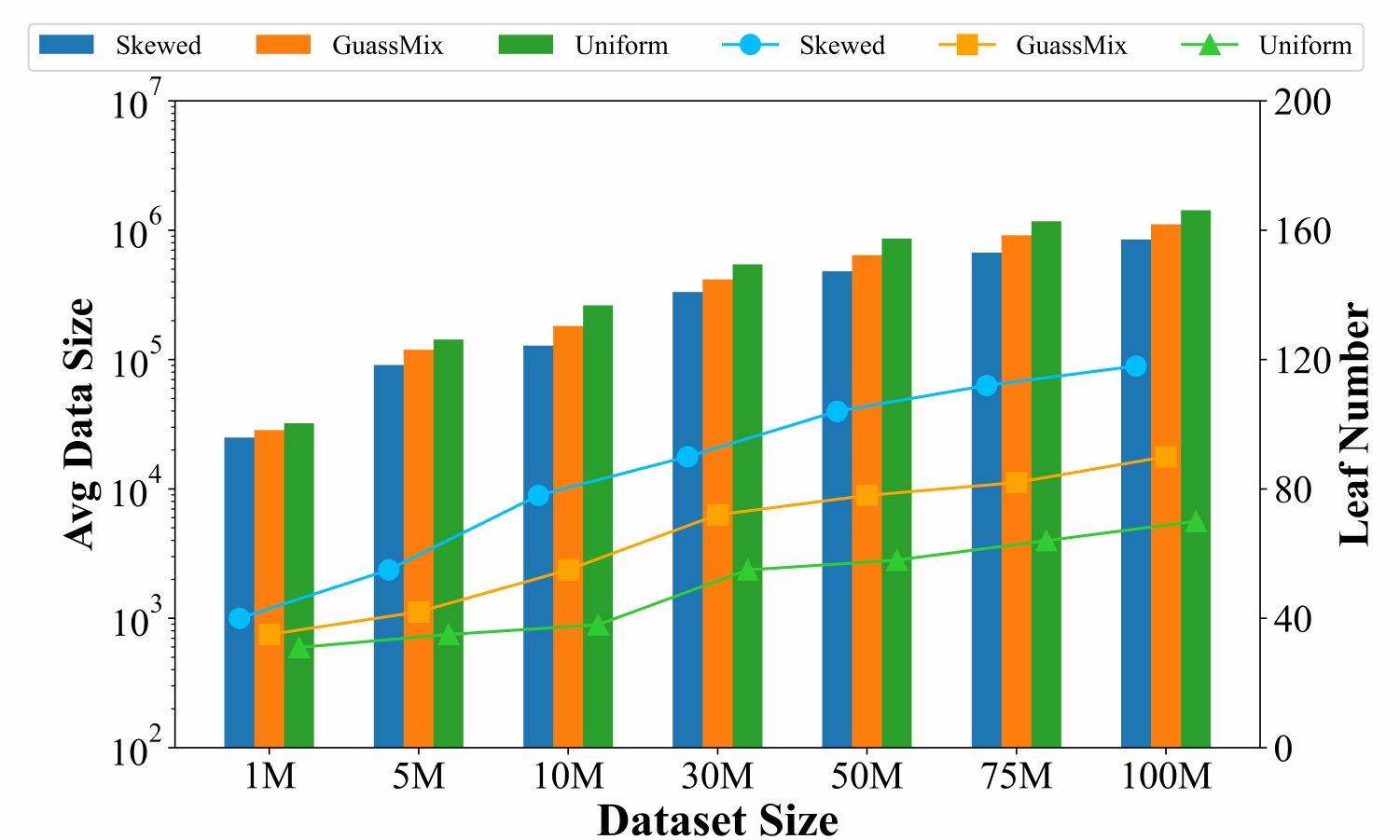}}
	 
		\centerline{(c)}
	\end{minipage}
 
	\caption{Calculation of $\delta$(a), depth of index tree(b) and leaf node characteristics for different dataset sizes(c). The results from (b) and (c) demonstrate that as the dataset size increases, there is a gradual convergence in both the depth of our index tree and the number of leaf nodes, rather than a significant increase.}
	\label{fig-ziEx}
\end{figure}

Following the step of obtaining initial sub-clusters $\{\hat{D_i}\}$, we assess whether the data points in each sub-cluster are well-distributed to support accurate "last-mile" training and the corresponding efficient searching, thereby determining whether further subdivision of the cluster is necessary. We represent the centroid of $\hat{D_i}$ as $C_i$ using the Average Mass method \citep{mass}, and calculate the distance from each point $p$ in $\hat{D_i}$ to $C_i$. If the points in $\hat{D_i}$ are sufficiently uniform, distances of all points in it should be evenly spread, so that a simple linear regression model is sufficient to fit their distribution. Therefore, we predict the search position $v$ of a given lookup point $p$ using a model $v=F(p)*|\hat{D_i}|$, where $F(p)$ is the estimated CDF obtained through linear regression, which estimates the likelihood that a point's distance to $C_i$ is less than or equal to the distance from the lookup point $p$. 
The accuracy of this model prediction validates the goodness that the CDF fits and further reflects the distribution of the points in $\{\hat{D_i}\}$. 
Therefore, we want the hit ratio of prediction for all points in the cluster to be greater than a given threshold $\delta$. 
Let $v(p)$ be the predicted searching position of point $p$ whose actual position is $v_a(p)$, our requirement is formulated as: 
\begin{equation}
\frac{1}{|\hat{D_i}|}\sum_{p\in \hat{D_i}}{IsEqual(v_a(p),v(p))}\geq \delta    
\end{equation}
where the parameter $\delta$ is set to the optimal value 95.1\% based on tuning with considerations of tree depth and query time (see Fig \ref{fig-ziEx}(a)). 
If this requirement is fulfilled, the points in $\hat{D_i}$ are uniform enough and do not require further partitioning. 
In this scenario, the model $F(p)$ will be stored for predicting the search position within $\hat{D_i}$. Otherwise, $\hat{D_i}$ must be further subdivided to achieve a better data distribution within its sub-clusters. 
  


Through experimental verification, we set $\delta$ to 95.1\%. The experiments also demonstrate that the depth of our tree does not significantly increase with the rapid growth of data volume. Detailed results are shown in Fig \ref{fig-ziEx}(a)-(c).

Fig \ref{fig-ziEx}(a) shows the cluster tree depth (we set the depth of the root node to 0) and average query time for various $\delta$ values. It can be observed that the best average query performance is achieved when $\delta$=95.1\%. Although smaller $\delta$ values can further reduce the cluster tree depth, they lead to larger leaf nodes, requiring the scanning of more irrelevant data to obtain query results. Conversely, higher $\delta$ values increase the cluster tree depth, resulting in more node traversals during the query process.

Fig \ref{fig-ziEx}(b) presents experimental results on the depth of the cluster tree for Uniform, GuassMix, and Skewed datasets with different dataset sizes. The depth of the cluster tree increases gradually with the data volume, but the overall growth trend of the tree layers becomes more gradual. From the perspective of data distribution, the Skewed dataset results in the deepest cluster tree, reaching five layers. The maximum depths of the cluster trees for the GuassMix and Uniform datasets are the same, but the average depth of the cluster tree for GuassMix is significantly greater than that for Uniform. Though data skewness causes some tree branches to be more deeply divided, the depth does not grow excessively using our optimized feature representation strategies.

Fig \ref{fig-ziEx}(c) shows the number of leaf nodes, and the average data volume in the leaf nodes for Uniform, GuassMix, and Skewed datasets with different dataset sizes. As the data volume increases, both the number of leaf nodes and the average data volume within the leaf nodes increase. Observing the overall trend in the growth of leaf nodes, the data volume within each leaf node increases significantly, while the number of leaf nodes does not increase significantly. This is because the data-aware partitioning strategy controls node splitting based on the average error value, which is independent of the data volume.

\begin{algorithm}
\caption{Divisive Hierarchical Clustering}
\label{clusterAlg}
\renewcommand{\algorithmicrequire}{\textbf{Input:}} 
\renewcommand{\algorithmicensure}{\textbf{Output:}} 
\begin{algorithmic}[1] %
    \REQUIRE dataset $\hat{D}$ to be divided
    \ENSURE an array $A$ containing all subsets of $S$
    \STATE Initialize a queue of datasets $DatasetQueue = \{\}$
    \STATE $\{\hat{D_i}\}=DPC(\hat{D})$, $DatasetQueue.push(\{\hat{D_i}.resort()\})$
    \WHILE{ $!DatasetQueue.isEmpty()$}
    \STATE $S = DatasetQueue.pop()$
    \STATE $\{\hat{S}\},\{\hat{C}\} = DPC(LPGF(S))$ /*$\{\hat{S}\}$ is set of sub-clusters of S, and $\{\hat{C}\}$ is set of centroids of $\{\hat{S}\}$*/
    \STATE $C_p = getCentroid(\{\hat{C}\})$ /*computing parent cluster S's centroid using sub-clusters' centroid set $\{\hat{C}\}$*/
    \STATE $\{\hat{S}'\},\{\hat{C}'\} = resort(\{\hat{S}\},\{\hat{C}\},C_p)$
    \FOR {$\text{$\hat{S_i}$ in $\{\hat{S}'\}$}$}
    \STATE $\{k\}=dist(C_i, p\in \hat{S_i})$
    \STATE construct position prediction model $v_i = F_i(k)*|\hat{S_i}|$, where $F_i$ is the CDF model obtained from linear regression, formulated as $F_i(k)=a*k+b$
    \STATE $A.add(\hat{S_i})$
    \IF{$\frac{1}{|\hat{S_i}|}\sum_{p\in \hat{S_i}}{IsEqual(v_a(p),v(p))}\leq \delta$}
    \STATE $DatasetQueue.push(\hat{S_i})$
    \ENDIF
    \ENDFOR
    \ENDWHILE
    \RETURN $A$
\end{algorithmic}
\end{algorithm}

\subsubsection{Cluster Tree Construction}\label{section:ClusterTree}

Following the divisive hierarchical clustering stage, the dataset is partitioned into nested sub-clusters, establishing a hierarchical structure of data points. We construct a cluster tree based on this partition result, encapsulating the order of subsets at each hierarchical level and the organization of each subset. At the top of this tree is the root node, representing the unique cluster that contains the entire dataset. As we traverse downward through the tree, each internal node represents a cluster that has been subdivided into smaller sub-clusters. An internal node with $n$ children signifies that its corresponding cluster has been partitioned into $n$ distinct sub-clusters. Furthermore, for each parent cluster $D_p$ with $n$ sub-clusters $\{D_s\}$, we calculate $D_p$'s centroid $C_p$ by Average Mass method using $\{C_s\}$, which is a set of centroids for $\{D_s\}$. The sub-clusters $\{D_s\}$ are then sorted based on the distances from their centroids $\{C_s\}$ to $C_p$ in increased order. This sorting is crucial for determining the search order among these sub-clusters, which is an important aspect when constructing the cluster tree. The leaf nodes represent the final, and undivided subsets. This hierarchical organization enables efficient data retrieval by progressively narrowing down the search space from the root to the pertinent leaf nodes.

To efficiently traverse the cluster tree, each node needs to store specific information that facilitates data searches within the node or navigation to a child. Based on this requirement, we record a tuple $t$ for each node. For non-leaf nodes, $t=\{C,R,L\}$, where $C$ is the centroid of the cluster corresponding to the node, and $R$ represents the cluster’s radius, defined as the Euclidean distance from $C$ to the farthest point in the cluster. Together, $C$ and $R$ define the search range for the data points within the node. $L$ is a pointer that directs to a list of its child nodes where these nodes are sorted and can be scanned based on the spatial order based on the sub-clusters' distances to the parent node's centroid.
For leaf nodes, $t=\{C,R,M\}$. Again, $C$ and $R$ limit the search range of each leaf node. $M$ points to the search model trained in the previous step, which is used to predict the search position of an input lookup point, thereby facilitating rapid query responses. 

\begin{algorithm}
\caption{Tree Node Ordering Optimization}
\label{al:3}
\renewcommand{\algorithmicrequire}{\textbf{Input:}} 
\renewcommand{\algorithmicensure}{\textbf{Output:}} 
\begin{algorithmic}[1] %
\REQUIRE initial tree node list $L_i$, query workload $Q$
\ENSURE optimized bucket order $\hat{L_i}$
\STATE /*$B$ is an array of scanned times of each node after executing query workload $Q$, $t$ is query time*/
\STATE $B, t = ExecuteQuery(Q,L_i)$
\STATE $\hat{L_i}$=sortInDescendingOrder($L_i$, B)
\STATE startIter=0, endIter=0
\WHILE{endIter<|$\hat{L_i}$|}
\STATE size=0
\WHILE{endIter<|$\hat{L_i}$| and B[startIter]==B[endIter]}
\STATE size+=1;endIter+=1;
\ENDWHILE
\IF{size>1}
\STATE $L_{sub}=Permutations(\hat{L_i}, \text{startIter}, size-1)$
\FOR{$l$ in $L_{sub}$}
\STATE $L_{l}=Reconstruct(\hat{L_i}, \text{startIter}, size-1)$
\STATE $t_l = ExecuteQuery(Q, L_l)$
\IF{$t_l$<$t$}
\STATE $t=t_l$,$\hat{L_i}=L_l$
\ENDIF
\ENDFOR
\ENDIF
\STATE startIter=endIter
\ENDWHILE
\RETURN $\hat{L_i}$
\end{algorithmic}
\end{algorithm}

\subsection{Index Optimization}\label{section:IndexOptimize}
Regarding query efficiency, different index structures are suited to different query workloads. In this section, we focus on optimizing and reordering sibling nodes (sub-clusters) that share the same parent node (parent cluster) based on the query-aware mechanism. This optimization aims to refine the index structure, thereby improving query performance and efficiency. Note that in our optimization strategy, we focus exclusively on the arrangement of all child nodes under each parent node while keeping the inheritance relationships unchanged. For instance, consider the scenario in Fig \ref{fig:index} where there are three non-leaf nodes ($N_1$,$N_4$,$N_5$) in the left branch of the cluster tree, each pointing to a list of their respective child nodes: $L_1=\{N_{4}, N_{5}\}$, $L_4=\{N_{12}, N_{13}, N_{14}\}$, $L_5=\{N_{15}, N_{16}, N_{17}\}$. Our method is to optimize the order of nodes within each list (e.g. $L_1$ can be reordered as $L_1=\{N_{5}, N_{4}\}$), without altering the parent-child relationships between nodes (e.g. $N_{5}$ cannot be moved to $L_4$).

This optimization is based on our observation that query patterns can vary with different query workloads, leading to certain nodes being accessed more frequently than others. By reordering the sibling nodes based on their access frequency, the index can be made more efficient. For example, as illustrated in Fig \ref{fig:index}, if our lookup point ultimately resides in $N_{18}$, with the initial tree structure six nodes are required for scanning, while after optimization, the number of nodes scanned could be reduced to three, significantly saving query time.

Now we delve into the details of our optimization algorithm. Each non-leaf node $N_i$ stores a pointer to a sorted list which we denote as $L_i$. $L_i$ comprises all child nodes of the parent node $N_i$, initially ordered by their distance to the parent node’s centroid. We iteratively traverse all non-leaf nodes and, in $i^{th}$ iteration, our optimization goal is to adjust the ordering of $N_i$'s child node list $L_i$ to obtain a new list $\hat{L_i}$, so that the query time of a given query workload $Q$ is minimized:
\begin{equation}
\hat{L_i}=argmin\sum_{q\in Q}{QueryTime(q,L_i)}    
\end{equation}

The optimization algorithm is shown in Algorithm \ref{al:3}. Its key idea is to move frequently accessed nodes to the beginning of the list to reduce the number of cross-leaf scans needed to find these "hot" nodes. We record the visit frequency of each node and sort them in descending order, placing the most frequently visited bucket at the head of the list. For nodes with the same visit frequency, we employ a brute-force approach to test each possible order and select the one that results in the minimum query time.



\section{Experiments}\label{section:experiments}
We implement our MQRLD index in Scala and apply it to the data lake and vector database respectively. The data lake is deployed in a distributed cluster environment consisting of three nodes, using Apache Spark as the computing engine and Apache Hudi for data lake management. Each node is equipped with 64-bit Ubuntu 18.04, Intel(R) Core(TM) i7-11700F CPU @ 2.50 GHz, and 16 GB RAM. The vector database is deployed on a single-node server using Milvus as its management framework. The server's hardware configuration includes 64-bit Ubuntu 20.04, two Intel(R) Xeon(R) Gold 6226 CPUs @ 2.90GHz, 256 GB RAM, and an NVIDIA RTX A5000 GPU.

\subsection{Experimental Setup}
\subsubsection{DataSets}
Table \ref{table:dataset} shows the seven real-world datasets with different characteristics and three synthetic datasets we used in our experiments. The real-world datasets are also used in the previous work.

\begin{table}[width=.7\linewidth,cols=5,pos=h]
\caption{Summary of datasets.}
\label{table:dataset}
\begin{tabular*}{\tblwidth}{m{2cm}<{\centering}m{2cm}<{\centering}m{2cm}<{\centering}m{2cm}<{\centering}m{2cm}<{\centering}}
\toprule
Datasets & Type &Cardinality & Dim. & Size(GB) \\
\midrule
OSM & real & 105M & 6 & 5.04 \\
Taxi & real &184M & 5 & 11.8 \\
Stocks & real &210M&4 & 13.2  \\
Yelp & real &6.99M &64 & 13.2  \\
Color & real & 1.28M&32 & 4.2   \\
Forest & real & 0.56M&12 & 1.5   \\
AI Challenger & real & 0.30M&N.A. & 38  \\
SIFT1B & real & 0.1-100M & 128 & 16.2 \\
LAION400M&real&1-100M&1026&83.9\\
Uniform & synthetic & 1-100M&3-16 & N.A.  \\
GuassMix & synthetic &1-100M &3-16 & N.A.  \\
Skewed & synthetic &1-100M &3-16 & N.A.  \\
\bottomrule
\end{tabular*}
\end{table}

\textbf{OSM}\footnote{https://download.geofabrik.de} is a spatial dataset consisting of 105 million records randomly sampled from North America in the OpenStreetMap dataset. Each record includes six attributes, such as ID, timestamp, GPS coordinates, etc.
\textbf{Taxi}\footnote{https://www1.nyc.gov/site/tlc/about/tlc- trip- record- data.page} is randomly sampled from records of yellow taxi trips in New York City in 2018 and 2019. We select five attributes from the dataset: the number of passengers, trip distance, pickup, drop-off locations, and total fare, to form a five-dimensional dataset.
\textbf{Stock}\footnote{https://www.kaggle.com/datasets/ehallmar/daily-historical-stock-prices-1970-2018}  contains 210 million records, consisting of daily historical stock price data for over 6,000 stocks from 1970 to 2018. We focus on four features: trading volume, opening price, high price, and adjusted closing price.
\textbf{Yelp}\footnote{https://www.yelp.com/dataset}
consists of 6,990,280 user reviews of business places (e.g., cafes, restaurants, hotels, shops) from 11 metropolitan areas. For each review, we extract a set of 64-dimensional semantic vectors as the experimental dataset.
\textbf{Color}\footnote{https://image-net.org/download-images}
is a 32-dimensional image feature dataset extracted from the ImageNet dataset, which contains 1,281,167 records.
\textbf{Forest}\footnote{https://www.kaggle.com/c/forest-cover-type-prediction/data} is collected by the US Geological Survey and the US Forest Service, which includes 565,892 12-dimensional records.
\textbf{AI Challenger}\footnote{https://challenger.ai} is a multimodal dataset that contains more than 300,000 images with text descriptions. In order to simulate the rich hybrid queries scenario, we define two numeric attributes for each image: the length of the text description and the size of the image file.
\textbf{SIFT1B}\footnote{http://corpus-texmex.irisa.fr/} is a standard benchmark dataset that contains 1 billion of 128-dimensional feature vectors. We select 0.1M-100M pieces of data from it.
\textbf{LAION400M}\footnote{https://laion.ai/}  is a multimodal dataset containing 400 million image-text pairs. To generate rich hybrid queries, we concatenate two 512-dimensional feature vectors representing the image and text, along with a 2-dimensional scalar vector denoting the original dimensions of the image, resulting in 1026-dimensional vectors.
\textbf{Synthetic Datasets} include datasets with three different distributions, which are Uniform, GuassMix, and Skewed. The dimensions and data volumes of these datasets are adjustable. By default, we use a 3-dimensional dataset with a volume of 10 million records.

\subsubsection{Query Generation}
In the experiments, we assess the performance of range queries using the OSM, Taxi, and Stock datasets, and KNN queries using the Yelp, Color, and Forest datasets. Additionally, we evaluate the performance of rich hybrid queries based on the four basic query types proposed in Section \ref{section:hybridQuery}, using the AI Challenger dataset.

The range queries and KNN queries, as common query operations, are each constituted by a record and its corresponding parameter, with the records being randomly selected from the dataset intended for querying. For range queries, the parameter is the query selectivity (i.e., the percentage of records that the query is expected to cover), which is set to a default of 10\%, with a range between 0.1\% and 20\%. For KNN queries, the parameter is the number of query results $K$, which defaults to 1,000, ranging from 1 to 10,000.

For rich hybrid queries, we select three typical types of query among the rich hybrid queries shown in Fig \ref{fig4}, which are V.R $\bigoplus$ V.K, V.R $\bigoplus$ N.R and N.R $\bigoplus$ V.K. To verify the ability of MQRLD to cope with more complex queries, we also involve a query type of V.R $\times$ N, which is a combination of N instances of V.R (N $\in$ [2,5]). These queries are all generated from the AI Challenger dataset, whose image and text data can be embedded in multiple vector attributes for performing V.R or V.K queries.
The parameters involved in rich hybrid queries include the query radius $R$ for V.R query, which defaults to 4\% with a range from 0.1\% to 10\%; the parameter $K$ for V.K query, which defaults to 100 with a range from 1 to 1,000; and the selectivity for N.R query, which defaults to 10\% with a range from 1\% to 20\%.

\subsubsection{Competitors}
We compare our index with other multi-dimensional and high-dimensional indexes, as shown in Table \ref{table:competitor}. All methods store data using DataFrame structures with data buckets as the smallest physical storage unit and apply the same optimizations where applicable.

\begin{table}[width=.8\linewidth,cols=4,pos=h]
\caption{Competitors.}
\label{table:competitor}
\begin{tabular*}{\tblwidth}{m{5.0cm}<{\centering}m{3cm}<{\centering}m{1.8cm}<{\centering}m{1.8cm}<{\centering}}
\toprule
Index & Type & Range & KNN \\
\midrule
ZM \citep{ZM} & multi-dimensional & \large\ding{51} & \XSolidBrush \\
ML \citep{Ml-index} & multi-dimensional & \large\ding{51} & \large\ding{51} \\
LISA \citep{LIMS} & multi-dimensional & \large\ding{51} & \large\ding{51} \\
Qd-tree \citep{Qd-tree} & multi-dimensional & \large\ding{51} & \XSolidBrush \\
LIMS \citep{LIMS} & multi-dimensional & \XSolidBrush & \large\ding{51} \\
Flood \citep{Flood}& multi-dimensional & \large\ding{51} & \XSolidBrush \\
Tsunami \citep{Tsunami}& multi-dimensional & \large\ding{51} & \XSolidBrush \\
R*-tree \citep{r_tree2}& multi-dimensional & \large\ding{51} & \large\ding{51} \\
SPB-tree \citep{spb_tree} & multi-dimensional & \XSolidBrush & \large\ding{51} \\
M-tree \citep{m_tree} & multi-dimensional & \XSolidBrush & \large\ding{51} \\
HNSW\citep{HNSW}&high-dimensional&\XSolidBrush & \large\ding{51}\\
IVF\citep{IVF}&high-dimensional&\XSolidBrush & \large\ding{51}\\
LSH\citep{LSH}&high-dimensional&\XSolidBrush & \large\ding{51}\\
DB-LSH\citep{DB_LSH}&high-dimensional&\XSolidBrush & \large\ding{51}\\
\bottomrule
\end{tabular*}
\end{table}

\subsubsection{Evaluation Metrics}
To evaluate the performance of the retrieval, we adopt the following two key metrics.

(1) \textit{Query time.} To more accurately assess the performance of each index, we perform five repeated experiments for each group of generated queries and take the average query time as the experimental result. 

(2) \textit{Recall.} Recall is the proportion of true relevant neighbors retrieved by a KNN query out of all the actual relevant neighbors. It measures the completeness of the search in finding the true nearest neighbors.

(3) \textit{CBR.} We use CBR to measure the dispersion of data access during the query process. A detailed definition refers to Section \ref{section:queryAware}. A lower CBR indicates that the query is more concentrated within fewer data buckets, implying higher query efficiency and better indexing performance.

\subsection{Evaluation of Range Query for Multi-dimensional Indexes}
\begin{figure}[pos=!h]
    \centering
    \includegraphics[width=\textwidth]{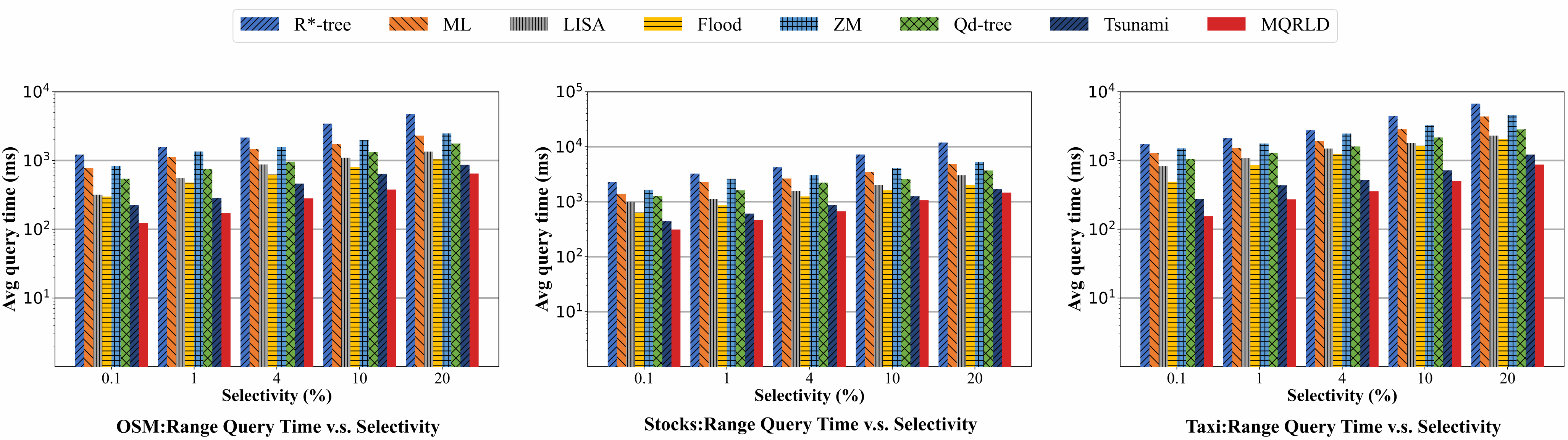}
    \caption{Range query evaluation results for multi-dimensional indexes on real datasets.}
    \label{fig:Exp-range}
\end{figure}

In this section, we compare MQRLD with our competitors for multi-dimensional range queries. Fig \ref{fig:Exp-range} shows the query time for each optimized index on the OSM, Taxi, and Stocks dataset. 
MQRLD uses the multimodal data representation and high-dimensional learned index construction process in Section \ref{section:dataPreparation} and Section \ref{section:highDimensional}, while we tune the other methods as much as possible (e.g., ordered dimensions by selectivity and tuned the page sizes). 
This represents the best-case scenario for the other indexes, that the database administrator had the time and ability to tune the index parameters. 
To ensure the generalizability of our experimental results, we test the average query time of various index structures under different query selectivities. 
Our observations indicate that while the average query time for all indexes increases with higher query selectivity, MQRLD consistently outperforms other index structures across different datasets:

(1) Compared to tree-based indexes (R*-tree and Qd-tree), MQRLD demonstrates a significant improvement in query efficiency, with average performance gains of 8.1× and 3.9×, respectively. This is attributed to MQRLD's ability to better capture the data distribution in high-dimensional space during its construction, retaining the advantages of tree-based indexes while ensuring similar data points are stored in the same or neighboring nodes. 

(2) Against ZM, ML, and LISA, MQRLD shows average query efficiency improvements of 5.8×, 5.1× and 3.1×, respectively. A common feature of MQRLD and these indexes is the use of one-dimensional value mapping to expedite data lookup. For instance, LISA uses a mapping function based on data distribution to map similar data into the same segment. However, our data representation strategy goes a step further by clustering similar data based on their characteristics, resulting in a more efficient index structure. 

(3) MQRLD achieves 2.4× and 3.3× speedup on query time compared to Flood and Tsunami, respectively. Although Flood’s cell division and Tsunami's grid augmentation strategy can learn from query workloads to adjust its data layout, it is evident that the improvements from our multimodal data representation process are substantially more effective.

\subsection{Evaluation of KNN Query for Multi-dimensional Indexes}
\begin{figure}[pos=!h]
    \centering
    \includegraphics[width=\textwidth]{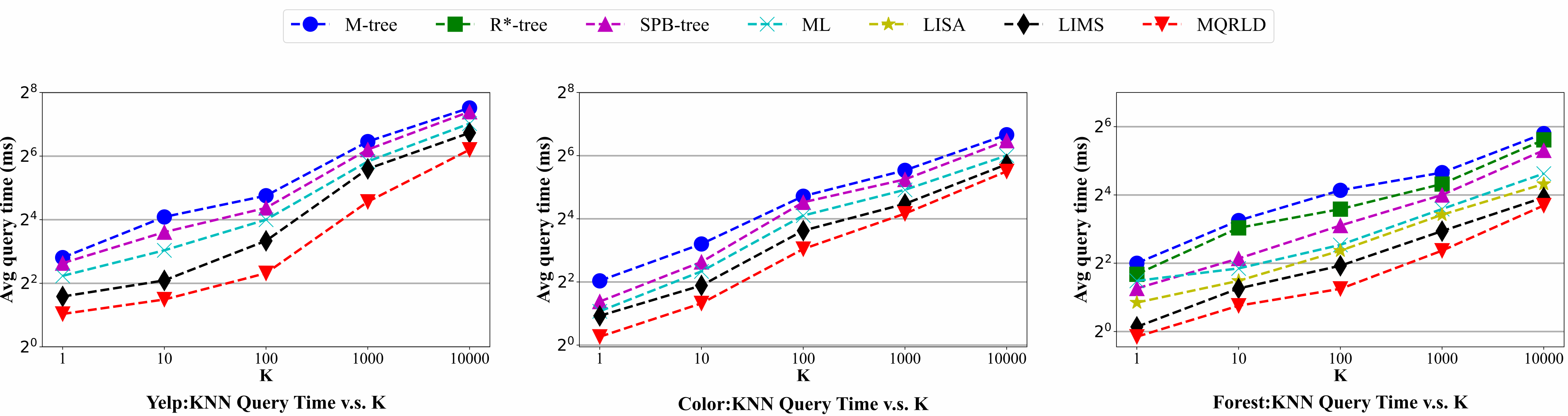}
    \caption{KNN query evaluation results for multi-dimensional indexes on real datasets. Due to incomparable results beyond 12 dimensions for R*-tree and LISA, we disregard them on the Yelp and Color.}
    \label{fig:Exp-KNN}
\end{figure}

Next, we evaluate MQRLD's performance on KNN queries using the Yelp, Color, and Forest datasets. Similarly, under the same query workload, we optimize other indexes as much as possible. It can be observed from Fig \ref{fig:Exp-KNN} that MQRLD maintains relatively low average query times across different $K$ values:

(1) Compared with non-learned multi-dimensional indexes, MQRLD shows average query efficiency improvements of 4.2×, 3.1×, and 3.3× over M-tree, R*-tree, and SPB-tree respectively. This is because the leaf nodes of MQRLD contain highly similar data, and have a prediction model to improve the query speed.

(2) Compared with learned multi-dimensional indexes, MQRLD improves query efficiency by an average of 2.2×, 1.9×, and 1.5× over ML, LISA, and LIMS datasets. ML uses clustering for data mapping but suffers from prediction errors and longer search times due to its global predictive model. LISA and LIMS utilize local predictive models to improve the query efficiency for "last-mile" search, but their models are complex and the mapping process is time-consuming. In contrast, MQRLD trains simple linear regression models at each leaf node, achieving more accurate and faster queries. This is enabled by MQRLD's data-aware and query-aware data representation and high-dimensional learned index construction strategy, which keep prediction model errors within a low threshold as well as improve MQRLD’s query efficiency.
\subsection{Evaluation of CBR for Multi-dimensional Indexes}
\begin{figure}[pos=!h]
    \centering
    \includegraphics[width=\textwidth]{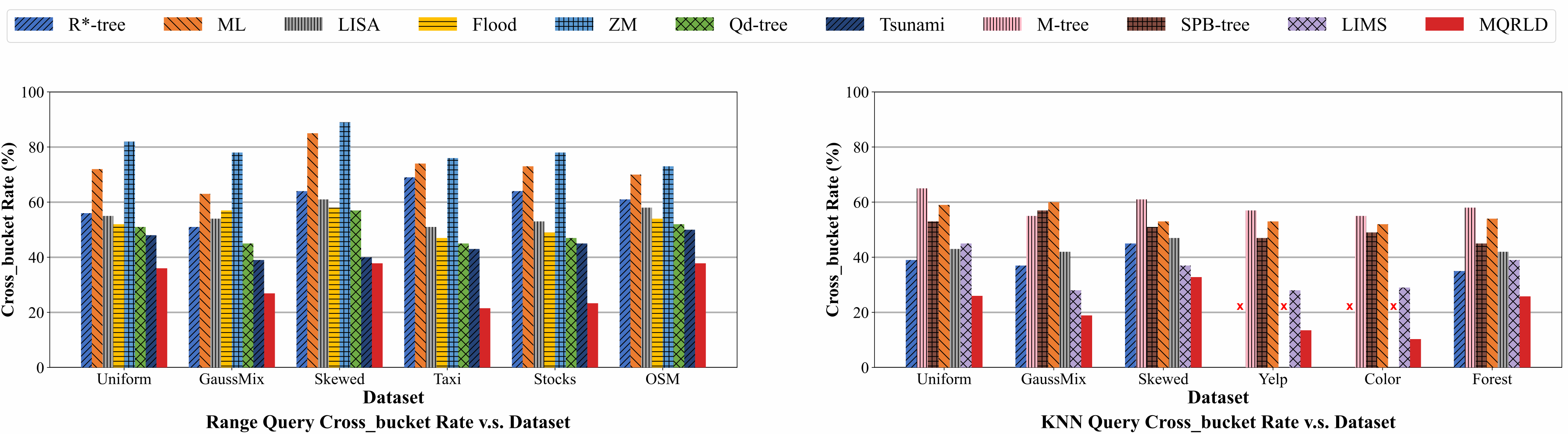}
    \caption{Cross-bucket rate evaluation results of range query and KNN query for multi-dimensional indexes on different datasets. Due to incomparable results beyond 12 dimensions for R*-tree and LISA, we disregard them on Yelp and Color.}
    \label{fig:Exp-CBR}
\end{figure}

In this section, we investigate the CBR of several multi-dimensional indexes on both real and synthetic datasets. 
Fig \ref{fig:Exp-CBR} reports the CBR for range queries across different datasets. 
We observe that MQRLD exhibits the lowest CBR across various datasets. The tree-based indexes, R*-tree and Qd-tree, show relatively less CBR, demonstrating that tree structure has the advantage of pruning a significant amount of irrelevant data regions. 
Flood uses grid indexes that also effectively filter out irrelevant data through grid partitioning, while Tsunami, which combines tree structures and grid indexes, further leverages these advantages to minimize CBR during queries. 
ZM and ML, however, show higher CBR across all datasets. This is due to the sequential scanning required on one-dimensional mapped values, compounded by the drawbacks of Z-order curves and prediction errors in learning models. Compared to these indexes, MQRLD combines the structural advantages of tree indexes with our index optimization strategy that ensures frequently accessed nodes are kept in positions that are more cache-friendly, allowing most queries to be completed by scanning only a few leaf nodes.

\begin{figure}[pos=!h]
    \centering
    \includegraphics[width=\textwidth]{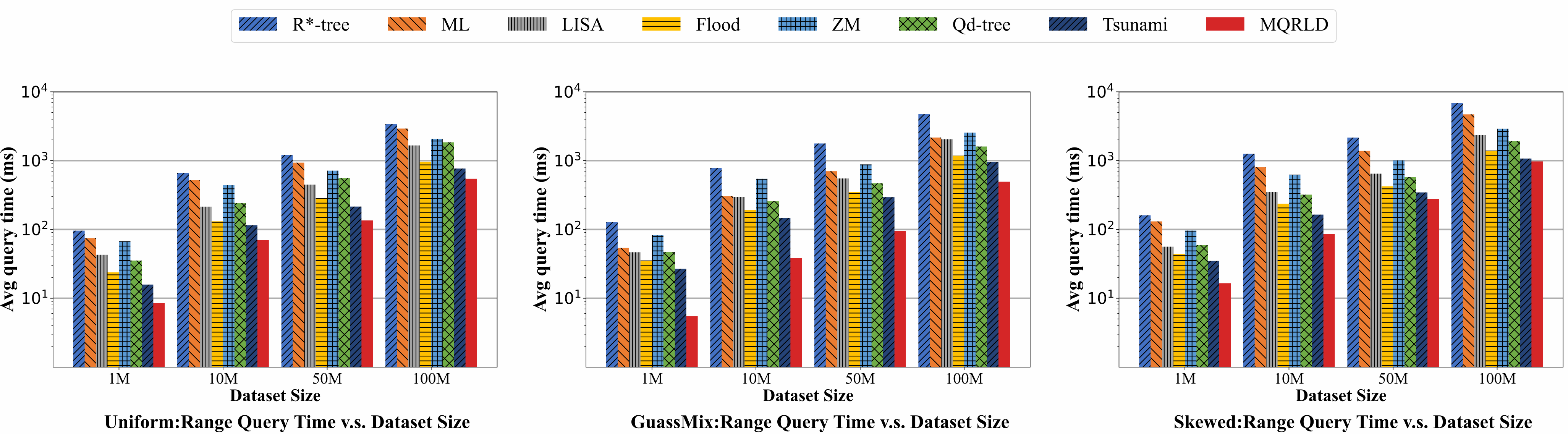}
\end{figure}
\vspace{-2em}
\begin{figure}[pos=!h]
    \centering
    \includegraphics[width=\textwidth]{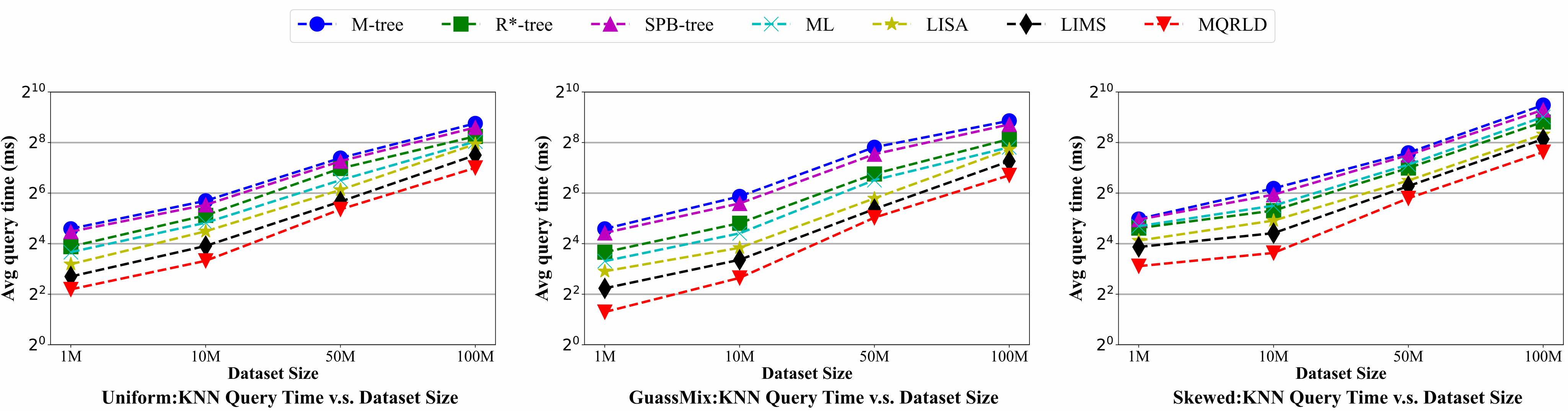}
    \caption{Scalability evaluation results w.r.t. dataset size for multi-dimensional indexes.}
    \label{fig:Exp-DataSize}
\end{figure}

\subsection{Evaluation of Scalability for Multi-dimensional Indexes}
\noindent\textbf{Dataset Size} 

To show how MQRLD scales with dataset size, we create synthetic datasets at sizes of \{1M,10M,50M,100M\}. We train and evaluate these datasets with the same train and test workloads as the full dataset. Fig \ref{fig:Exp-DataSize} shows the results of each index on different-sized datasets, where average query time for all indexes exhibits a nearly linear relationship with dataset size, with MQRLD demonstrating the best query performance overall. 

In the range query experiments, for the Uniform and Skewed datasets, which lack distinct clustering characteristics, the query efficiency decreases. However, MQRLD's data representation strategy, which involves repositioning data points in hyperspace, enhances clustering characteristics and mitigates the impact of data distribution.

In the KNN query experiments, we can observe that the query efficiency of MQRLD outperforms other indexes by 1.6× to 5.4×. Unlike other tree indexes, MQRLD's tree depth and the number of leaf nodes do not significantly increase as the data volume increases. This allows MQRLD to effectively prune a large amount of irrelevant data during KNN queries. Additionally, the prediction models within its leaf nodes further speed up locating the lookup data points.\\

\begin{figure}[pos=!h]
    \centering
    \includegraphics[width=\textwidth]{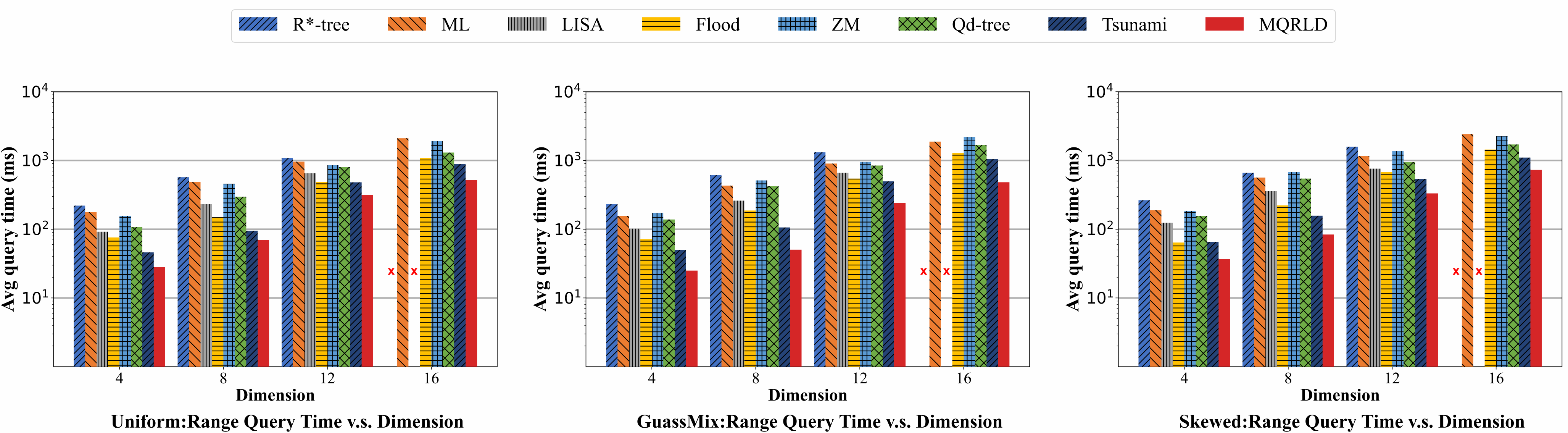}
\end{figure}
\vspace{-2em}
\begin{figure}[pos=!h]
    \centering
    \includegraphics[width=\textwidth]{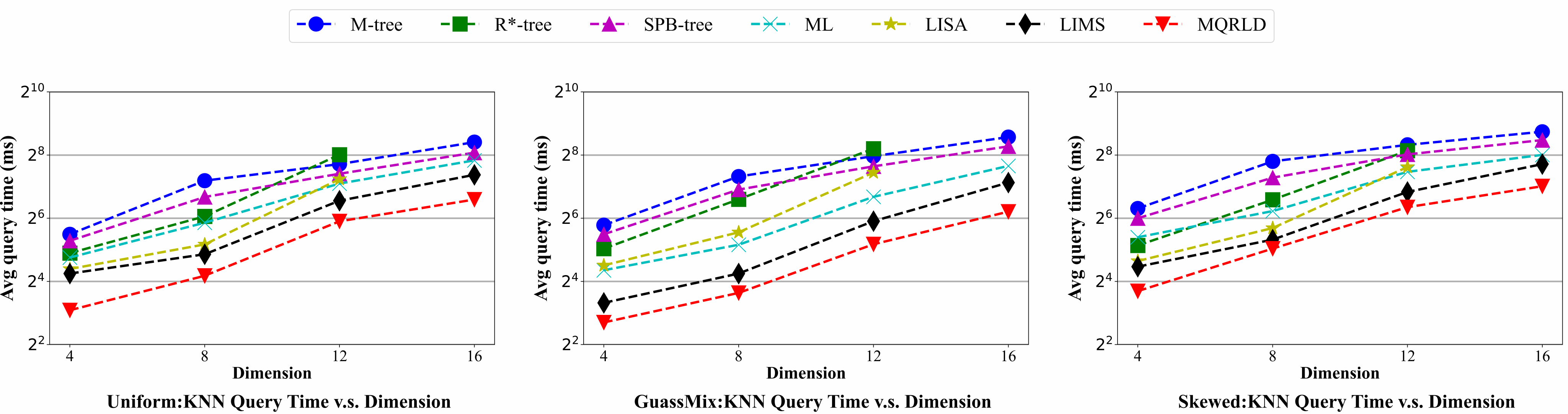}
    \caption{Scalability evaluation results w.r.t. dimension for multi-dimensional indexes. Due to incomparable results beyond 12 dimensions for R*-tree and LISA, we disregard them on experiments of 16 dimensions.}
    \label{fig:Exp-Dim}
\end{figure}

\noindent\textbf{Number of Dimensions} 

To show how MQRLD scales with dimensions, we create synthetic $d$-dimensional datasets $(d\leq16)$ with 10 million records whose values in each dimension are distributed uniformly at random. Fig \ref{fig:Exp-Dim} shows that MQRLD continues to outperform other indexes at each dimension. The growth rates of MQRLD, LIMS, and ML are lower compared to other indexes, indicating that clustering-based data techniques effectively mitigate the curse of dimensionality phenomenon. R*-tree and LISA maintain efficient query performance in low-dimensional queries. However, their query time increases exponentially with dimensionality, and thus, their results are not recorded beyond 12 dimensions.

Both range and KNN queries on the GaussMix dataset show that the average query time for ML, LIMS, and MQRLD is lower than on other datasets, primarily due to the dataset's inherent clustering-friendly nature. MQRLD's data representation strategy further enhances this clustering characteristic, thereby improving its query efficiency. In range query experiments, the increase in dimensionality significantly raises the number of grids in Flood and Tsunami, which in turn escalates grid filtering time during queries. In contrast, the number of leaf nodes in MQRLD remains unaffected by the dimensionality, ensuring consistent query performance. In KNN query experiments, LIMS shows higher average query times in low-dimensional scenarios. This is due to the limited pruning information available in low dimensions, which increases query costs. Conversely, MQRLD's pruning strategy, based on distance determination, remains stable across various dimensionality, demonstrating more robust performance.

\subsection{Evaluation of Rich Hybrid Queries for Multi-dimensional Indexes}\label{sec7.6}
In this section, we evaluate the performance of rich hybrid queries for multi-dimensional indexes.
Except for MQRLD, other indexes in our experiment do not support rich hybrid queries and can only search a single vector, so we combine them to simulate the execution of rich hybrid queries. These combinations include six different hybrid indexes: M-tree $\times$ N, SPB-tree $\times$ N, ML $\times$ N, LIMS $\times$ N, LIMS + Flood, and LIMS + Tsunami. $\times$ N means a combination of N instances of the same index (N $\in$ [2,5]). Note that these combined indexes execute queries sequentially. For example, if ML $\times$ 2 is used to perform a V.K $\bigoplus$ V.R query, we will first use one ML index to execute the V.K query, then build another ML index to perform the V.R query.

Fig \ref{fig:Exp-hybrid} presents the results of the average query times for these rich hybrid queries. We can observe that MQRLD consistently achieves the lowest average query times across all experiments. In experiments for three typical rich hybrid queries (V.R $\bigoplus$ N.R, N.R $\bigoplus$ V.K, and V.R $\bigoplus$ V.K), MQRLD improves query performance by 2.8×, 1.5×, and 3.2× on average, respectively, compared to the next best-performing index, LIMS  $\times$ 2.
Notably, in the V.R  $\times$ N experiment, while the average query time for all indexes increases as the number of columns grows, MQRLD demonstrates the smallest increase. Its query performance is, on average, 4.7× better than that of the next best-performing index, LIMS  $\times$ N. This superiority is due to MQRLD's ability to maintain a single index for multiple feature vectors, whereas other indexes require combinations to complete the query tasks. Additionally, MQRLD's data representation strategy and index optimization algorithm are effective in rich hybrid query scenarios, optimizing data distribution and index structure based on query requirements, thus maintaining high query efficiency.

\begin{figure}[pos=!h]
    \centering
    \includegraphics[width=1.0\textwidth]{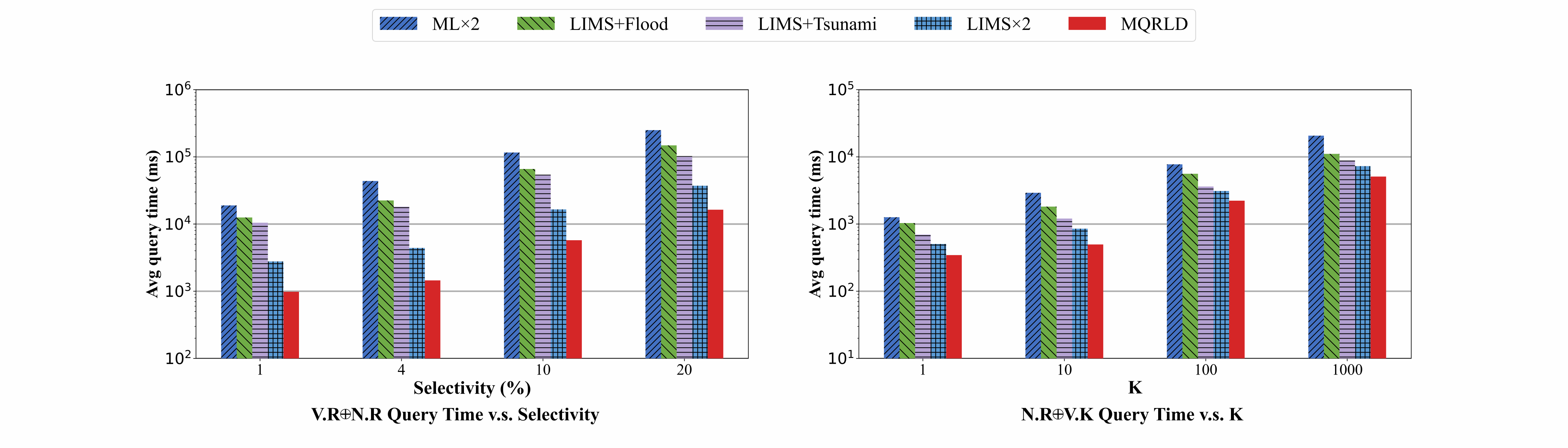}
\end{figure}
\vspace{-2em}
\begin{figure}[pos=!h]
    \centering
    \includegraphics[width=1.0\textwidth]{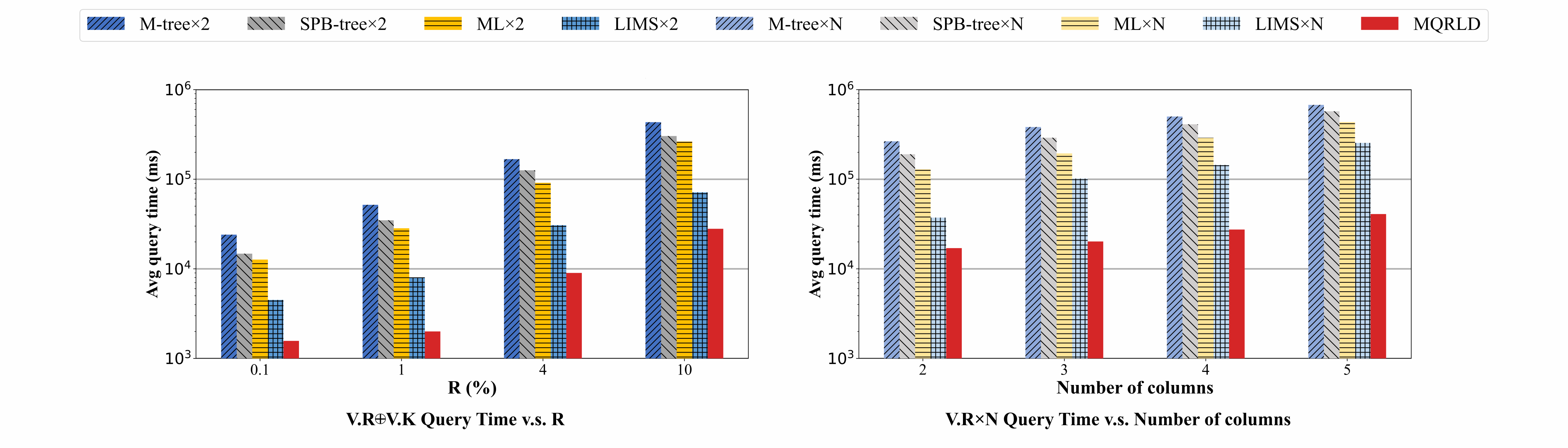}
    \caption{Rich hybrid query evaluation results for multi-dimensional indexes.}
    \label{fig:Exp-hybrid}
\end{figure}

\subsection{Evaluation of KNN Query for High-dimensional Indexes}
\begin{figure}[pos=!h]
	
	\begin{minipage}{0.41\linewidth}
		\vspace{3pt}
		\centerline{\includegraphics[width=\textwidth]{23_1.pdf}}
	\end{minipage}
	\begin{minipage}{0.58\linewidth}
		\vspace{3pt}
		\centerline{\includegraphics[width=\textwidth]{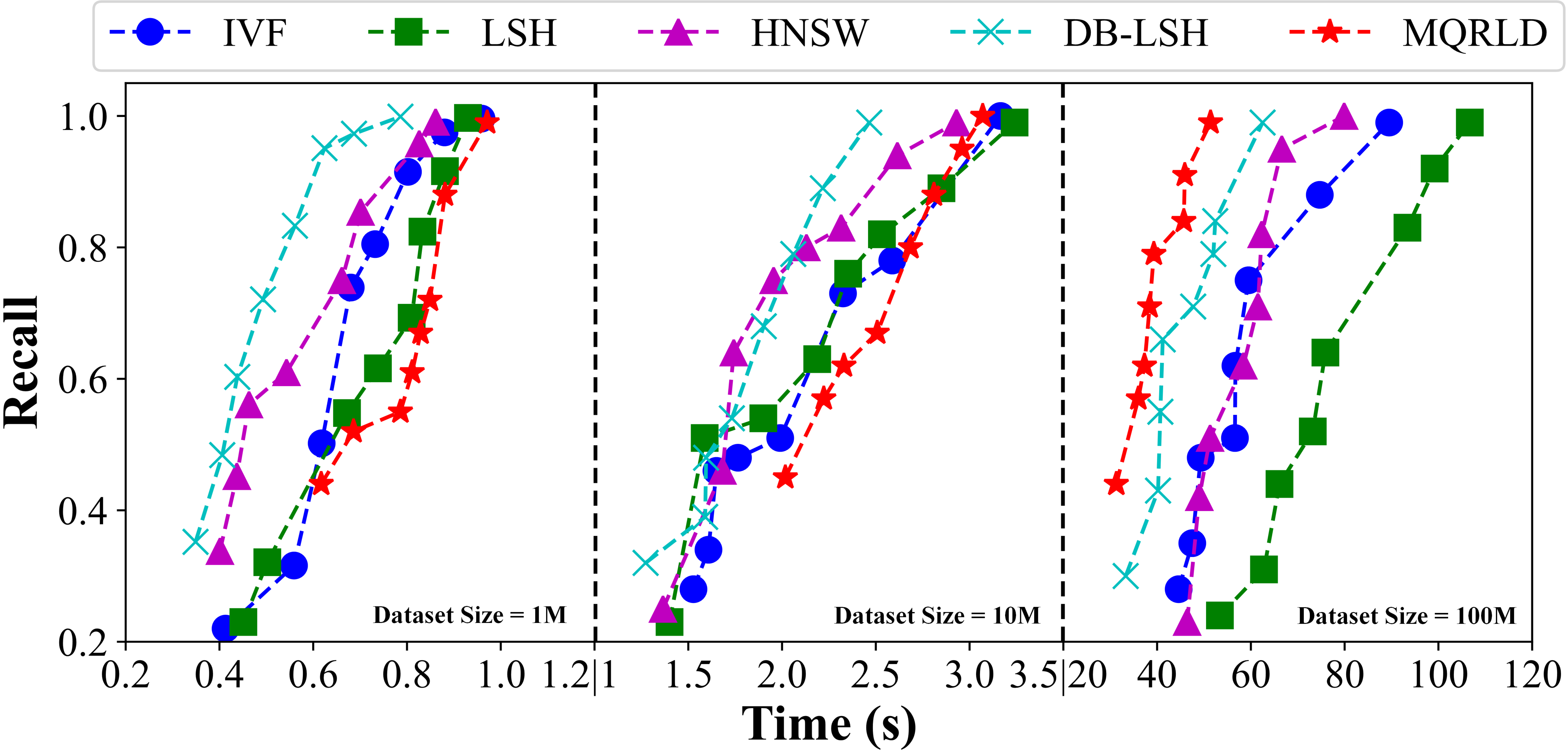}}
	\end{minipage}
 
	\caption{KNN query evaluation results for high-dimensional indexes.}
	\label{fig:KNNONHI}
\end{figure}

To fully evaluate the KNN query performance of MQRLD against other high-dimensional indexes on datasets of varying sizes, we select the SIFT1B dataset with sizes of {1M, 10M, 100M} for comparison. We compare the average query time and the recall-time curve of different high-dimensional indexes. As shown in the left figure of Fig \ref{fig:KNNONHI}, MQRLD's advantage in both query time and recall efficiency becomes increasingly apparent as the dataset grows larger. On the dataset of size 100M, MQRLD's average query time is 1.3× to 2.8× faster than other high-dimensional indexes, and it takes the least time to reach the same recall. Other high-dimensional indexes' data structures are sensitive to increasing data volumes, such as the DB-LSH and LSH hash table structures, the clustering structures included in IVF, and the HNSW single-layer graph structures, resulting in their query times increasing rapidly as the data size grows. In contrast, MQRLD's structure is robust in the face of increasing data volume, due to the predictive models maintained in the leaf nodes that allow for rapid filtering of large amounts of data during queries. At the same time, the high-precision predictive models combined with efficient multimodal data representation strategies give MQRLD a significant advantage over other vector indexes in the time to reach the same recall on large-scale datasets.

\subsection{Evaluation of Rich Hybrid Queries for High-dimensional Indexes} 
\begin{figure}[pos=!h]
	
	\begin{minipage}{0.41\linewidth}
		\vspace{3pt}
		\centerline{\includegraphics[width=\textwidth]{24_1.pdf}}
	\end{minipage}
	\begin{minipage}{0.58\linewidth}
		\vspace{3pt}
		\centerline{\includegraphics[width=\textwidth]{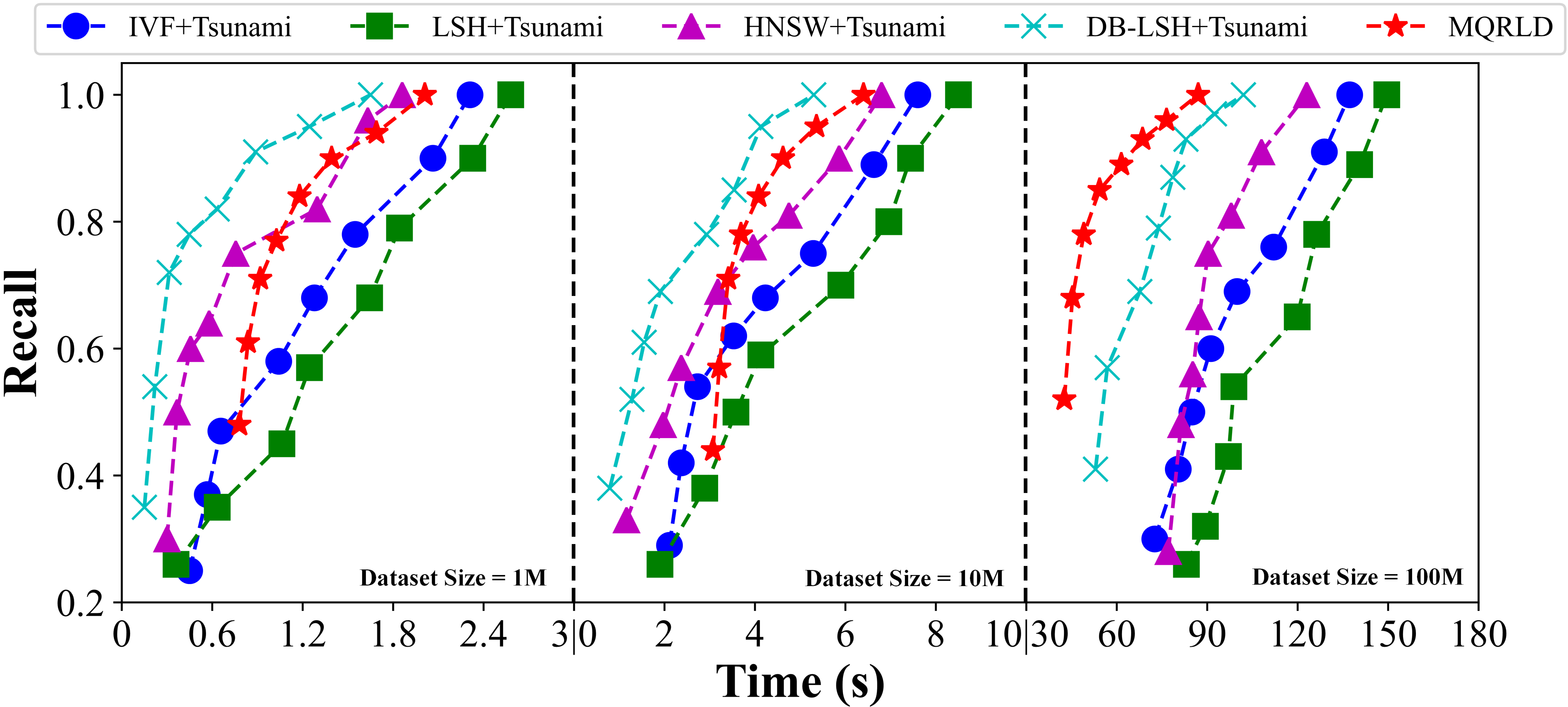}}
	\end{minipage}
 
	\caption{Hybrid query evaluation results for high-dimensional indexes.}
	\label{fig:HQONHI}
\end{figure}

In this section, we evaluate the performance of various high-dimensional indexes for rich hybrid queries. Except for MQRLD, other high-dimensional indexes only support vector queries. 
Therefore, we combine these indexes with Tsunami to evaluate rich hybrid queries, similar to Section \ref{sec7.6}. The experimental dataset is a subset of the LAION400M dataset with sizes of {1M, 10M, 100M}.
Fig \ref{fig:HQONHI} shows the average query time and recall-time curve for different indexes. It can be observed that MQRLD significantly outperforms other combined indexes on the 100M datasets. On average, MQRLD's query time is 1.2× faster than the sub-optimal index DB-LSH+Tsunami, and its recall-time curve demonstrates that it consistently achieves the same recall faster than other indexes. These experimental results highlight that MQRLD not only supports rich hybrid queries but also substantially enhances query performance for high-dimensional data, especially on the largest dataset.

\subsection{Discussions} \label{discussion}
In this section, to clarify some potential concerns that have not been mentioned before, we will discuss the trade-off between query workload and data appending, time and space complexity associated with MQRLD, and demonstrate the effectiveness of our proposed optimization strategies through ablation experiments. Furthermore, all experiments conducted to verify these concerns utilize the GuassMix dataset as the experimental data.\\

\noindent \textbf{The Trade-off Between Query Workload and Data Appending}

The trade-off between multimodal data query workload (considered as read operations) and data appending (considered as write operations) is a hot topic,  which is always analyzed in specific application scenarios. 
For multimodal data retrieval platforms, the performance of data queries has a direct impact on downstream tasks. As a result, ensuring the performance of the query workload is more critical than supporting data appending. The platform’s ability to handle query workload primarily depends on the multi-dimensional or high-dimensional indexing methods it employs, as discussed in Section \ref{RelatedWork}. 
Most existing methods focus on achieving fast and accurate retrieval without analyzing continuous data growing (e.g., Flood, Tsunami, LISA, ML, ZM, RSMI, LMSFC, etc). Only a few studies can support the functionality of data appending (e.g., HNSW, IVFADC, E2LSH, DB-LSH, etc). Given that large-scale multimodal data is accumulated over long periods, with the data appending frequency being low and primarily occurring in batch forms, we offline established a high-dimensional learned index for MQRLD to support high-performance retrieval and preserve its functionality of data appending. Since the learned index structure is based on a cluster tree, data appending can also be achieved through search operations. By querying new data and identifying its nearest neighbor (i.e., equivalent to a KNN search, K=1), we can determine the most likely insert location for it. As shown in Fig \ref{fig-ziEx} (b) and (c), the index structure changes only when the data grows exponentially. Therefore, even when data is frequently appending, the index structure remains stable. However, it is undeniable that if new excessive data accumulates, the retrieval performance will deteriorate, requiring index reconstruction when the query accuracy falls below a predefined threshold. \\

\noindent \textbf{The Time Cost of QBS Table Construction}

In this paper, the QBS table has a significant impact on the query-aware mechanism. However, since the calculation of statistics variables Recall@K and Query Accuracy in the QBS table is time-consuming, computing the statistics information corresponding to all query statements may result in significant time costs. Therefore, we sample query statements during the query process and calculate their corresponding statistics information to construct the QBS table. \\

\noindent \textbf{The Time Cost of Feature Measurement}

In feature measurement, we select the embedding model with the highest score by calculating the measurement metrics across different embedding models. However, due to the computations of SC, FID, and extrinsic measurement metric (see details in Section \ref{section:eval}) being quite slow, performing feature measurement on the entire dataset would also result in a heavy workload. Therefore, similar to QBS table construction, we calculate the measurement metrics by sampling the data (for calculating SC and FID) and queries (for calculating extrinsic measurement metrics) to select the embedding model with the highest score.\\

\noindent \textbf{The Time Complexity of A Single Query}

Assuming a multimodal dataset contains $N$ records, the complexity of the traverse time from the root node to the leaf node is bounded by \( O(logN) \), while the complexity of the "last-mile" search using the linear regression model inside the leaf node is \( O(1) \). Therefore, the time complexity of a single query is bounded by \( O(logN) \).\\

\noindent \textbf{The Time Complexity of Index Construction and Experiments}

In each iteration of the Divisive Hierarchical Clustering process, DPC calculates the density and distance of each data point to others in order to find cluster centers, resulting in a complexity of \( O(N^2) \). The number of iterations is bounded by \( O(logN) \), so the overall time complexity for the Divisive Hierarchical Clustering phase is bounded by \( O(N^2logN) \). Then, for each leaf node, a linear regression model is trained. The training step for all leaf nodes has a time complexity of \( O(N) \). The overall time complexity of the indexing construction is dominated by the Divisive Hierarchical Clustering process, with a complexity bounded by \( O(N^2logN) \). Fig \ref{fig:ablation}(a) illustrates the index construction times for multi-dimensional and high-dimensional indexes. The multi-dimensional indexes are evaluated using the GaussMix dataset, where we observe that MQRLD's index construction time is comparable to that of other multi-dimensional indexes. As for the high-dimensional indexes, the SIFT10M dataset is employed, and we observe that the index construction times of high-dimensional indexes are lower than multi-dimensional indexes in general. \\



\noindent \textbf{The Space Complexity of Index and Experiments}

Since our clustering algorithm ensures that intermediate nodes in the cluster tree have at least two child nodes, the cluster tree will, in the worst case, degrade into a full binary tree with $N$ leaf nodes. The total number of non-leaf nodes in such a tree is $N-1$.
Leaf nodes store a tuple containing its centroid, 
radius, and a linear regression model, requiring \( O(N) \) space.
Non-leaf nodes store a tuple consisting of its centroid, radius, and a pointer, also requiring \( O(N) \) space. Therefore, the total space complexity of our index is bounded by \( O(N) \).
However, due to the efficiency of our iterative and divisive approach, the tree depth is low, and the total number of nodes is significantly less than $2N-1$ (as shown in Fig \ref{fig-ziEx} (b) and (c)), making our index structure practically very small. Fig \ref{fig:ablation}(b) shows the index size of the index structure we constructed on datasets with varying data volumes. We can see that MQRLD saves an average of 65.7\%-90.1\% space compared to multi-dimensional indexes and an average of 81.4\%-99.7\% compared to high-dimensional indexes.\\

\begin{figure}[pos=!h]
	
	\begin{minipage}{0.32\linewidth}
		\centerline{\includegraphics[width=0.9\textwidth]{25_a.pdf}}
		\centerline{(a)}
	\end{minipage}
	\begin{minipage}{0.32\linewidth}
		\raggedright{\includegraphics[width=0.9\textwidth]{25_b.pdf}}
		\centerline{(b)}
	\end{minipage}
        \begin{minipage}{0.32\linewidth}
		\raggedright{\includegraphics[width=0.9\textwidth]{25_c.pdf}}
		\centerline{(c)}
	\end{minipage}
 
	\caption{The time cost of different index construction(a), The space cost of different index construction(b), and ablation study(c).}
	\label{fig:ablation}
\end{figure}
\noindent \textbf{Experiments of Deploying MQRLD on Vector Database}

\begin{figure}[pos=!h]
	
	\begin{minipage}{0.41\linewidth}
		\vspace{3pt}
		\centerline{\includegraphics[width=\textwidth]{25_1.pdf}}
	\end{minipage}
	\begin{minipage}{0.58\linewidth}
		\vspace{3pt}
		\centerline{\includegraphics[width=\textwidth]{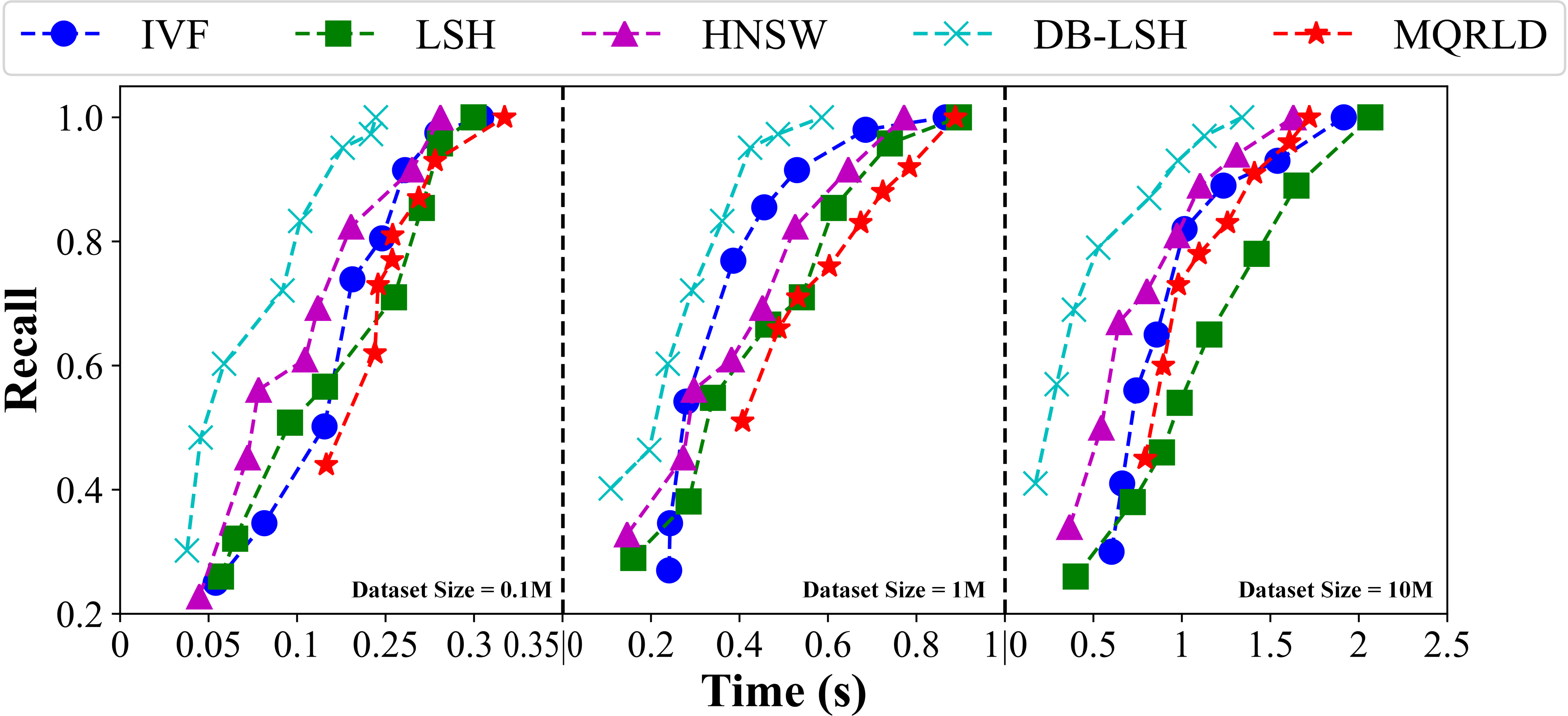}}
	\end{minipage}
 
	\caption{Evaluation results for high-dimensional indexes on vector database.}
	\label{fig:KNNONVD}
\end{figure}

In this section, we apply MQRLD to the vector database Milvus and compare its KNN query performance with high-dimensional indexes HNSW, IVF, LSH, and DB-LSH. The experimental dataset uses the SIFT1B dataset with sizes of {0.1M, 1M, 10M}. Due to the limitations of the Milvus database in our experiment environment, we cannot use larger datasets to evaluate query performance.
The left figure of Fig \ref{fig:KNNONVD} shows the average query time for MQRLD and different high-dimensional indexes at various dataset sizes. We can see that as the data scale increases, the advantage of MQRLD gradually becomes apparent. On the dataset of size 10M, MQRLD's average query time is significantly lower than IVF and LSH. The increase in data scales of the same order of magnitude does not significantly impact the tree depth of MQRLD's cluster tree, thereby resulting in a negligible increase in MQRLD's query time. The right figure of Fig \ref{fig:KNNONVD} shows the recall-time curves for MQRLD and different high-dimensional indexes at various dataset sizes. Compared to other indexes, MQRLD takes the least time to achieve a 100\% recall.\\

\noindent \textbf{Ablation Experiments}

To explore the impact of different components on MQRLD performance, we evaluate the performance of range queries using MQRLD in various states. Fig \ref{fig:ablation}(b) provides a detailed illustration of the average query times for different query optimized methods, in which "Full Scan" refers to scanning the entire dataset to implement range query, "Initialized\_MQRLD" indicates querying with the initialized MQRLD without index optimization, "Optimized\_T" indicates querying with MQRLD optimized after feature representation, and "Optimized\_Index" indicates querying with MQRLD optimized using index structures.  It is noteworthy that the sampling frequency used at different optimization stages is the minimum effective value, which is 10\%. We believe that as the sampling frequency increases, the performance of MQRLD will continue to improve.

From the experimental results, we can draw the following conclusions: 

(1) The results of "Full Scan" and "Initialized\_MQRLD" shows that our data representation strategy and index structure significantly reduce query times, even in the initial state. 

(2) The results of "Initialized\_MQRLD" and "Optimized\_T" validate the effectiveness of our feature representation method, with the optimized matrix T further enhancing query performance.

(3) The results of "Optimized\_T" and "Optimized\_Index" validate the effectiveness of our index structure, with the optimized index structure effectively reducing the scanning of irrelevant data buckets and reducing query time.

(4) Overall experimental results demonstrate the effectiveness of our query-aware mechanism, with the optimized matrix T and index structure further enhancing query performance.

\section{Conclusion}


In this study, we introduce a multimodal data retrieval platform MQRLD. This platform builds on a data lake that offers transparent data storage and integrates a multimodal open API for rich hybrid queries. In addition to these capabilities, its multimodal data representation strategy transforms data into effective feature vectors through feature embedding, measurement, and enhancement, aiding accurate data retrieval. To further improve query efficiency, high-dimensional learned indexes are built using divisive hierarchical clustering and cluster tree construction, with the query-aware mechanism optimizing the index structure. Overall, MQRLD not only supports transparent data storage and rich hybrid queries but also ensures effective and efficient retrieval.

Although our proposed MQRLD addresses the functionalities of transparent storage and rich hybrid queries, and enhances query performance, as mentioned in Section \ref{section:intro}, it still faces additional challenges: (1) The MQRLD platform mainly consists of two modules: feature representation and high-dimensional learned index. Experiments have shown that these modules can effectively improve the retrieval performance of data lakes. However, the scalability of them on other platforms still needs further discussion. (2) Despite the large accumulation of multimodal data in the form of historical records, the development of Internet of Things (IoT) technologies supports multimodal data in a streaming format. Furthermore, user query behaviors often exhibit varying preferences over time. Therefore, to maintain the platform retrieval performance, whether it can support dynamic index reconstruction when facing frequent data appending and different query workloads  remains a challenge. (3) Large language models (LLMs) have demonstrated their high scalability in different fields. It remains to be explored whether MQRLD can leverage LLMs to achieve more efficient and high-quality feature representation. (4) Although multimodal data are primarily accumulated in batch form, it is reasonable to assume that more recent data are likely to exhibit more similar characteristics. Additionally, query workloads within a specific time period are likely to exhibit similar preferences. Therefore,  it is worth exploring whether temporal information can be utilized by the platform to enhance feature representation and index construction.

We can further optimize our platform in the future, including: (1) We consider applying our multimodal data feature representation strategy and high-dimensional learned index to other platforms to enhance their multimodal data retrieval performance while preserving their inherent characteristics as discussed in Section \ref{section:intro}. (2) Existing researches typically maintain retrieval performance under frequent data appending and different query workloads by monitoring query results to determine whether the index needs to be reconstructed \citep{MTO, SAT}. So how to determine the threshold for triggering index reconstruction is an opportunity worth exploring. Moreover, to avoid the expensive cost of reconstructing the entire index, it is worth considering whether partial index optimizations can help balance the trade-off among online data appending, query workloads, and index reconstruction. (3) LLMs demonstrate significant potential in representation learning. Currently, the feature embedding models pool in our feature representation includes the popular CLIP-based models \citep{clip, CLIP4Clip, Xclip}. However, these models do not support a unified training representation across various modalities. Exploring the possibility of leveraging LLMs to offer a unified embedding strategy of different modalities presents an interesting opportunity for further investigation. (4) \cite{TMac} shows that incorporating temporal information into multimodal graph learning can effectively aid multimodal data classification tasks, which may help with feature representation (aims to gather similar multimodal data). Additionally, \cite{S2T} indicates that temporal information can enhance the prediction of interactions, supporting the implementation of online query workloads on our platform (allowing the platform to dynamically adjust the index in advance based on predicted query workload \citep{workloadPredict}). Therefore, leveraging temporal information to optimize our platform functionalities and performance presents a valuable opportunity.

\printcredits
\section*{Acknowledgements}
The work is funded by the National Natural Science Foundation of China under Grant 42371480.
\bibliographystyle{cas-model2-names}

\bibliography{refs}





\end{document}